\documentclass{aa}

\usepackage{natbib}
\bibpunct{(}{)}{;}{a}{}{,} 
\usepackage{ae}
\usepackage{bm}
\usepackage{amsmath}
\usepackage{amssymb}
\usepackage{latexsym}
\usepackage{epsfig,graphicx}
\usepackage{tabularx}
\usepackage{subfig}
\usepackage{float}
\usepackage{color}
\usepackage{hyperref}
\usepackage{enumitem}
\usepackage[utf8]{inputenc}
\usepackage{cleveref}
\crefname{section}{\S}{\S\S}
\usepackage[T1]{fontenc} 
\usepackage[utf8]{inputenc}
\usepackage{csvsimple}
\usepackage{aecompl}
\graphicspath{{}{figures/}}
\definecolor{purple}{rgb}{0.5, 0.0, 0.5}

\def\reff@jnl#1{{\rm#1\/}}

\def\aj{\reff@jnl{AJ}}                  
\def\araa{\reff@jnl{ARA\&A}}            
\def\apj{\reff@jnl{ApJ}}                
\def\apjl{\reff@jnl{ApJ}}               
\def\apjs{\reff@jnl{ApJS}}              
\def\apss{\reff@jnl{Ap\&SS}}            
\def\aap{\reff@jnl{A\&A}}               
\def\aapr{\reff@jnl{A\&A~Rev.}}         
\def\aaps{\reff@jnl{A\&AS}}             
\def\baas{\reff@jnl{BAAS}}              
\def\jrasc{\reff@jnl{JRASC}}            
\def\memras{\reff@jnl{MmRAS}}           
\def\mnras{\reff@jnl{MNRAS}}            
\def\physrep{\reff@jnl{Phys.Rep.}}
\def\pra{\reff@jnl{Phys.Rev.A}}         
\def\prb{\reff@jnl{Phys.Rev.B}}         
\def\prc{\reff@jnl{Phys.Rev.C}}         
\def\prd{\reff@jnl{Phys.Rev.D}}         
\def\prl{\reff@jnl{Phys.Rev.Lett}}      
\def\pasp{\reff@jnl{PASP}}              
\def\pasj{\reff@jnl{PASJ}}              
\def\skytel{\reff@jnl{S\&T}}            
\def\solphys{\reff@jnl{Solar~Phys.}}    
\def\sovast{\reff@jnl{Soviet~Ast.}}     
\def\ssr{\reff@jnl{Space~Sci.Rev.}}     
\def\nat{\reff@jnl{Nature}}             

\def\farcs{\hbox{$.\!\!^{\prime\prime}$}}

\newcommand{\beq}{\begin{equation}}
\newcommand{\eeq}{\end{equation}}
\newcommand{\beqa}{\begin{eqnarray}}
\newcommand{\eeqa}{\end{eqnarray}}

\newcommand{\rmd}{\mathrm{d}}

\newcommand{\newtext}[1]{\textcolor{black}{#1}}

\newcommand{\sersicn}{S\'{e}rsic $n$}
\newcommand{\sersic}{S\'{e}rsic}
\newcommand{\avg}[1]{\ensuremath{\left\langle #1 \right\rangle}}
\newcommand{\euclid}{\emph{Euclid}}
\newcommand{\lensfit}{\emph{lens}fit}

\newcommand{\sextractor}{\textsc{{\large SE}xtractor}}
\newcommand{\swarp}{\textsc{{\large SW}arp}}

\newcommand{\school}{SCHOo\emph{l}}
\newcommand{\college}{COllege}
\newcommand{\galsim}{\textsc{GalSim}}
\newcommand{\galfit}{\textsc{GalFit}}
\newcommand{\imtshape}{{\sc im3shape}}

\defcitealias{FC17}{FC17}
\defcitealias{Hildebrandt17}{H17}
\defcitealias{Hildebrandt2019}{H18}
\defcitealias{Zuntz2018}{Z18}
\defcitealias{Mandelbaum2018a}{M18a}

\date{\today}
\begin{document}

   \title{Towards emulating cosmic shear data: Revisiting the calibration of the shear measurements for the Kilo-Degree Survey}

   \author{Arun Kannawadi\inst{1}
          \and
          Henk Hoekstra\inst{1}\and      
          Lance Miller\inst{2} \and
          Massimo Viola\inst{1}\and
          Ian Fenech Conti\inst{3,4} \and
          Ricardo Herbonnet\inst{5,1}\and
          Thomas Erben \inst{6} \and
          Catherine Heymans\inst{7}\and
          Hendrik Hildebrandt\inst{8,6} \and
          Konrad Kuijken\inst{1}\and
          Mohammadjavad Vakili\inst{1} \and
          Angus H. Wright\inst{6,8}
          }

   \institute{Leiden Observatory, Leiden University, PO Box 9513, 2300 RA, Leiden, The Netherlands\\ \email{arunkannawadi@strw.leidenuniv.nl}
         \and
Department of Physics, University of Oxford, Keble Road, Oxford OX1 3RH, UK
        \and
Institute of Space Sciences and Astronomy, University of Malta, Msida, MSD 2080, Malta
        \and
Department of Physics, University of Malta, Msida, MSD 2080, Malta
        \and
Department of Physics and Astronomy, Stony Brook University, Stony Brook, NY 11794, USA
    \and
Argelander-Institut f\"{u}r Astronomie, Auf dem H\"{u}gel 71, 53121 Bonn, Germany
    \and
Institute for Astronomy, University of Edinburgh, Royal Observatory, Blackford Hill, Edinburgh EH9 3HJ, UK
    \and
Astronomisches  Institut,  Ruhr-Universit\"{a}t  Bochum,  Universit\"{a}tsstr.  150,  44801,  Bochum,  Germany
             }
             
\titlerunning{Emulating cosmic shear data and calibrating shear for KiDS}
    
\abstract{
Exploiting the full statistical power of future cosmic shear surveys will necessitate improvements to the accuracy with which the gravitational lensing signal is measured. We present a framework for calibrating shear with image simulations that demonstrates the importance of including realistic correlations between galaxy morphology, size, and more importantly, photometric redshifts. This realism is essential to ensure that selection and shape measurement biases can be calibrated accurately for a tomographic cosmic shear analysis. We emulate Kilo-Degree Survey (KiDS) observations of the COSMOS field using morphological information from {\it Hubble} Space Telescope imaging, faithfully reproducing the measured galaxy properties from KiDS observations of the same field. We calibrate our shear measurements from \lensfit, and find through a range of sensitivity tests that \lensfit\ is robust and unbiased within the allowed two per cent tolerance of our study. Our results show that the calibration has to be performed by selecting the tomographic samples in the simulations, consistent with the actual cosmic shear analysis, because the joint distributions of galaxy properties are found to vary with redshift. Ignoring this redshift variation could result in misestimating the shear bias by an amount that exceeds the allowed tolerance. To improve the calibration for future cosmic shear analyses, it will also be essential to correctly account for the measurement of photometric redshifts, which requires simulating multi-band observations.

}

\bibliographystyle{aa}
\maketitle



\section{Introduction}
The observed distribution of matter in the Universe is determined by the interplay between the expansion history, its composition, and the laws of gravity that govern the evolution of cosmic structure. Consequently, the growth of large-scale structure encodes key information about the origin and nature of the key ingredients in the Universe. One complication is that most of the matter is invisible, and can only be inferred indirectly through its gravitational pull. One observable consequence is the distortion of space-time, which results in correlations in the ellipticities of distant galaxies, a phenomenon called `weak gravitational lensing' \citep[see e.g.][for recent reviews]{Kilbinger15,Mandelbaum2018}. 

The cosmological lensing signal is now routinely measured \citep[e.g.][]{Heymans13,Jee2013,Jee16, Hildebrandt17,Troxel2018, Hikage2018}. Moreover, thanks to the significant increase in survey area and improvements in the determination of photometric redshifts (or photo-$z$s) of the sources, cosmic shear results are starting to yield competitive constraints on cosmological parameters \citep[e.g.][]{Joudaki18, vanUitert18, Baxter2019}. The amplitude of the lensing signal is largely determined by a combination of $\sigma_8$, the normalisation of the matter fluctuations, and $\Omega_{\rm m}$, the mean matter density. Ongoing lensing surveys have therefore reported the constraints on $S_8\equiv\sigma_8\sqrt{\Omega_{\rm m}/0.3}$ as their main result. Once completed, these surveys will constrain $S_8$ with a precision that is comparable to the most recent measurements from the cosmic microwave background (CMB) radiation \citep{Planck18}. 

Compared to the CMB constraints, the weak lensing results favour somewhat lower values for $S_8$~\citep{Joudaki2017c,Hildebrandt17,Leauthaud2017,Troxel2018,Hikage2018}. A relevant question is whether this could be caused by biases in the estimates of the shear signal. An important step in shear measurement is to correct for the blurring by the atmosphere and telescope optics, which modifies the shapes of the faint galaxies that are used to infer the lensing signal. In particular, the  finite width of the point spread function (PSF) makes the images rounder, thus lowering the signal~\citep[e.g.,][]{KSB95}. If this is not correctly accounted for, the resulting cosmological parameter estimates will be biased. Moreover, anisotropy in the PSF introduces alignments in the shapes that can dwarf the cosmological signal. A straightforward correction for the blurring by the PSF is not possible because the images are noisy.

To exploit the full potential of current cosmic shear surveys such as the Kilo-Degree Survey\footnote{http://kids.strw.leidenuniv.nl/} ~\citep[KiDS;][]{DeJong2012}, the Dark Energy Survey\footnote{https://www.darkenergysurvey.org/}~\citep[DES;][]{Diehl2014,Flaugher2015}, and the Hyper-Suprime Cam survey\footnote{https://hsc.mtk.nao.ac.jp/ssp/}~\citep[HSC;][]{Aihara2018},
the improvements in the statistical uncertainties are to be matched by a better understanding of observational and astrophysical sources of bias. This is even more important for the future surveys (Stage IV) such as the Large Synoptic Survey Telescope\footnote{https://www.lsst.org/}~\citep[LSST;][]{ivezic08}, Euclid\footnote{http://sci.esa.int/euclid/}~\citep{EuclidReport} and the Wide-Field Infra-Red Space Telescope\footnote{https://wfirst.gsfc.nasa.gov/}~\citep[WFIRST;][]{Spergel2015}. Fortunately,  the various sources of observational bias are well understood and can be characterised using the available data. Importantly, the resulting (residual) biases can be studied and quantified by applying  the shape measurement algorithm to simulated data, where the galaxy images are sheared by a known amount. 

A number of blind community challenges using galaxy image simulations \citep{STEP1,STEP2,GREAT08,GREAT10,great3} have compared the performance of algorithms, thus improving our understanding of the measurement process. However, with such generic approaches, the complexity and the level of realism present in the real data is limited and not all sources of biases can be captured.  In order to remove the biases in the shear estimated from a specific data set, it is essential that the performance of the algorithm is determined using mock data that are sufficiently realistic \citep{Miller13, Hoekstra15, Samuroff2018}, such that the inferred bias is robust against the uncertainties in the input parameters \citep[e.g.][]{Hoekstra2017}. The fidelity of the image simulations is therefore crucial, not only to quantify biases in the shape measurements but also to correctly capture the selection of galaxies. For instance, \citet[][FC17 hereafter]{FC17} show that selection biases can be comparable to other sources of bias.

One approach is to match the observed properties of the simulated images to those of the real data by modifying the input distributions in case differences are found~\citep[e.g.,][]{Bruderer16}. In this case, the bias can be determined directly from the simulated data. The result, however, depends on the input parameters considered, and different combinations of input parameters may result in observed distributions that are difficult to distinguish but yield different biases. Alternatively, the simulated output can be used to account for differences with the actual data by parameterising the bias as a function of observed galaxy properties, provided there is an overall agreement between the data and the simulations. This approach has been used by a number of weak lensing studies \citep[e.g.,][]{Hoekstra15,Hildebrandt17,Hikage2018}.  Another approach that is gaining traction is metacalibration~\citep{Huff2017,Sheldon2017,Zuntz2018}, which in principle allows any shear measurement method to obtain an unbiased estimate of shear from the data, without requiring image simulations. As we discuss in more detail in~\cref{sec:metacal}, metacalibration cannot quantify all sources of biases however, and we argue that it should be considered somewhat complementary to the image simulations approach we employ here.

In this paper, we revisit the shear calibration for the cosmic shear analysis of the Kilo-Degree Survey \citep[KiDS;][]{deJong2015,Kuijken15}, with an emphasis on creating realistic tomographic samples within the simulations. The cosmological parameter constraints presented in~\citet{Hildebrandt17} (H17 hereafter) were based on the first $\sim 450$ square degrees of observed data. The biases in the shape measurements from \lensfit~\citep{Miller2007,Miller2013} were calibrated using image simulations, described in \citetalias{FC17}, where the input galaxy catalogue was constructed to be consistent with the \lensfit\ priors. The shear biases for the different tomographic bins were determined by resampling the simulated catalogues so that the output distributions matched the observed signal-to-noise ratio and size distributions. \citetalias{FC17} assumed, however, that the galaxy ellipticities do not correlate with other parameters and that those galaxy parameters do not depend explicitly on redshift.

In this paper, we largely follow~\citetalias{FC17}, but introduce a number of (minor) improvements to better reflect the actual data analysis steps. The main difference is the use of a catalogue of galaxies for which structural parameters were determined from {\it Hubble} Space Telescope (HST) imaging, and for which individual redshifts were measured using multi-band photometry. Specifically, we use data from the Cosmic Evolution Survey \citep[COSMOS;][]{Scoville07} with the aim to emulate KiDS observations. Comparison to actual KiDS observations in the same field not only enables us to evaluate the fidelity of our simulated data for the different tomographic bins. In fact, as explained in more detail below, the quality of our simulated data allowed us to identify errors in the implementation of the weighting schemes used in both \citetalias{FC17} and \citetalias{Hildebrandt17}. Under the {reasonable} assumption that the COSMOS galaxy sample is representative of the full survey, we can construct our mock KiDS lensing survey by varying the seeing conditions.

This paper is structured as follows: In \cref{sec:theory}, we present a mathematical framework for shear calibration using image simulations, and discuss the common pitfalls in this context. This motivates several tests carried out throughout the rest of the paper. In \cref{sec:data}, we describe the data and the shape measurement algorithm for which we wish to calibrate the lensing shear. We also briefly describe our image simulations setup and highlight the improvements from the prior work~\citepalias{FC17}. In~\cref{sec:imsim}, we describe the input catalogue for the simulations, which is one of the main focus\newtext{es} of this paper, and show that the simulations match the data very well.~\cref{sec:bias_in_sims} deals with the different sources of selection bias. In~\cref{sec:calibration_results}, we derive the calibration corrections for shear. The sensitivity of our main results to various choices made in our simulations is explored in~\cref{sec:sensitivity_analysis} and we conclude in~\cref{sec:discussion}.

\section{Theory and overview}
\label{sec:theory}
\subsection{An estimator for shear}
The differential deflection of light rays caused by inhomogeneities in the intervening mass distribution results in a distortion in the observed images of distant galaxies. In the limit of weak gravitational lensing, the quantity of interest is the lensing shear $\gamma$, which can be estimated by averaging the ellipticities of a sample of galaxies.

If we denote the intrinsic ellipticity of galaxies by a complex number $\epsilon^\text{int}$, then the lensed ellipticity~\citep{Bartelmann2001} is
$\epsilon^\text{lens} = {(\epsilon^\text{int}+\gamma)}/{(1+\gamma^* \epsilon^\text{int})} \approx \epsilon^\text{int} + \gamma -\gamma^* \left(\epsilon^\text{int}\right)^2$, where the approximation to first order\footnote{{Strictly speaking, the observable is not the shear but another quantity called as reduced shear, but to first order in the lensing potential, they are the same.}} in $\gamma$ holds for small values of $|\gamma|$. If we treat the unknown intrinsic ellipticity as a source of noise, the value of $\epsilon^\text{lens} $  is an unbiased\footnote{This assumes that the galaxies are randomly oriented. Local tidal effects are known to cause intrinsic alignments of galaxies that bias the shear estimate.}, but a noisy estimate of $\gamma$. The challenge for any weak lensing study is thus to obtain accurate measurements of $\epsilon^\text{lens}$.

The shear due to the large-scale structure is typically $\sim 10^{-3}$, while the strength of the shape noise, $\left\langle |\epsilon^{\text{int}}|^2 \right\rangle \sim 0.3$. To reduce the statistical uncertainty in the shear estimate in order to be of any use, the (weighted) average ellipticity for an ensemble of galaxies is used instead. Thus, an estimator for the gravitational shear is
\begin{equation}
\hat{\gamma} = \frac{\sum_g w_g \hat{\epsilon}_g }{\sum_g w_g},
\label{eq:avshear}
\end{equation}
where $g$ labels the galaxies, $\hat{\epsilon}_g$ is the ellipticity measured by a shape measurement algorithm, and $w_g$ is the weight assigned to the galaxy $g$, based on its signal-to-noise ratio, ellipticity, etc.

If $\hat{\gamma}$ is an ideal estimator of the lensing shear $\gamma$, then, by definition, $E(\hat{\gamma}) = \gamma$, where $E$ stands for taking the expectation value over all possible noise realisations. However, simple practical estimators suffer from biases~\citep[see][for some examples]{HS03,Viola2011}. The estimator is then not only a function of the shear, but also depends on the distribution of various intrinsic parameters pertinent to the sample, which we denote as $q_\text{obs}$. In the case of weak gravitational lensing, one can linearise the estimator in $\gamma$ to obtain the standard linear bias model~\citep{STEP1} to obtain
\begin{equation}
E(\hat{\gamma} | q_\text{obs}) = \gamma\left(1+m[q_\text{obs}]\right) + c[q_\text{obs}].
\label{eq:linear_bias_model}
\end{equation}
Here, $c[q_\text{obs}] \equiv E(\gamma = 0 \, | \,  q_\text{obs})$ is the value of the estimator for zero input and is referred to as additive bias and $m[ q_\text{obs}]$
is the linear response of the estimator to the shear and is referred to as multiplicative bias. {Strictly speaking, the multiplicative bias is a $2\times 2$ tensor, but in practice, it is approximately a scalar matrix and is treated as a scalar. For simplicity, we will treat $m[q_\text{obs}]$ as a scalar as well.} Thus, given a biased estimator $\hat{\gamma}$, one can construct an ideal estimator $\hat{\tilde{\gamma}}(\gamma \, |  \, q_\text{obs})$ by calibrating out the biases as
\begin{equation}
\hat{\tilde{\gamma}}(\gamma \  | \,  q_\text{obs} )  = \frac{ \hat{\gamma}(\gamma \, | \, q_\text{obs}) - c[q_\text{obs}]} { ( 1 + m[q_\text{obs}] ) }
\end{equation}
such that $E(\hat{\tilde{\gamma}}) = \gamma$, provided one knows the bias terms precisely.

The presence of an additive bias can be inferred by stacking the shear estimates in an appropriate coordinate frame. For instance, the mean shear across a large survey should vanish. Moreover, the magnitude of the bias can be determined directly from such data combinations. In contrast, the multiplicative bias cannot be determined directly\footnote{See~\cref{sec:metacal} for recent methods that aim to obtain them from data.} since it requires the knowledge of the magnitude of the shear.

Traditionally, the performance of shape measurement algorithms has therefore been evaluated using simulated galaxy images, where the ground truth is known. A series of blind community challenges have benchmarked the performance of various shape measurement methods \citep{STEP1,STEP2,GREAT08,GREAT10,great3}. While such efforts have helped to improve our understanding of the various sources of bias, the results cannot be applied directly to the actual survey data \citep{Hoekstra15}. 
Thus, even after an internal calibration, residual biases may still be present. The true magnitude of this systematic error depends on three factors:
 \begin{enumerate}
     \item the difference in the distributions of parameters that affect the bias between the observations and the simulations (c.f.~\cref{sec:compare_sims_data})
    \item the selection criterion in the simulations (c.f~\cref{sec:bias_tomo}) and
     \item the sensitivity of the bias to the galaxy population (c.f.~\cref{sec:sensitivity_analysis})
 \end{enumerate}
 
We make this mathematically exact in the following sub-section. The equations in the following discussion are not meant to provide a computational advantage in estimating multiplicative bias for any sample in any observed data from an arbitrary simulation. Rather, they provide a useful mathematical framework to understand the limitations of the state-of-the-art shear calibration methods, and highlight where further work needs to be done. We will occasionally refer back to this framework, placing our calibration results (c.f.~\cref{sec:calibration_results} and~\cref{sec:sensitivity_analysis}) in this context.

\subsection{A mathematical framework for calibrating shear with image simulations}
\label{sec:calibration_math}

The multiplicative bias in the shear estimator $\hat{\epsilon}_g$ for a single galaxy can be characterised in terms of the various properties that can be measured; the size and signal-to-noise ratio (S/N) are generally the most important. The latter is a measure of the importance of noise, whereas the former captures how resolved a galaxy is with respect to the PSF. The shear bias for a sample of galaxies estimated using Eq.~(\ref{eq:avshear}) may then taken to be the ensemble weighted mean of the individual biases. However, in practice, individual galaxy properties alone are insufficient to determine the bias of the sample. Due to observational effects such as blending with background galaxies, contamination due to nearby galaxies etc., the true bias in the shear is more complicated~\citep[see][for example]{Hoekstra2017} and any residual biases are estimated using realistic simulations that mimic the particular survey in hand.

We denote by $\vec{S}$ the collection of all the variables that determine the bias in the measured ellipticity. Some examples of $\vec{S}$ are S/N, galaxy size, ellipticity, size and ellipticity of the PSF, and additional parameters as well, such as galaxy morphology and population dependent properties such as the distance to the nearest neighbour, size and the brightness of the nearest object. It is useful to think in terms of a galaxy population rather than a sample selected based on some criterion, because this provides a natural way to account for biases due to blending, selection effects, etc. 
The set (continuous) of all such $\vec{S}$ is denoted by $\mathcal{S}$. 
The data (from observations or from simulations) are then described by a probability distribution function $p$.
We will denote the function space of such probability distributions by $\mathcal{P}$. 

 
 The observed population of galaxies normally spans a wide range in the set of observables $\mathcal{S}$. Often, the dimensionality of $\vec{S}$ is reduced empirically by using a combination of two or more quantities (e.g., S/N instead of galaxy magnitude and pixel noise and the ratio of galaxy size and PSF size). We express the collections of variables $\vec{S}$ as a union of two mutually exclusive sub-collections, that is, $\vec{S} = (\vec{D},\vec{h})$, where $\vec{D}$ is the set of observables over which we explicitly characterise the bias (S/N and resolution in this paper) and $\vec{h}$ is the rest of the `hidden variables' (some of which may still be observables). We know that the bias in the shear estimator (ellipticity) depends on many parameters~\citep[see for e.g.,][]{Pujol2017} including the intrinsic ellipticity itself~\citep{Viola2014}, for which we must not characterise the bias to avoid selection effects, and hence $\vec{h}$ is non-empty. We also know that nearby faint (undetected) galaxies can affect the bias~\citep[see for e.g.,][]{Hoekstra15,Martinet2019}, and hence $\vec{h}$ is non-local. The realism in the image simulations is implicitly expected to naturally account for the biases that arise from $\vec{h}$. By construction, $\vec{S}=(\vec{D},\vec{h})$ includes information about adjacent galaxies as well, and thus completely determines the bias. We express the exact per-object contribution to the shear multiplicative bias as $b(\vec{D},\vec{h})$. This might be thought of as a per-object responsitivity to shear.
 
We imagine selecting a sample $q$ from the overall population $p$ by means of a selection function $s(\vec{D},\vec{h})$ which is binary\footnote{Our convention is such that the non-zero value of the selection function is chosen so that it normalises the distribution, so that $\int\rmd\,\vec{x}s(\vec{x})p(\vec{x})=1$.} in nature. In practice, the selection function can be an implicit one, such as objects lying above the detection threshold, or explicit, for instance resulting from redshift cuts in cosmic shear tomography. To correct the shear estimate obtained from a galaxy sample $q \in \mathcal{P}$ selected from $p$ using a selection function $s$,  we treat $m[q] \equiv m[p;s]$ as a functional that can take in a probability distribution $p$ and a selection function $s$. Because the shear estimator $\hat{\gamma}$ in Eq.~(\ref{eq:avshear}) is a linear combination of individual shear estimators ($\hat{\epsilon}_g$; galaxy ellipticities), we can write the bias of the sample $q$ as
 \begin{equation}
 \begin{split}
     m[q] & = \int\int\rmd \vec{D}\,\rmd \vec{h}\, q(\vec{D},\vec{h}) b(\vec{D},\vec{h}) \\
    & = \int\int\rmd \vec{D}\,\rmd \vec{h}\, s(\vec{D},\vec{h}) p(\vec{D},\vec{h}) b(\vec{D},\vec{h})
\end{split}
\label{eq:b_to_m}
\end{equation}
for an appropriate measure $\rmd \vec{D}\,\rmd \vec{h}$ in the set $\mathcal{S}$. We assume without any loss of generality that the measure is separable. We can turn around Eq.~(\ref{eq:b_to_m}) and formally define (up to a constant of integration)
\begin{equation}
    b(\vec{D},\vec{h}) := \frac{\delta m[q]}{\delta q(\vec{D},\vec{h})}
\end{equation}
as the sensitivity of the multiplicative bias of a galaxy sample to a small change in the sample. 

Since the bias is characterised only as a function of $\vec{D}$ in practice, we marginalise over $\vec{h}$ as follows. We first express $p(\vec{D},\vec{h}) = p'(\vec{D})r(\vec{h}|\vec{D})$ and the selection function $s(\vec{D},\vec{h}) = s'(\vec{D}) t(\vec{h}|\vec{D})$. We note that although $t$ is not a (conditional) probability distribution, we choose to denote the argument of the function as $\vec{h} | \vec{D}$ for convenience. The bias of the sample after marginalising over $\vec{h}$ is
\begin{align}
    m[p;s] &= \int\rmd\vec{D} s'(\vec{D}) p'(\vec{D}) \int\rmd\vec{h} t(\vec{h}|\vec{D}) r(\vec{h}|\vec{D}) b(\vec{D},\vec{h})\\
            &= \int\rmd\vec{D}\, w(\vec{D}) \, b[r;t](\vec{D}),
\end{align}
where
\begin{equation}
    w(\vec{D}) = s'(\vec{D}) p'(\vec{D}) = \int\rmd h s(\vec{D},\vec{h}) p(\vec{D},\vec{h})
\end{equation}
and
\begin{equation}
b[r;t](\vec{D}) := \int\rmd\vec{h}\, t(\vec{h}|\vec{D}) r(\vec{h}|\vec{D}) b(\vec{D},\vec{h}),
\end{equation}
with $b[r;t](\vec{D})$ to be interpreted as the (mean) bias\footnote{This term is the generalisation of bias as a function of $\log$(S/N) and resolution parameter $\mathcal{R}$.}
of the sub-sample of the galaxies that lie at $\vec{D}$. The multiplicative bias of the sample is expressed as the average over the sub-samples. 

We specifically draw the reader's attention to the fact that the bias surface depends on both the selection function and on the overall population of the galaxies. Because of this dependence of the population, it is incorrect to think of the shear bias $b[r,t](\vec{D})$ estimated for the sub-population of galaxies as something that can be associated with the individual galaxies without any reference to the sample of galaxies. We therefore advise against splitting a galaxy sample after having calibrated the shear from the sample. The dependence on the selection function in the bias surface will turn out to be crucial in our analysis, and we will show its importance in~\cref{sec:calibration_results}. 

If $p_\text{real}$ is the galaxy population corresponding to the real Universe, and $q_\text{real}$ is a sample selected from it using a selection function $s_\text{real}$, we would like to evaluate $m[p_\text{real};s_\text{real}] := m_\text{real}$ to provide an accurate estimate of the bias.
The goal of image simulations is to estimate the multiplicative bias using a simulated population $p_\text{sims}$. Unfortunately, $p_\text{sims}$, and therefore the corresponding sample $q_\text{sims}$, obtained from simulations with a selection function $s_\text{sims}$ generally do not  match those of the observations perfectly, even if $s_\text{sims} \equiv s_\text{real}$. Therefore, while $b[r_\text{sims},t_\text{sims}](\vec{D})$ may be used to correct for raw biases of a sub-population of galaxies, $m_\text{sims}$ estimated from the simulations alone is often not a good estimate of $m_\text{real}$. The difference between the two is then given by
 %
\begin{equation}\begin{split}
\Delta m :=& m_\text{real} - m_\text{sims} \\
	       =& \int\rmd \vec{D} \rmd\vec{h}\, b(\vec{D},\vec{h}) \; \times \\ & [ w_\text{real}(\vec{D}) t_\text{real}(\vec{h}|\vec{D}) r_\text{real}(\vec{h}|\vec{D}) \\ &- w_\text{sims}(\vec{D}) t_\text{sims}(\vec{h}|\vec{D}) r_\text{sims}(\vec{h}|\vec{D}) ].
\end{split}\end{equation}
We have considered $b(\vec{D},\vec{h})$ to be the same since the same shape measurement algorithm is executed on both simulations and the real data. Defining $\Delta w = w_\text{real} - w_\text{sim}$ and with similar definitions for $\Delta t$ and $\Delta r$, we can expand $m_\text{real}$ as (with the arguments suppressed)
\begin{equation}\begin{split}
	     m_\text{real} = &\int\rmd\vec{D}\rmd\vec{h}\, b(\vec{D},\vec{h}) (w_\text{sims} + \Delta w)  (t_\text{sims} r_\text{sims} + \Delta (t r)),
\end{split}\label{eq:master_correction}\end{equation}
where 
\begin{equation} \begin{split}
\Delta (tr) (\vec{h}|\vec{D}) = &  r_\text{sims}(\vec{h}|\vec{D})\Delta t(\vec{h}|\vec{D}) + t_\text{sims}(\vec{h}|\vec{D})\Delta r(\vec{h}|\vec{D}) \\ & + \Delta t(\vec{h}|\vec{D}) \Delta r(\vec{h}|\vec{D}).\end{split}
\end{equation}
The term involving $w_\text{sims} t_\text{sims} r_\text{sims}$ is the same as $m_\text{sims}$ by definition. If the simulations were statistically identical to the real data in all aspects, then the correction term $\Delta m$ would be zero. Another guaranteed way to ensure $\Delta m = 0$ would be to have $b(\vec{D},\vec{h}) \equiv 0$ in the range of interest; in fact, it would guarantee $m_\text{real} = m_\text{sims} = 0$. So far, no method has been demonstrated to achieve this in practice, especially on the measured quantities $(\vec{D},\vec{h})$. Methods with small bias $b(\vec{D},\vec{h}) \, \forall (\vec{D},\vec{h})$ are preferable as they help in keeping all the correction terms small, providing robust calibration. For a given function $b(\vec{D},\vec{h})$, we must aim to keep the differences between the simulations and the real data as small as possible to estimate accurate bias values.

Since the simulations and the real data will inevitably differ from each other, the correction $\Delta m$ is estimated by post-processing the simulations. Although the correction to the bias has many terms, the term that has received the most attention in the literature so far is the one involving $\Delta w t_\text{sims} r_\text{sims}$. This term is estimated by re-weighting the simulations as in~\citetalias{FC17,Mandelbaum2018a, Zuntz2018} or by resampling, as in~\citetalias{FC17}. So far, the other terms contributing to $\Delta m$ have been assumed to be negligible and hence ignored. A proper marginalisation over these terms must lead to an increased systematic uncertainty, as it does in this work. We will return to these other terms after discussing how the $\Delta w t_\text{sims} r_\text{sims}$ term is evaluated in practice.

In order to evaluate the $\Delta w r_\text{sims} t_\text{sims}$,  the set of observables is partitioned (or binned) arbitrarily into several subsets $\mathcal{D}_i$ such that
 $\mathcal{D} = \bigcup\limits_{i} \mathcal{D}_i$ and $\mathcal{D}_i \cap \mathcal{D}_j = 0$ if $i \ne j$. As mentioned in the beginning of this sub-section, a typical choice for $\vec{D}$ in practice is a measure of the signal-to-noise ratio and a resolution parameter. A practical estimator (see Appendix~\ref{app:estimator} for a full derivation) for the bias term is
 \begin{equation} \begin{split}
 m[p_\text{real};s_\text{real}] &\approx \int\rmd\vec{D} w_\text{real} (\vec{D}) b[r_\text{sims};t_\text{sims}](\vec{D}) \\ &= \sum_i \int\limits_{\mathcal{D}_i} \rmd \vec{D}  w_\text{sims}(\vec{D}) b[r_\text{sims};t_\text{sims}](\vec{D}) \frac{w_\text{real}(\vec{D})}{w_\text{sims}(\vec{D})}\\
 & \longrightarrow \sum_i \avg{b_i[r_\text{sims};t_\text{sims}]} \frac{w_{i,\text{real}}}{w_{i,\text{sims}}},
 \end{split} \label{eq:zeroth_estimator} \end{equation}
 where 
 \begin{equation} 
 \avg{b_i[r_\text{sims};t_\text{sims}]} := \int\limits_{\mathcal{D}_i} \rmd \vec{D}  w_\text{sims}(\vec{D}) b[r_\text{sims};t_\text{sims}](\vec{D}),
 \end{equation}
  is the average value of $b[r_\text{sims};t_\text{sims}](\vec{D})$ in the $i^\text{th}$ partition, and 
  \begin{equation}
  w_{i,\text{real}} := \int\limits_{\mathcal{D}_i} \rmd \vec{D} w_\text{real}(\vec{D}),
  \end{equation}
  is the number (or total weight) of galaxies in the $i^\text{th}$ partition in the observed data and $w_{i,\text{sims}}$ is defined similarly for the simulated data. The division by $w_\text{sims}(\vec{D})$, or equivalently by $w_{i,\text{sims}}$ assumes that the simulations have covered all the regions of interest in the parameter space of observables $\mathcal{D}$ sufficiently.
  
  We refer to the ratio of the two as the re-weighting factor. For a pre-defined set of partitions, the averaged quantities in Eq.~(\ref{eq:zeroth_estimator}) are themselves noisy; in particular, $w_{i,\text{sims}}$ is. As this term appears in the denominator, the estimator itself may be slightly biased.
Thus, the bias in the estimator is present even if $w_\text{sims} = w_\text{real}$, and arises because the simulations represent a different sample from the observations. If we instead define the partitions such that $w_{i,\text{sims}}$ is the same in each of the partitions, the bias in the estimator may be partly mitigated, but not completely eliminated, since the partitions themselves are correlated with $w_{\text{sims}}$. The bias in the estimator depends on the actual distributions, but as long as the partitions (bins) contain fairly large numbers of galaxies, the bias in the estimator must be small, and may be neglected compared to the uncertainty in the estimate itself (c.f.~\cref{app:estimator}). 

\citetalias{FC17} show that the shear bias for a sub-population of galaxies defined by their noisy observables is different from that for a sub-population of galaxies defined by their intrinsic parameters (see Fig. 6 of~\citetalias{FC17} for example), and that the shear bias is more sensitive to the observed parameters. This was referred to as `calibration selection bias' in~\citetalias{FC17}. This makes $b(\vec{D},\vec{h})$ and $b[r;t](\vec{D})$ steeper functions when based on the observed parameters, which increase the sensitivity to the simulated population of galaxies. Moreover, the practical estimator is an approximation to the integral, whose validity relies crucially on the smoothness assumption of the integrand within the partitions. As we know from~\citetalias{FC17},~\citetalias{Zuntz2018} and~\cite{Samuroff2018}, $b[r_\text{sims};t_\text{sims}]$ for \lensfit\ and \imtshape\ does not appear to be smooth enough to lend itself amenable to even fairly sophisticated interpolation schemes.  To circumvent the difficulty of calibrating shear robustly when $b[r_\text{sims};t_\text{sims}]$ is noisy or not smooth enough,~\citetalias{FC17} suggested using a resampling approach (see Sect.~5 of~\citetalias{FC17}) to obtain a resampling weight $w_{\text{res}}$ based on the number of times a galaxy in the simulation is matched to galaxies in the observed data in a `nearest-neighbour' sense. The resampling weights, along with the lensing weights, are used to obtain a shear estimate for the desired sample, and the bias in that estimate is then the bias assigned to the sample. Within our framework, it corresponds to inducing the notion of a metric in the parameter space of observables and then partitioning $\mathcal{D}$ into Voronoi tessellations instead of regularly spaced bins. Each tessellation by construction contains `one' galaxy ($w_{i,\text{sims}}=1$) and $w_{i,\text{real}}$ is the `number' of the galaxies in the observed data that occupy the $i^\text{th}$ region, which is also the resampling weight for those observed galaxies. One could consider replacing the number by their lensing weights, but~\citetalias{FC17} chose to use the number of galaxies instead for practical reasons. This is therefore mathematically equivalent to Eq.~(\ref{eq:zeroth_estimator}), with the advantage that it avoids having to calculate the bias surface explicitly for each of the sub-samples. This description is strictly true for the nearest neighbour matching alone, and for a generic $k$-nearest neighbour search as used by~\citetalias{FC17}, it amounts to repeating this procedure $k$-number of times, with previously assigned matches discarded. If the simulations are a good representation of the observed data, the re-weighting factor is close to unity and the resampling factor is the same for all partitions. In such a case, we expect the inferred bias to be insensitive to the calibration methodology and hence expect both the re-weighting and resampling methods to give identical results.~\citetalias{FC17} also demonstrated that the resampling approach and the re-weighting approach yield consistent bias values within their statistical uncertainty.

In the first year HSC shape catalogue~\citep[HSC-DR1;][M18a hereafter]{Mandelbaum2018a}, galaxies are assigned baseline multiplicative (and additive) bias corrections\footnote{The multiplicative bias was offset by an undisclosed constant to aid blinded analysis.} based on the simulations using a somewhat sophisticated interpolation of $b[r_\text{sims};t_\text{sims}](\vec{D})$.  The multiplicative bias for a sample, $m[q_\text{real}]$ is then calculated as the weighted average of the per-object multiplicative bias~\citep{Hikage2018}. For the \imtshape\ catalogue of first year results of DES~\citep[DESy1][Z18 hereafter]{Zuntz2018} and KiDS-450~\citepalias{FC17} as well, a grid-based scheme is implemented owing to slightly better performance, compared to fitting an analytical function to $b[r_\text{sims};t_\text{sims}](\vec{D})$. Galaxies are assigned a multiplicative bias depending on which bin in the grid they belong to. In all three analyses, the sample selection was made after the per-object multiplicative bias were calculated.


We now return to the other correction terms. If the selection function on the real data and simulations are the same, then the terms involving $\Delta t$ vanish. If object detection, star-galaxy separation and shape measurements are carried out on the simulations as done for the data, we expect the $\Delta t$ term to be small. However, there is an additional selection function based on galaxy colours introduced in cosmic shear tomography through photometric redshifts. Such selection cuts have not been applied explicitly to the simulations in~\citetalias{FC17}, \citetalias{Mandelbaum2018a}, and \citetalias{Zuntz2018}. The correction terms involving $t_\text{sims} r_\text{sims }\Delta w$ were computed for each tomographic sample, but the contributions from $\Delta t$ terms were ignored. In this work, we explicitly apply redshift cuts on the simulations as we do in the data, and therefore assume that the terms involving $\Delta t$ are truly negligible. The importance of including redshift information in the simulations for calibrating shear were highlighted in~\citetalias{Zuntz2018} and we demonstrate it in this paper as well (c.f.~\cref{sec:calibration_results}).

If the simulations are representative of the real data, in terms of image quality, galaxy populations etc., then we can expect $\Delta r$ (and $\Delta w$) to be small. The difference between the two populations may be bridged either through a Monte-Carlo control loop as in~\cite{Bruderer16} on the joint probability distribution or by starting with a deep catalogue from a space-based telescope as in this work, \citetalias{Mandelbaum2018a} and \citetalias{Zuntz2018}. We note that $r(\vec{h}|\vec{D})$ is a conditional probability distribution, implying that correlations between $\vec{D}$ and $\vec{h}$ (for example, between size and ellipticity; c.f.~\cref{sec:input_cor} and Fig.~\ref{fig:size_e_comparison}) have to be captured correctly. It may not suffice for the simulations to match only the marginal distributions of parameters in the real data, as $\Delta w=0$ does not guarantee that $\Delta m=0$. We explicitly show the importance of capturing these correlations in this paper (c.f.~\cref{sec:input_dependence}). The error introduced by neglecting any residuals in these terms are quantified approximately by performing various sensitivity tests, which place an upper bound on the terms that are ignored~\citepalias[see Sec. 6 of][for example]{FC17}. For instance, \cite{Hoekstra2017} studied the sensitivity of the bias with \euclid -like simulations for the classic KSB shape measurement algorithm~\citep{KSB95,Luppino1997,Hoekstra1998} to various parameters such as the distribution of the galaxy sizes and ellipticities, galaxy density, limiting magnitude, etc.

To summarise our framework, we argue that the image simulations used to calibrate the shear must mimic the observed data as closely as possible to avoid any deviation from the required $m[p_\text{real};s_\text{real}]$. The necessity for good agreements also holds true for sub-samples of galaxies for which we wish to estimate the shear bias, as in galaxies within a redshift bin for cosmic shear tomography. If $p_\text{real}$ varies significantly among the different sub-samples, then re-weighting the simulations to match the distribution of the sub-samples  is not guaranteed to obtain accurate calibration for shear estimated from that sub-sample. The other terms may no longer be negligible when $p_\text{sims}$ is substantially different from $p_\text{real}$. The selection function used on the data must also be applied to the simulations to  match the population, and for cosmic shear tomography, this implies that the simulations must include photometric redshifts for galaxies explicitly.
 
 As an application of this mathematical framework, we propose a test that can verify the validity of the image simulations. Although it is possible in principle to tune the input parameters of the simulations to match some of the observed data $w_\text{real}(\vec{D})$, as in~\cite{Bruderer16}, the shear biases may still depend on other variables, $\vec{h}$, that are inaccessible to us. As we show later in this work (c.f.~\cref{sec:sensitivity_analysis}), the shear biases depend on several assumptions we make about the Universe through the input catalogue, although their observed distributions look similar. A consistency check to validate the simulations would then be useful, especially for future surveys, where accurate shear calibration is required. One such consistency check is to deploy two different shape measurement algorithms on the same data, and calibrate them using the same image simulations. 
 Two shape measurements methods with different biases ($m[p_\text{real};s_\text{real}]$, $b[r_\text{sims};t_\text{sims}](\vec{D},\vec{h})$ and hence $m[p_\text{sims};s_\text{sims}]$) that are calibrated using the same image simulations can only arrive at consistent cosmological results, when applied to the data, if the $\Delta$ terms are sufficiently small. Such an approach not only boosts the confidence in the methodology  and in the estimated cosmological parameters, but could possibly help identify and eliminate other potentially unknown sources of biases.
 

 \subsection{Calibrating shear without image simulations}
 \label{sec:metacal}
 
So far we have examined how limitations in the simulated data may result in biases in the estimate of the multiplicative bias. To circumvent the need for image simulations, a different approach, called metacalibration, has been proposed recently \citep{Huff2017,Sheldon2017}. The basic idea behind metacalibration is to find the shear response of each galaxy using the observed data. This is achieved by deconvolving the PSF first, shearing the galaxy and reconvolving it with a (slightly larger) PSF model and measuring the galaxy shape. This approach has been used to calibrate the \textsc{ngmix}\footnote{{http://github.com/esheldon/ngmix}} algorithm and was applied to the DESy1 shape catalogues~\citepalias{Zuntz2018}.
 
As the biases are derived from the actual data, one can treat the observed data as the simulation equivalent and treat the $\Delta w$, $\Delta r$ and $\Delta t$ terms to be almost zero, allowing for a more robust calibration than what may ever be possible with image simulations. The use of the actual catalogued data, however, naturally leads to a situation in which at least some selections have been applied at the detection step. The detection selection acts in one direction with only the detected galaxies being sheared.  These galaxies may not be detected after shearing but galaxies previously undetected before shearing, that may be detected afterwards, are not included in the metacalibration analysis.  This will lead to a small, but non-zero $\Delta t$ term.  The limitation of working with only the detected galaxies may be partly circumvented by injecting synthetic galaxies in real images~\citep{Suchyta2016,Huang2018}. \newtext{At a more fundamental level, the algorithm requires that the sources of degradation in image acquisition, such as pixelisation, noise and other detector imperfections, which may also be stochastic in nature, be reversible so as to be able to shear the intrinsic galaxy}. In the case of a wavelength-dependent PSF, the spatial variation of the colour across a galaxy profile leads to multiplicative bias \citep{Semboloni2013,Er2018}, which cannot be determined using metacalibration. This holds true for errors in PSF modelling as well, as is also the case with image simulations. For space-based surveys such as \euclid\ and WFIRST, as a result of the PSF not being Nyquist-sampled, metacalibration cannot capture the biases accurately (Rosenberg et al., in prep). Whether or not these limitations can be ignored will depend on the desired level of accuracy, but it is not evident that such an approach would be able to account for biases arising at the object detection stage. Moreover, any residual effects after the images have been corrected for detector imperfections~\citep[see][for a general discussion]{mandelbaum14b} such as Charge Transfer Inefficiency (CTI), brighter-fatter effect \citep{Antilogus2014,Guyonnet2015,Coulton2018}, and read-out effects such as binary offset effect \citep{Boone2018} may introduce residual biases that are captured better through image simulations. Such residual biases will definitely be significant for the next generation of cosmic shear surveys. Hence it seems likely that a forward-modelling approach using image simulations, perhaps using  metacalibration to minimise raw biases at the first step, may be the best way forward for accurate and robust shear calibration.

Another approach to calibrate shear without using image simulations was suggested by~\cite{Zhang2018} using field distortion introduced by the optical setup of the telescope. As a result, galaxies on average have a preferred orientation depending on their position on the detectors. By evaluating a so-called field distortion shear as a function of the CCD position from the astrometric solution and comparing it with the local shear measured using any method, one can obtain the multiplicative and additive biases from the data itself. In a way, this approach is not fundamentally different from the metacalibration approach, in that, the artificial shear that we apply in the metacalibration is applied naturally by the telescope, albeit after the PSF convolution step.~\cite{Zhang2018} demonstrated this approach on the publicly available data from the Canada-France-Hawaii Telescope Lensing Survey~\citep[CFHTLenS;][]{Heymans2012,Erben2013} for two shape measurement methods and estimated their multiplicative biases to within 4 per cent (limited by survey volume). Although that is a commendable achievement, this is well above the requirements needed for the ongoing surveys. Moreover, the field distortion shear in KiDS is expected to be tiny, due to the modified Ritchey-Chr\'{e}tien design of the VLT Survey Telescope (VST)~\citep{Arnaboldi1998}. Therefore, while the larger volume of data in KiDS compared to CFHTLenS implies that the constraint on the multiplicative bias could become tighter in principle, the low amount of camera distortion on the VST compared to the Canada-France-Hawaii Telescope (CFHT) means that it would be much harder to estimate the multiplicative bias robustly using camera distortions.

Finally, lensing of the CMB by foreground mass, also known as CMB lensing, can also help constrain a combination of shear estimation and photo-$z$ biases. Current studies yield uncertainties on the shear calibration at the 10 per cent level~\citep{Baxter2016,HarnoisDeraps2017,Singh2017}. However,~\cite{Schaan2017} predict that even with the Stage 4 CMB experiments, shear calibration biases can be at best be constrained to within 0.5 per cent for future surveys. In particular, CMB lensing enables robust calibration at high redshifts, where our knowledge about the galaxy population is the poorest to include them correctly in our simulations and where the galaxies are typically fainter and noisier to reliably calibrate with metacalibration. CMB lensing can therefore potentially act as an independent way to validate shear measurement biases estimated with simulations and metacalibration. Realistic image simulations are nevertheless required to verify the validity of these approaches. We defer the exploration of metacalibration in combination with simulations to future studies and use realistic image simulations in this work to calibrate shear for KiDS.

\section{Simulating KiDS+VIKING-450}
\label{sec:data}

The first cosmic shear constraints from KiDS, presented in \citetalias{Hildebrandt17} used data from the third data release (DR3), described in \cite{deJong2017}. Further details about the survey can be found in \cite{deJong2015} and \cite{Kuijken15}. DR3 comprises 454 deg$^2$ of imaging data in the  $ugri$ filters obtained with OmegaCAM  on the VST. The shapes were measured from the $r$-band images using an updated version of \lensfit\ described and tested in \citetalias{FC17}. The multi-band data allowed \citetalias{Hildebrandt17} to divide the sources into four tomographic bins using point estimates of the photo-$z$, $z_{\rm B}$, obtained with the Bayesian Photometric Redshift code \citep[BPZ;][]{Benitez2000}. 

The area surveyed by KiDS is complemented by $ZYJHK_{\rm s}$ observations from the VISTA\footnote{VISTA is short for Visible and Infrared Survey Telescope for Astronomy} Kilo-degree INfrared Galaxy survey \citep[VIKING;][]{Edge2013}. We refer to the combination of the $\sim450$ deg$^2$ from KiDS-DR3 \citep{deJong2017} and the overlapping VIKING data~\citep{Wright2019} as KiDS+VIKING-450, or KV-450 for short. The improved wavelength coverage reduces the outlier rate in the photo-$z$s and improves the overall precision, resulting {in an improvement in the tomographic bin selection}. Furthermore, dedicated observations of fields with extensive spectroscopy provide a more robust training set for the calibration of the underlying source redshift distribution. As a consequence the analysis can be extended to include sources with $z_{\rm B}>0.9$, provided we can also improve the calibration of the shape measurements for these distant galaxies.

To measure the cosmic shear signal, galaxy shapes are determined using the same version of \lensfit~\citep{Miller2007, Miller2013} that was used in \citetalias{Hildebrandt17}. Here we revisit the calibration of the shear measurement pipeline by increasing the realism of the image simulations, so that an updated cosmic shear analysis~\citep[][H18 hereafter]{Hildebrandt2019} can take advantage of the KV-450 data set. We also improve the analysis of the simulated data to better reflect the steps in the data analysis pipeline.

To do so, we create simulated KiDS-like observations of the COSMOS field in the $r$-band using a publicly available catalogue of galaxies with structural parameters determined from images taken with the Advanced Camera for Surveys (ACS) on-board HST \citep[][see \cref{subsec:input_catalogue} for more details]{Griffith2012}. We also make use of VST and VISTA observations of the COSMOS field, which were taken in the same way as the nominal survey\footnote{The COSMOS field is not part of the nominal area covered by VIKING, and thus lacks the $Z$-band imaging. It has however been observed in the other filters by VISTA. Instead, we use the $z$-band data from CFHT as a proxy to obtain the equivalent nine-band $z_B$ estimate. 
}. This allows us to determine photo-$z$ estimates and shapes using the same pipeline used for the cosmic shear analysis. We use the same tomographic bin definitions as~\citetalias{Hildebrandt2019}, namely, $0.1 < z_B \le 0.3$, $0.3 < z_B \le 0.5$, $0.5 < z_B \le 0.7$, $0.7 < z_B \le 0.9$ and $0.9 < z_B \le 1.2$, where $z_B$ is the peak of the posterior distribution from BPZ \citep{Benitez2000} using the available photometry in the nine bands. We label the tomographic bins, starting from low redshift as B1 through B5 for convenience.

Our new simulation setup is based on that used in~\citetalias{FC17}. The main difference is that by basing the input catalogue on actual multi-wavelength observations, we can assign photo-$z$s to the sources, thus reproducing the definition of the tomographic bins. Moreover, the HST observations allow us to naturally include correlations between the key input galaxy parameters. Another benefit is that we can compare the output directly to actual VST $r$-band observations of the same field, because the image simulations are based on real data. 

\subsection{Simulation setup}
\label{sec:sim_setup}
The basic simulation setup is similar to the~\citetalias{FC17} SCHOol pipeline (Simulations Code for Heuristic Optimization of \lensfit). The most important change is that the simulations are made to match the observed COSMOS field as observed by the VST in the $r$-band, by using input parameters based on HST observations (see \cref{subsec:input_catalogue} for more detail). To reflect the higher level of sophistication, we refer to the new pipeline as the COllege pipeline ({CO}SMOS-{l}ike {l}ensing {e}mulation of {g}round {e}xperiments). We describe  the image simulation pipeline only briefly in order to highlight the improvements made, and their significance. For more details, we refer the interested reader to \citetalias{FC17}.

For the simulated images to be realistic, in addition to a realistic galaxy catalogue, the simulations must reflect the instrumental setup of the KiDS $r$-band data and follow the same data acquisition procedure. The camera consists of 32 e2v CCDs that are arranged in four rows of eight chips, each with $2048\times 4080$ pixels sampling the focal plane at a uniform pixel scale of $0\farcs{213}$ per 15$\mu$m pixel. The field-of-view covers approximately 1~deg$^2$, but the chips  have gaps (up to 1.35\,mm) between them. To ensure that the gaps are exposed in any given pointing, five dithers are taken in a staircase pattern, with dither steps of $25''$ and $85''$ along right ascension and declination respectively~\citep{DeJong2012}. The five exposures in a single tile are taken in succession, so that the seeing conditions are fairly homogeneous. The PSF is robust and its position-dependent ellipticity can be modelled well by cubic polynomials \citep[see Fig. 5 of][]{Kuijken15}. 

The KiDS-450 data are analysed tile-by-tile, meaning that the data from the overlap of tiles is ignored. It is thus sufficient to simulate individual tiles as opposed to continuous patches obtained from multiple pointings. The galaxy surface brightness profiles are assumed to be given by \sersic\ profiles and are simulated using \galsim\footnote{{https://github.com/GalSim-developers/GalSim}}~\citep{Rowe2015}. A significant improvement from~\citetalias{FC17} is the input galaxy catalogue. We use morphological parameters based on HST observations that are described in detail in~\cref{subsec:input_catalogue}. Within each tile, a constant shear $\gamma^\text{true}$ is applied to all galaxies. To minimise the contribution of shape noise to the bias estimates, three additional tiles are generated per galaxy for each applied shear and PSF, where the galaxies are rotated by $45^\circ$ (prior to the shearing operation) in successive tiles. The applied shear values $\gamma^\text{true}$ are equally spaced on a ring corresponding to $|\gamma^\text{true}|=0.04$, with $\gamma_1=0.04$ being one of them. 

We assign a PSF set containing five different spatially-constant PSFs for the five subsequent exposures for a given tile. We consider 13 such unique PSF sets in total.  We describe the PSFs using Moffat profiles. \newtext{Our choice of Moffat parameters is the same as those used in~\citetalias{FC17}. In Appendix~\ref{app:PSF_modelling}, we show how the simulated PSF parameters compare to the KV-450 dataset}.

The sheared galaxies are convolved with the five PSFs (corresponding to five exposures) and are rendered with the sub-pixel offsets that the dither pattern introduces, thereby closely mimicking the survey specifications. The stellar magnitude distribution is obtained from the Besan{\c c}on model\footnote{{http://model.obs-besancon.fr/}}~\citep{Robin2003,Czekaj2014} corresponding to right ascension $\alpha$ of $175^\circ$ and declination $\delta$ of $0^\circ$. The stars in the simulations are PSF images, also rendered with the appropriate sub-pixel offsets.
The galaxy catalogue is the same for each tile, but the locations of the stars are varied randomly from one tile to another. The background noise is assumed to be Gaussian, whose strength is adjusted to correspond to a magnitude limit $m_\text{lim} = 26$ as in~\citetalias{FC17}.


Each of the simulated exposures is chopped into 32 pieces, each of size $2048\times 4080$ pixels, corresponding to the $8\times 4$ CCD chips in the VST/OmegaCam, with a gap of 70 pixels between them. Based on the dither offsets of the exposures and chip positions, a flat WCS is assigned to the chopped images and a co-added image is obtained using \swarp~\citep{Bertin2010}. \sextractor~\citep{bertin1996sextractor} is run on the co-added image with the same parameter settings as those used for the analysis of the KiDS data. The co-added image is only used to detect galaxies using \sextractor, and not for measuring galaxy shapes. 
 
\begin{figure*}
\centering
	\includegraphics[width=0.49\textwidth]{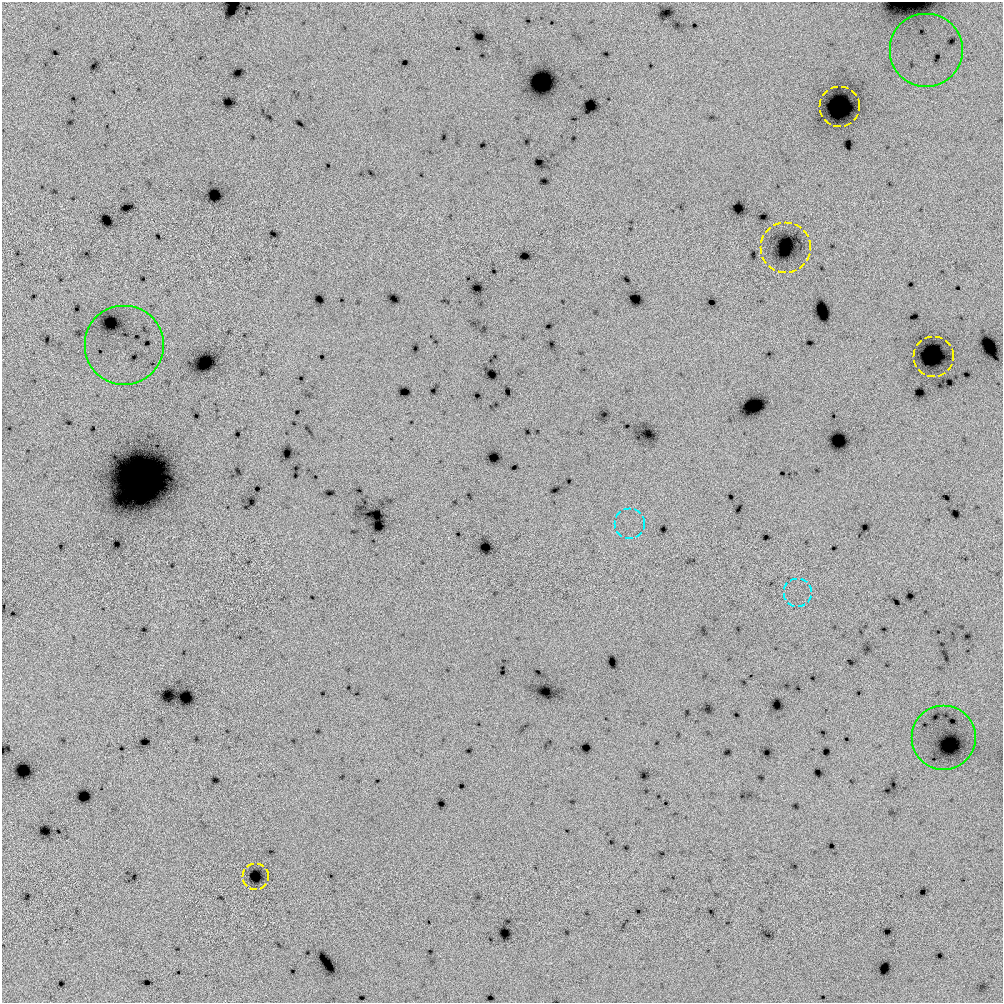}
	\includegraphics[width=0.49\textwidth]{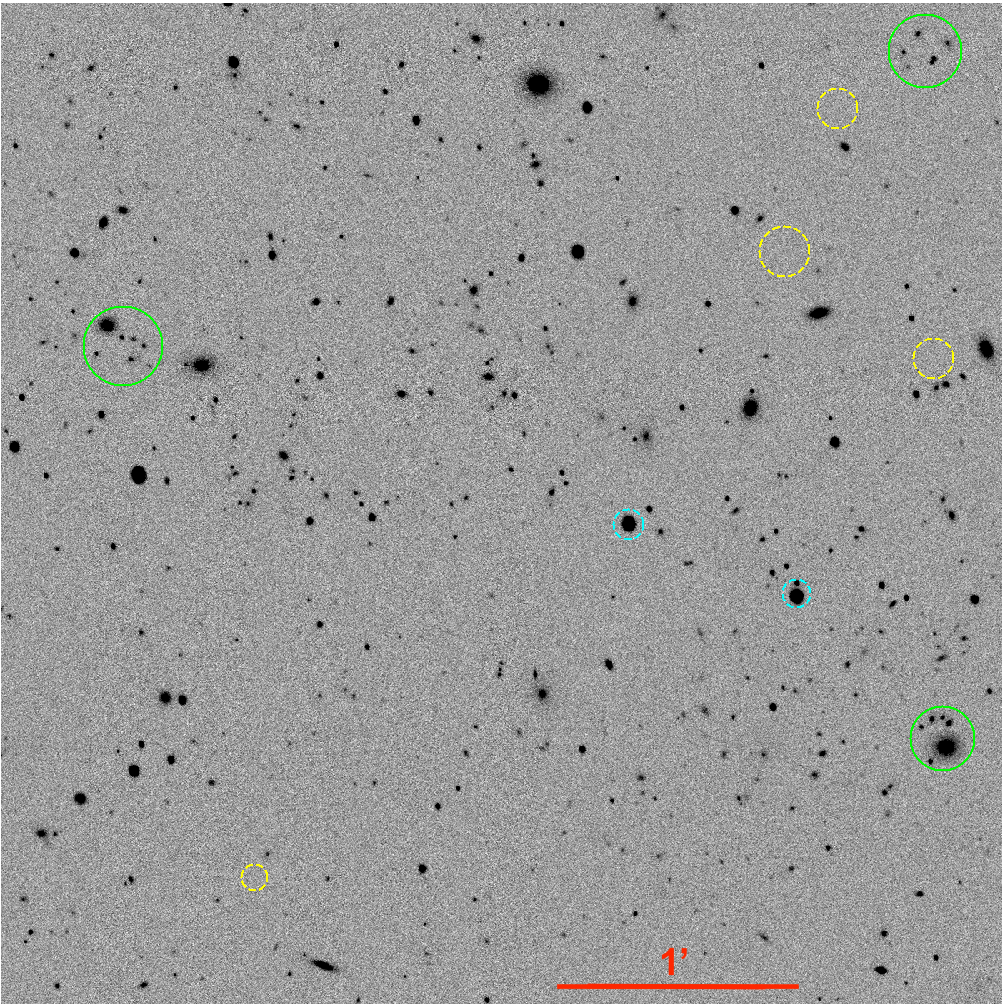}
\caption{{\it Left:} Cutout from the co-add of the COSMOS field observed with VST in the $r$-band as a part of KiDS.  {\it Right:} The corresponding region, but now simulated under similar seeing conditions with morphological parameters of the galaxies taken from  the HST COSMOS catalogue described in~\cref{sec:imsim}, simulated using the setup described in~\cref{sec:sim_setup}. The images are 1200 pixels, roughly equivalent to $4\farcm{28}$, on the side and are rendered in \textsc{ds9} with \texttt{zscale} colour scale. {We do not simulate the bright saturated stars that can be seen in the VST image, and choose to place additional stars at random locations.  The position angles of the galaxies,  as measured by \galfit\ are noisy, which can be seen from the differences between the galaxy orientations in the left and right panels. The solid green circles indicate some examples of regions with distinctive patterns on the sky involving close pairs of galaxies, which we are able to replicate fairly well. The broken circles in yellow and cyan respectively highlight some objects that are not included in our simulations or not present in the original data.}}
\label{fig:emulation}
\end{figure*}

A small region of the observed (left) and simulated co-added image (right) is shown in Fig.~\ref{fig:emulation}. The images agree well, with the main differences caused by bright stars; galaxies are simulated at their observed location, but stars are placed at random positions. {More specifically, we highlight with green circles, some distinctive patterns on the sky and show our ability to replicate them. We also indicate objects missing from the simulations with yellow circles}. 

By using 13 realisations of observing conditions (13 PSF sets), each with four rotations for shape noise cancellation and eight lensing shears, we simulate a total of 416 square degrees of the survey which, due to the shape noise cancellation, is equivalent to 3750 square degrees, which is more than eight times the size of KV-450 footprint (see section 3.2 of~\citetalias{FC17} for this calculation).

In this work, we ignore instrumental effects, such as the brighter-fatter effect \citep[e.g.][]{Antilogus2014}. Although we have detected this and other low-level detector effects during the course of this work, their impact on multiplicative bias appears to be minimal for the current cosmic shear analysis (Hoekstra et al., in prep.).

\subsection{Shape measurements with \lensfit}
The shapes of the galaxies detected by \sextractor\ are measured using the self-calibrating version of \lensfit\ that was used in~\citetalias{FC17} and in~\citetalias{Hildebrandt17}. It is a likelihood-based model-fitting algorithm that describes galaxies as the sum of an exponential disc (\sersicn\ $ =1$) and a bulge  component (\sersicn\ $ = 4$). The model parameters are determined by \lensfit\ from a joint fit of the PSF-convolved galaxy model to the individual exposures. To reduce the model complexity, the ratio of disc and bulge scale-lengths is a fixed parameter and the ellipticities of the disc and bulge are set equal, resulting in seven free parameters (flux, size, complex ellipticity, 2D position and bulge-to-total flux ratio). The resulting ellipticity parameters are deduced from the likelihood-weighted mean parameter value, marginalised over the other parameters, adopting priors for their distributions. To counter the bias in the ellipticities introduced by noise in the images, \lensfit\ employs a self-calibration scheme, which was described in detail in~\citetalias{FC17}. Metacalibration, discussed in~\cref{sec:metacal}, may be seen as a generalisation of this self-calibration approach, and performs better than self-calibration. A \lensfit\ version with built-in metacalibration is currently under development. We use the version that was described in~\citetalias{FC17} for the KV-450 dataset and refer the interested reader to this paper for details on the overall performance of the self-calibrating \lensfit\ \citep[also see][for its performance on the GREAT3 challenge]{Mandelbaum2015}.

To estimate the shear, the ellipticities of the galaxy models are combined with a weight that accounts for the uncertainty in the ellipticity measurement. Galaxies with intermediate ellipticities exhibit a tendency to have larger weights compared to galaxies with either low or high ellipticities, but with similar sizes and signal-to-noise ratios. This leads to a bias in the shear estimate that is sensitive to the distribution of galaxy ellipticities. To reduce the bias in the shear estimate, the weights are therefore re-calibrated at the catalogue level.~\citetalias{FC17} determined this correction to the weights from
the catalogues of each pointing, which exhibit a coherent shear. During the course of this work, we realised that this approach was in fact incorrect because the adjusted weights account for the shear as well, leading to increased multiplicative bias. Here we derive this correction using a combined catalogue for each PSF set (so that the net shear is zero). This change from~\citetalias{FC17} {alters the} shear calibration, and we refer to the corrected catalogues from~\citetalias{FC17} as `FC17cor'. Comparing the simulated data to the actual observations revealed that the final weight recalibration procedure was also done incorrectly in the KiDS-450 analysis~\citepalias{Hildebrandt17}, due to a different error. This has been rectified for KV-450 and the impact of this error appears to be minimal for the overall cosmic shear signal~\citepalias{Hildebrandt2019}. It was, however, the origin of the low-level detection ($2.7\sigma$) of non-cosmological B-modes~\citep{Asgari2018}. In the corrected KV-450 cosmic shear analysis~\citepalias{Hildebrandt2019}, we find the B-modes to be consistent with zero.

Following the shape measurement step, we determine the multiplicative and additive biases. To do so, we use Eq.~(\ref{eq:avshear}) to compute the average (reduced) shear $\hat\gamma_i$ for a given selection of galaxies from the catalogue of  \lensfit\ ellipticities and re-calibrated weights. We adopt a linear model as in~\cite{STEP1} to relate the true (reduced) shear $\gamma^{\rm true}$ and the measured value $\hat{\gamma}$ {for each component separately.}
%
%
{The slope of the best-fit line} yields the multiplicative bias\footnote{We assume that the multiplicative bias $m_1$ does not depend on $\gamma_2$, and vice versa. Moreover, we find consistent values for both multiplicative bias estimates, and thus take the mean when we report the final multiplicative bias estimates.} and the offset is the additive bias. The bias parameters are obtained by a simple linear regression to each component of the shear. We have explicitly verified the validity of this linear model by simulating images with $|\gamma^\text{true}|=0.04$ and with $|\gamma^\text{true}|=0.03$ and found no difference in the bias values, indicating that the linear model is adequate.

{During the course of this work, inspired by the metacalibration approach,~\cite{Pujol2019} proposed an alternative way of precisely estimating the bias in shear from simulations, devoid of any shape noise. We note that the uncertainties in our final calibration are driven by systematic errors and not by statistical errors, and therefore, the increased precision obtained by adopting this alternative approach does not add much value at this stage. It remains to be seen if this method can capture the detection bias, as is the case for metacalibration. Moreover, as we mention later in~\cref{subsec:input_catalogue}, our input catalogue happens to have a preferred orientation that is not due to a coherent lensing shear. If the galaxies are not isotropic in the absence of shear, the \lensfit\ weights themselves will be biased. Rather than randomising the galaxy orientations to eliminate this source of bias,  we prefer to rotate all galaxies to achieve isotropy, while preserving the relative orientations between any pair of galaxies. }

\subsection{Improvements since FC17}
\label{sec:comparison}
\begin{table*}
\caption{Summary of the differences between the \school~\citepalias{FC17} and the \college\ simulations}
\label{tab:difference_summary}
\begin{tabular}{ l  p{0.38\textwidth}  p{0.375\textwidth} }
\hline \hline
{} & \school\ (\citetalias{FC17}) & \college\ (this work) \\
\hline
Input distributions & Input quantities (size, morphology etc.) correspond to the \lensfit\ priors and a power law magnitude distribution & Input quantities are taken from  \citet{Griffith2012} based on HST ACS observations\\
Analytical models & Bulge+Disc models, with scalelengths coupled & \sersic\ models \\
Object detection & \sextractor\ is run on one rotation and the same detection catalogue is used in all four rotations & \sextractor\ is run on all rotations and corresponding detection catalogues are used \\
Correlations & Ellipticity is uncorrelated with size, magnitude, morphology, etc. & {The dependence between ellipticity, size, morphology, etc., is automatically included by using the measurements} from \citet{Griffith2012} on HST data \\
Coaddition & \swarp\ is run with pixel resampling turned off & \swarp\ is run with pixel resampling turned on \\
Depth: & Extends to $29^\text{th}$ magnitude & Relatively shallow and somewhat incomplete beyond the detection limit \\
Sample variance & The input catalogue changes for each shear and PSF realisation & The input galaxy catalogue is identical for all shear and PSF realisations \\
Clustering & Galaxies are placed randomly on the sky  & Galaxies are placed at their true locations \\
Intrinsic alignment & Galaxies have random orientation & Observed complex ellipticities from \citet{Griffith2012} are used\\
Redshifts & No explicit redshift information for galaxies & A nine-band photo-$z$ is assigned to every simulated galaxy based on matching to the KiDS observations of the COSMOS field\\
Weight bias correction & Separately calculated for each shear and each rotation
& Calculated for a combined catalogue for every PSF set \\
\hline
\end{tabular}
\end{table*}

Although our basic setup is largely unchanged with respect to \citetalias{FC17}, there are a number of important differences. In this section we therefore highlight the main differences between the \school\ and \college\ pipelines. Apart from the change in the input catalogues, which are discussed in  \cref{subsec:input_catalogue}, several improvements have been implemented. We detail only some of those that require a description below and provide a more exhaustive list in Table~\ref{tab:difference_summary}.

Firstly, the two pipelines use different parameterisations of the galaxy surface brightness to generate the galaxy images. The \school\ pipeline in~\citetalias{FC17} used bulge+disc models, with the distributions of size, bulge-to-disc ratio, ellipticity following the priors used by \lensfit. In this work, we describe the galaxies by \sersic\ models, with the parameters determined from the HST images themselves (c.f.~\cref{subsec:input_catalogue}). No assumptions about the distributions of the parameters are made; in particular, we do not assume that the different quantities are uncorrelated. Tests in the GREAT3 challenge~\citep{Mandelbaum2015} and in~\citetalias{FC17} show that \lensfit\ has negligible model bias, and therefore this change in the morphology is not expected to affect the calibration. Moreover, \lensfit\ uses a fixed ratio between the bulge and disc scalelengths, thereby having the same number of free parameters as the \sersic\ model. Consequently the \sersic\ model is similar to the bulge+disc model\footnote{\tiny{{http://strw.leidenuniv.nl/$\sim$arunkannawadi/internal/RadialProfiles.html}}}.

In contrast to the KiDS data analysis, in \citetalias{FC17} the individual exposures were not resampled while being combined to produce the co-added image. This changes the noise properties, and thus the detection catalogue. This does not affect the shape measurement itself, because \lensfit\ is run on the individual exposures and generates its own segmentation map to mask neighbouring sources. {In this work, we run \swarp\ with the resampling option turned on, although this has a negligible impact on the shear estimation}.

In \citetalias{FC17}, \sextractor\ analysed one co-added image and this detection catalogue was used in the other three realisations where the galaxies were rotated. The rationale behind this choice was to ensure that the shape noise cancellation was effective. However, as the weights are assigned independently and differ slightly due to pixel noise, the shape noise cancellation was never perfect. Running \sextractor\ on all images captures the selection bias that would be present in the real data. We therefore re-ran the \school\ pipeline used in \citetalias{FC17} with \sextractor\ on each of the rotations. {After this change,} we find that the multiplicative biases become more negative, by about 0.005. We return to this in~\cref{sec:sextractor_bias}.

In the \college\ pipeline, for each of the PSF sets, the 32 \lensfit\ catalogues for the eight lensing shears (averaging to zero, pairwise) and four rotations are first combined, before the weight recalibration script is run on this combined catalogue. This ensures that there is no anisotropy in the catalogues. In the \school\ pipeline, the weight recalibration procedure was incorrectly applied to each catalogue, with a net lensing shear, separately. This difference leads to a change of about 0.03-0.04 in the multiplicative biases and by far, the biggest difference between the two analyses.

\section{Image simulations}
\label{sec:imsim}

The input parameters of the image simulations used in \citetalias{FC17} were based on the priors used by \lensfit\ \citep{Miller2007,Miller2013}, the shape measurement method used. A comparison to the KiDS observations showed, however, that the signal-to-noise ratio and size distributions obtained directly from the simulations exhibited some differences with those inferred from the data themselves, which had to be adjusted for by resampling or re-weighting the simulations. In particular, the simulations lacked low S/N objects relative to the data. As shown in ~\cref{sec:calibration_math}, it is important that the simulations match the data fairly well. 

Improving the agreement between simulations and data compared to~\citetalias{FC17} was one of the initial objectives of this work and the way we \newtext{achieved it is by improving the fidelity of the input catalogue}, which we describe in this section. Rather than using distributions of galaxy properties, we use morphological parameters determined from HST observations of the COSMOS field. These are used to simulate the COSMOS field under the same seeing conditions of the VST observations to show that we are able to recover the observations of that field. Under the assumption that the galaxies in the COSMOS field are representative of the whole population, we vary the PSF parameters to sample the seeing conditions of KiDS. 
\subsection{Input object catalogue}
\label{subsec:input_catalogue}
\begin{figure}
 \includegraphics[width=\columnwidth]{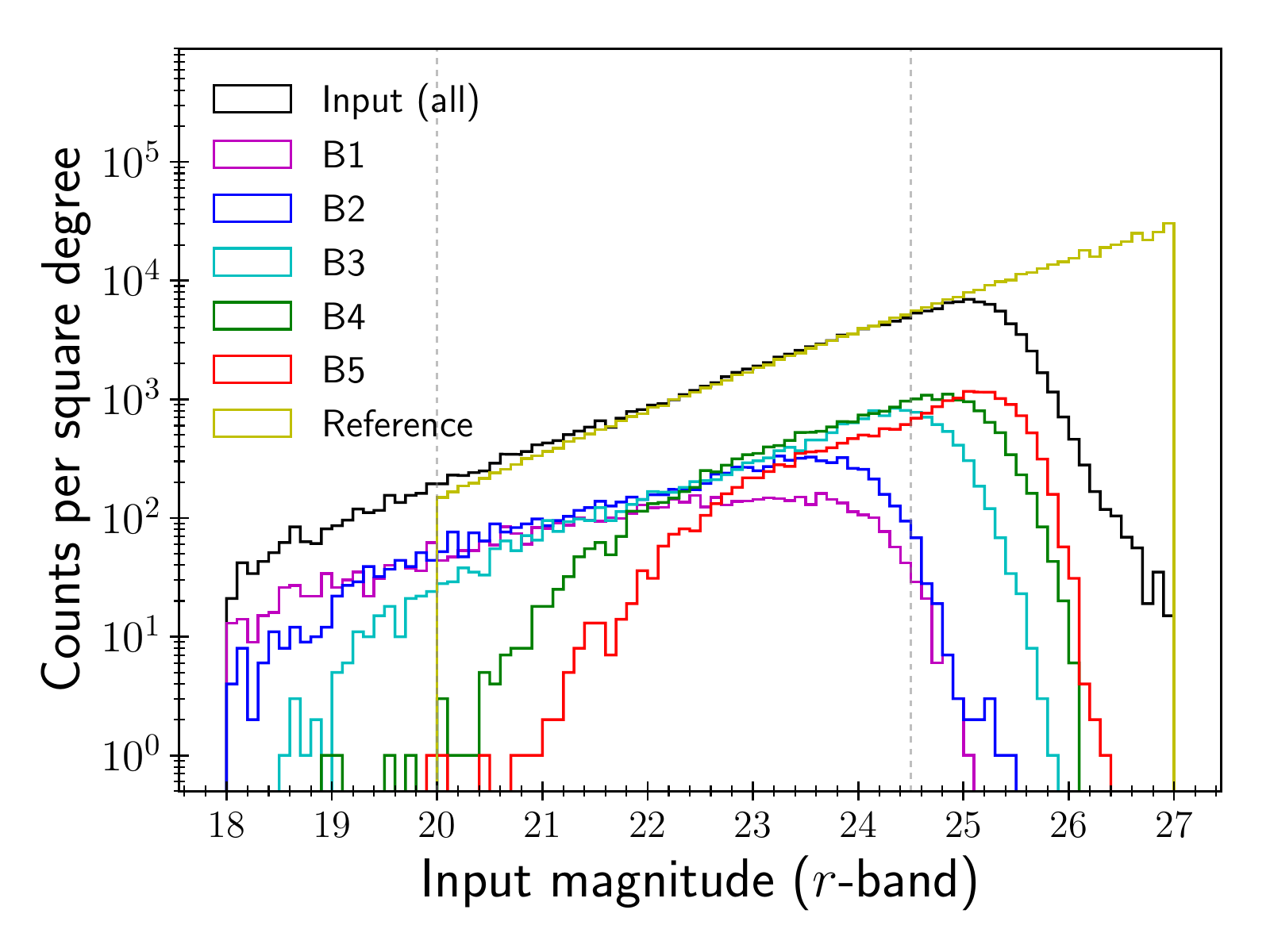}
 \caption{Distribution of input magnitudes for all the galaxies (black) and the distributions when the galaxies are divided into  the tomographic bins based on their `true' redshifts. The analytic magnitude distribution used in \citetalias{FC17} is given in yellow for reference. The region enclosed by the vertical lines denotes the range in the output magnitude for which shapes are measured.}
 \label{fig:input_mag_hist}
\end{figure}
A key improvement compared to \citetalias{FC17} is the use of observed positions and structural parameters of galaxies in the COSMOS field. 
For this we use the publicly available ACS-GC catalogue from \cite{Griffith2012}, based on \sersic\ model fits to ACS imaging data in the $F814W$ filter. The catalogue contains 304\,688 objects for which the best-fit \sersic\ parameters are reported (this includes stars as well). These were obtained by fitting PSF-convolved \sersic\ profiles to each source using \galfit~\citep{Peng2002}, which determines the best-fit parameters  using the Levenberg-Marquardt algorithm for $\chi^2$ minimisation. 
The best-fit \sersic\ parameters are based on the observations in the $F814W$ filter, whereas the shapes for KiDS are measured in the $r$-band. In our analysis, we implicitly assume that the galaxy shapes do not depend on the pass-band used. Given the relatively modest difference in wavelength this is not a major concern, although we note that galaxies do appear to be somewhat rounder at longer wavelengths \citep[e.g.][]{Schrabback2018}. 
We found the position angle measurements to be noisy and biased, with a preferred intrinsic orientation. The noise in the position angles is evident in Fig.~\ref{fig:emulation}. The bias in the position angles is not of a major concern, because we rotate the galaxies to cancel shape noise (see~\cref{sec:sim_setup}).

For every unmasked object in the KiDS observation of the COSMOS field, we find its `best' match in the ACS-GC catalogue using the positions and magnitudes in both catalogues. Since the two catalogues have different depths and hence different number densities, a symmetric match is not possible. The best match is identified by finding the four nearest neighbours in the ACS-GC catalogue in position and magnitude for every object in the KiDS catalogue. \newtext{Na\"{i}vely, every object in the KiDS catalogue may be expected to be matched with exactly one galaxy in the ACS-GC catalogue, although this is not the case}.  Due to differences in detection threshold and noise in the images, a small number of objects in the KiDS catalogue end up matched to the same object in the deeper ACS-GC catalogue. A unique one-to-one matching is obtained in the following manner. For every object in the ACS-GC catalogue with multiple matches, the pair with the smallest distance is retained and all other objects in the KiDS catalogue are matched with its next nearest neighbour. We iterate over the multiply matched objects 6 times, and if an object in the KiDS catalogue has found no unique match, it is discarded. Such discarded objects account for about 0.07\% by weight and hence do not affect our calibration in a significant manner.

We remove stars from this matched catalogue using the following criteria: {\sc SExtractor} {\tt CLASS STAR} $ <0.9$ and {\tt FLAG{\_}GALFIT{\_}HI}=0, indicating that no problems with the fit were reported by \galfit. Comparison to objects identified as stars in a  magnitude-size diagram using the HST values showed that this removed most\footnote{In order to quantify the impact of the remaining stars contaminating our galaxy catalogue, we later adopted a stricter star selection criterion and removed objects from the output catalogues post-simulation by imposing a magnitude-dependent size cut on the HST catalogue. We found the impact on the multiplicative bias to be negligible ($\sim 0.002$). } of the star candidates. We also discard matches between the catalogues if the corresponding entries are separated by more than $1$ arc second . Finally, we require that the reduced chi-square values for the \sersic\ fits satisfy $\chi_\nu^2 < 1.5$.

For objects in the ACS-GC, we assign the $r$-band magnitude measured in the KiDS survey if a match is found. For the remaining objects, which are typically galaxies below the magnitude limit of KiDS, we assign the galaxy the Subaru $r^+$ magnitude provided by \cite{Griffith2012}. We compared the magnitudes and found they agree fairly well: the mean difference\footnote{In order to ensure that the small difference in magnitudes is caused by the matching in magnitude-position space, we also matched only based on the sky positions and we found similar differences.} in the magnitude is about $0.07 \pm 0.35$, with the faint galaxies contributing to the majority of the scatter. 

Fig.~\ref{fig:input_mag_hist} shows that the resulting galaxy catalogue is complete to $m_r \lesssim 25$, after which the counts decrease rapidly. 
The bright end of the magnitude number counts is described well with the analytic magnitude distribution
\begin{equation}
\log N(m) = -8.85 + 0.71m - 0.008m^2
\end{equation}
used in~\citetalias{FC17}, where $N(m)$ refers to the number of galaxies per square degree with magnitudes between $m \pm 0.05$. We refer to this as the `reference' distribution in Fig.~\ref{fig:input_mag_hist}.

\begin{figure}
    \centering
    \includegraphics[width=\columnwidth]{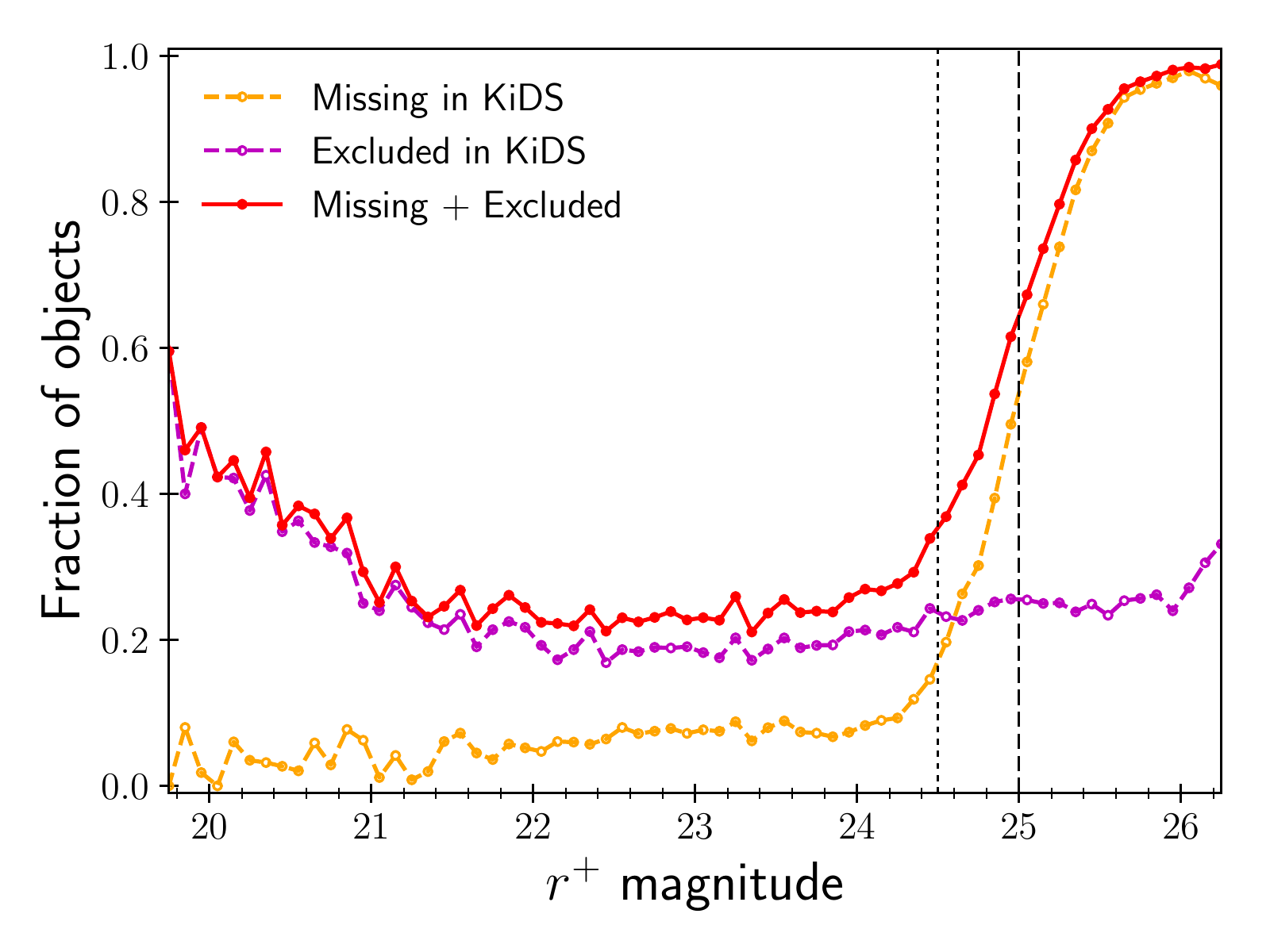}
    \caption{Fraction of the objects in the HST catalogue that are missing, or excluded from the simulations plotted against the Subaru $r^+$-band magnitude.}
    \label{fig:goodness_matching_MAGR}
\end{figure}

The orange line in Fig.~\ref{fig:goodness_matching_MAGR} shows the fraction of objects that are in the ACS-GC catalogue, but were not detected in the KiDS imaging of the COSMOS field. At bright magnitudes most objects are matched, and the differences can be attributed to blending, etc. The fraction increases rapidly beyond $m_{r^+}=24.5$, where the KiDS catalogue is incomplete. This does suggest, however, that the limiting magnitude of the input catalogue is sufficient to simulate the images of the brighter galaxies that are above the detection limit in the KiDS data. We test the sensitivity of \lensfit\ to these faint galaxies in \cref{sec:sensitivity_analysis}.

We find that 83\% of the objects detected in the KiDS data are matched to the catalogue from \cite{Griffith2012}. Of the matched objects 91\% are galaxies and 84\% of these have \sersic\ parameters that can be simulated by \galsim~\citep{Rowe2015}. \cite{Griffith2012} also report the reduced $\chi^2$ value for the best-fit \sersic\ model, and 94\% of the useful galaxies have reported values $<1.5$,  suggesting that the model is a decent fit to the images. The final sample of galaxies that we simulate comprises 142\,869 galaxies.

\begin{figure*}
    \centering
    \hbox{
    \includegraphics[width=0.49\textwidth]{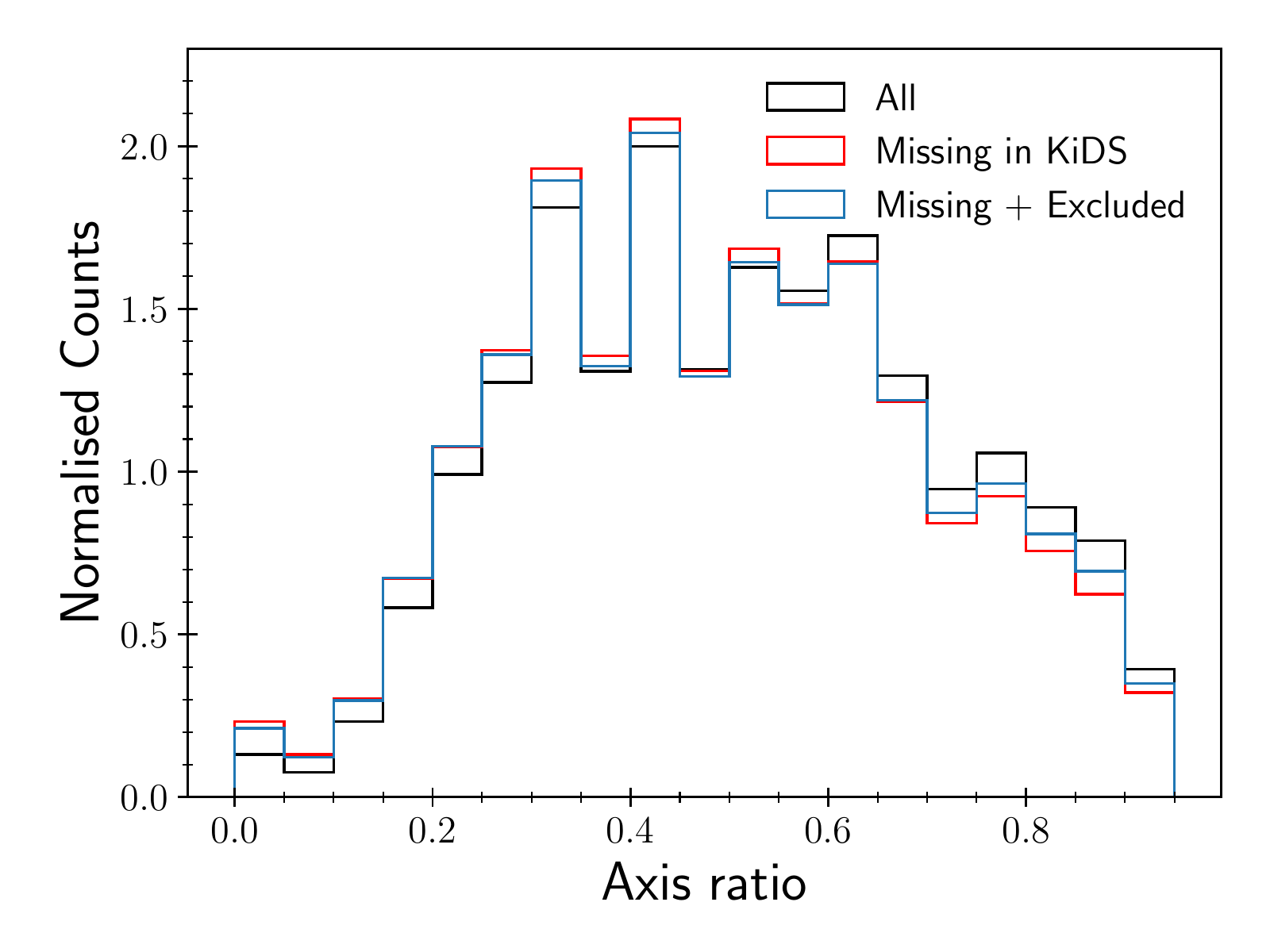}
    \includegraphics[width=0.49\textwidth]{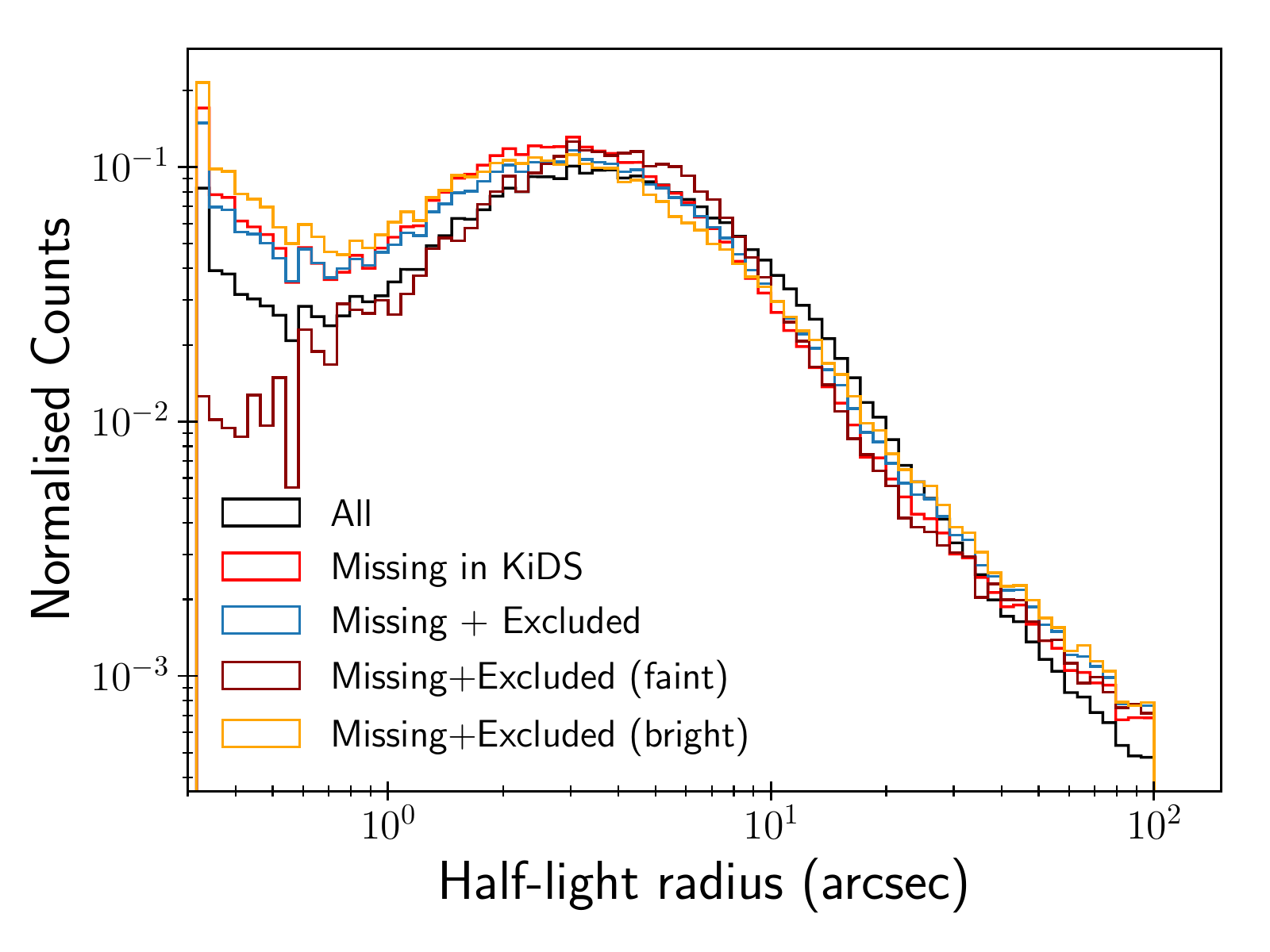}}
    \caption{{\it Left panel:} Distribution of axis ratios of galaxies determined by \galfit\ from ACS imaging of COSMOS. {\it Right panel:} Distribution of half-light radii of galaxies determined by \galfit\ from ACS imaging of COSMOS. Not all galaxies in the HST catalogue were detected in KiDS, but this does not appear to affect the ellipticity distributions significantly. Faint galaxies ($m_{r}>25$) that are also smaller in size are preferentially detected in KiDS, but no such trend is found among the bright galaxies ($m_{r}\le 25$) }
    \label{fig:goodness_matching}
\end{figure*}

By applying these exclusion criteria, we omit more than $20\%$ of the bright galaxies from the HST catalogue, as indicated by the purple line in Fig.~\ref{fig:goodness_matching_MAGR}. If these galaxies are not a representative population, removing them could introduce a selection effect in the multiplicative bias estimates. To verify that we do not exclude any particular sub-population of galaxies from the HST catalogue, and we show results for the two most important parameters, namely ellipticity and size defined as the half-light radius along the major axis in Fig.~\ref{fig:goodness_matching}. The left panel in Fig.~\ref{fig:goodness_matching} shows that the distribution of axis ratios\footnote{The ellipticity $\epsilon$ is related to the axis ratio $q$ via $|\epsilon|=(1-q)/(1+q)$.} remains unchanged after excluding the objects that did not meet the selection criteria. Similarly, the results in the right panel of Fig.~\ref{fig:goodness_matching} show that for the galaxies that are detected in the KiDS data, the size distribution also remains unchanged after exclusion (red vs. blue histograms). When we split the galaxies into broad magnitude bins as bright and faint galaxies, at the detection limit of KiDS, we see that the size distribution changes. This is caused by the incompleteness of the HST catalogue beyond $m_{r^+} > 25$, with preferentially compact galaxies ($r_e \lesssim 3\farcs{}$) making the cut for a given magnitude.

The HST COSMOS catalogue used for this work is substantially deeper, and has a higher number density {than the~\cite{COSMOS_Alexie} catalogue that is the default input catalogue in \galsim, and is used in both the DESy1 and HSC-DR1 image simulations (parent samples 1-3 in HSC-DR1). \citetalias{Mandelbaum2018a} discuss in detail what impact the HST selection and goodness criteria have on the input sample, and thereby on the shear biases. In particular, they mention that their goodness criteria exclude close pairs of galaxies separated by less than $2\farcs{4}$; effectively all galaxy pairs with a separation below $0\farcs{9}$ are removed. The impact of excluding such close pairs of galaxies was shown to change the multiplicative bias by $\sim 0.1$, which exceeds the systematic error budget for all ongoing surveys. The fiducial parent sample (parent sample 4 in the HSC-DR1 simulations) used in the HSC-DR1 simulations includes HST images without any such selection criteria. The parametric catalogue (parent sample 2 in the HSC-DR1 simulations) are obtained after such exclusion cuts are applied. In contrast, our input catalogue~\citep{Griffith2012} was processed differently and \sersic\ fits were obtained prior to any selection cuts. We therefore do not expect to suffer from the exclusion of close galaxy pairs. For every object in each of the two HST catalogues, we computed the distance to its nearest neighbour and found (not shown here) that there were many more close neighbours in the~\cite{Griffith2012} catalogue than in~\cite{COSMOS_Alexie} catalogue. A direct comparison of our simulated image with the corresponding VLT image in Fig.~\ref{fig:emulation} indicates that the simulations include realistic clustering of galaxies on the sky.}

\subsection{Correlations between input parameters}
\label{sec:input_cor}

We use observed parameters based on high-quality HST observations to simulate KiDS data. This approach naturally accounts for potential correlations between input parameters, such as size, magnitude, and importantly, ellipticity. Moreover, we have photo-$z$ estimates for the galaxies. This enables us to examine the shape measurement biases consistently for the different tomographic bins, which have different magnitude distributions as shown in Fig.~\ref{fig:input_mag_hist}.

Throughout this work, by size, we refer to the half-light radius measured along the major axis of the best-fit elliptical profile, and denote it as $r_e$. We will refer to the azimuthally averaged size, defined as $r_{ab} \equiv r_e\sqrt{q}$, where $q$ is the axis ratio as the circularised size. In \citetalias{FC17} the input parameters were drawn from the \lensfit\ prior distributions, which introduces a correlation between size and magnitude. The ellipticities, however, are drawn independently from the distribution provided by \cite{Miller2013}. This was motivated by noting that that there is no correlation expected between size $r_e$ and ellipticity for disc-like galaxies, as the ellipticity is caused due to the inclination angle relative to the line-of-sight. This is an important assumption, because at each step in the cosmic shear analysis cuts on size and magnitude are made (implicitly by dividing the sample into tomographic bins -- see Fig.~\ref{fig:input_mag_hist}). If the ellipticity distribution depends on any of these, a cut in one parameter results in an implicit cut on the galaxy shape, thus biasing the shear estimate. We therefore examined if the input parameters correlate with the ellipticity. Since the half-light radius $r_e$ is degenerate with the \sersic\ index $n$, we first look if the \sersic\ index correlates with the ellipticity.

\begin{figure}
\centerline{\includegraphics[width=0.49\textwidth]{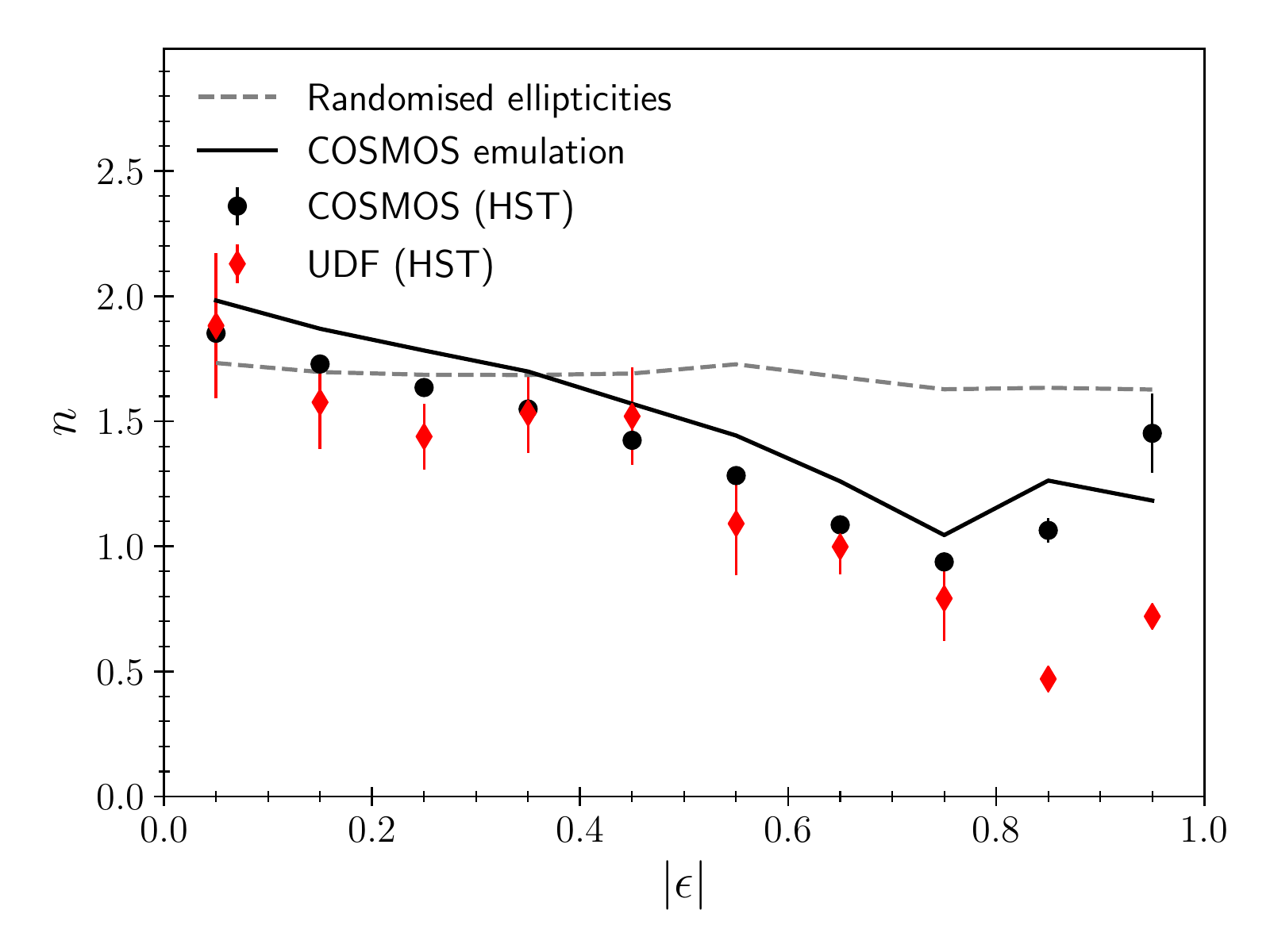}}
\caption{Average \sersic\ $n$ value as a function of ellipticity for galaxies with $20<F814W<24.5$  in COSMOS \citep[black points;][]{Griffith2012} and the UDF \citep[red points;][]{Coe2006}. The solid line indicates the average $n$ as a function of $\epsilon$ used in the image simulations (for galaxies with $20<m_r<25$), whereas the dashed lines is for the case we scramble the ellipticities to resemble the \citetalias{FC17} results more closely.}
\label{fig:n_vs_e}
\end{figure}

The black points in Fig.~\ref{fig:n_vs_e} show the average \sersic\ index $n$ for galaxies with $20<F814W<24.5$ as a function of ellipticity $\epsilon$ using the measurements from the ACS-GC catalogues by \cite{Griffith2012}. We observe a clear trend between $n$ and ellipticity, with more elliptical galaxies having a lower \sersic\ index.
Not surprisingly, the input catalogue, after the various cuts applied, shows the same trend (solid line), whereas the inputs from \citetalias{FC17} do not show any correlation (dashed line).

\begin{figure}
\centerline{\includegraphics[width=0.49\textwidth]{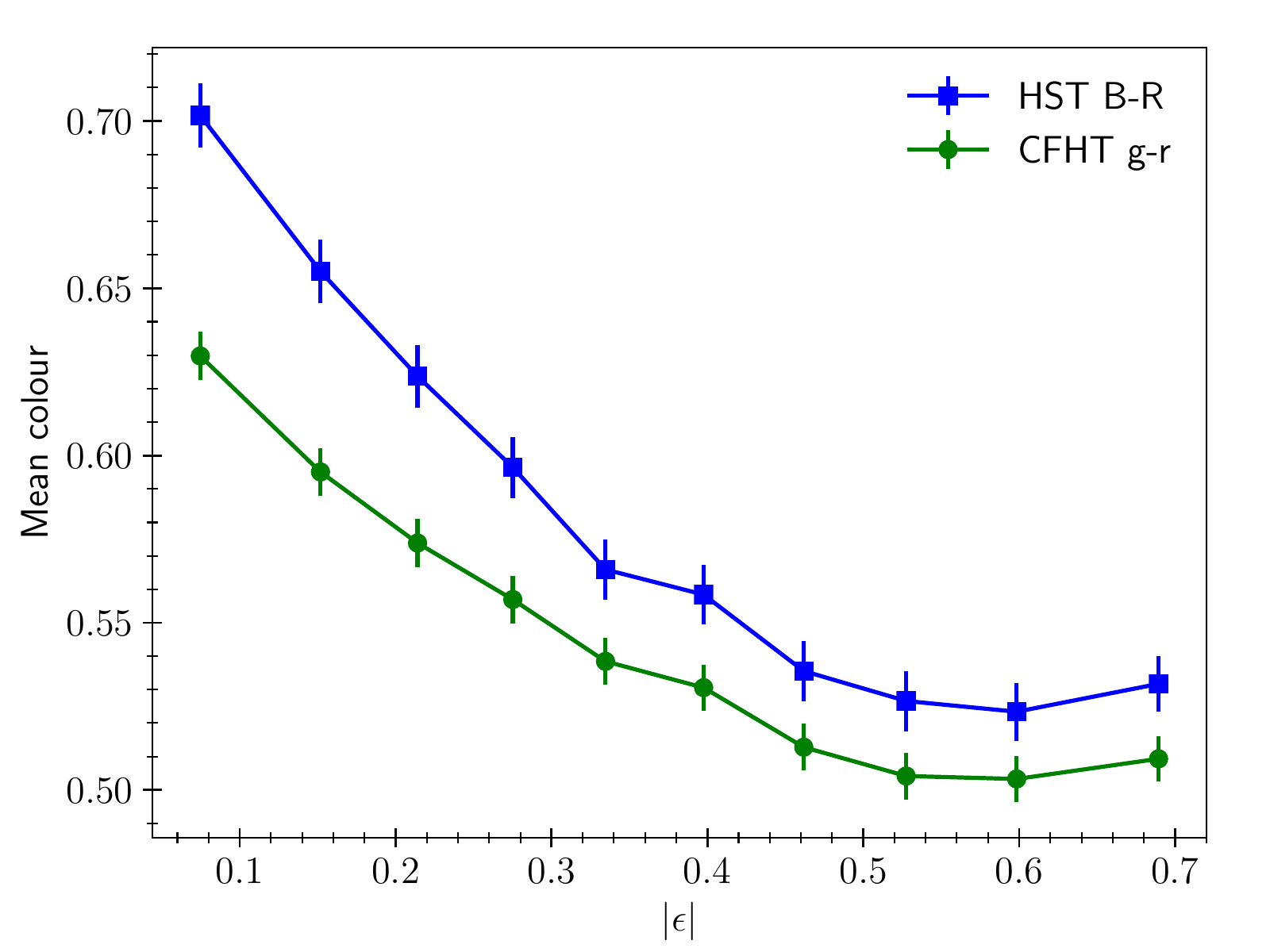}}
\caption{Mean galaxy colour as a function of ellipticity $\epsilon$. Using colour as a proxy for morphology, the results support the trend of \sersicn\ vs. ellipticity from Fig.~\ref{fig:n_vs_e}.}
\label{fig:e_vs_col}
\end{figure}

Although the structural parameters obtained by \cite{Griffith2012} are based on high-quality HST observations, we need to ensure that noise in the images does not introduce spurious correlations in the parameters found by \galfit. {This is a concern, given the evidently different orientations of the simulated galaxies in Fig.~\ref{fig:emulation}, despite the use of the position angles measured by \galfit.} After all, noise in the images is a major cause of parameter degeneracies and introducing such correlations in the image simulations affects the performance of \lensfit. To evaluate the robustness of the other  measurements in~\cite{Griffith2012}, we compare the results to \galfit\ measurements for galaxies in the Hubble Ultra Deep Field (UDF) presented by \cite{Coe2006}. These data are considerably deeper than the COSMOS observations, albeit over a much smaller area of only about 11 arcmin$^2$. We find that the distributions of \sersic\ index, effective radius and ellipticity are comparable between the two data sets for galaxies with $20<F814W<24.5$. Moreover, the red points in Fig.~\ref{fig:n_vs_e} are consistent with the COSMOS results. We therefore conclude that the trend is not caused by noise in the shallower COSMOS data, but reflects a real trend. \newtext{Similar trends are also observed by~\cite{Hill2019} for galaxies in CANDELS/3D-HST fields~\citep{Brammer2012,Skelton2014,Momcheva2016}, thereby confirming our conclusion. } 

To explain this trend, we hypothesise that in a mixed population of galaxies, old \newtext{(quiescent)} red galaxies with a dominant bulge component are typically more spherical (low ellipticity, high $n$) whereas young \newtext{(star-forming)}, blue disc-dominated galaxies appear highly elliptical when oriented edge-on to us. In this scenario, a correlation between $n$ and $\epsilon$ is expected. As a test of our hypothesis we show the mean galaxy colour as a function of ellipticity in Fig.~\ref{fig:e_vs_col} for two different filter combinations. The trend of bluer galaxies having larger ellipticities is consistent with our hypothesis. Since the photometry is measured independently from the galaxy profile fitting,
it seems unlikely that spurious dependencies are be introduced in the \sersic\ parameters due to noise alone.

\newtext{Our hypothesis is also supported by the findings of~\cite{Hill2019}, who split the galaxies into the quiescent and star-forming galaxies based on their rest frame $U-V$ and $V-J$ colours. They report that at fixed stellar mass, and for $z<2$, quiescent galaxies are rounder than their star-forming counterparts.}
Based on the physical argument and consistency check with the UDF (and CANDELS/3D-HST), {we conclude that the correlations between the various \sersic\ parameters are in fact real, and not merely an artefact of fitting a profile to noisy galaxy images. We find that $r_e$ of the best-fit \galfit\ model depends on the ellipticity of the model. Moreover, as discussed below, we are able to reproduce the correlations in the measured size and ellipticity very well (see Fig.~\ref{fig:size_e_comparison}).}

The correlation between galaxy colour and ellipticity also implies that when tomographic cuts are imposed based on photometric redshifts, which are essentially complex decision boundaries in multi-dimensional colour space, an implicit ellipticity selection occurs. In our mathematical framework of~\cref{sec:calibration_math}, different tomographic samples correspond to different selection functions $t(\vec{h}|\vec{D})$ which have a non-trivial dependence on $\vec{h}$ through the ellipticity distribution based on Fig.~\ref{fig:e_vs_col}. Because the intrinsic ellipticity distributions differ, we expect some contribution to the multiplicative bias from $\Delta t$ terms if this selection is not accounted for correctly. We discuss the impact of this selection on the multiplicative bias later in~\cref{sec:calibration_results}.

\subsection{Comparison to VST observations of COSMOS}
\label{sec:compare_sims_data}

As explained in \cref{sec:calibration_math}, the simulated data need to match the observations for a robust calibration of the shear measurement pipeline. \citetalias{FC17} could only compare output distributions. Although useful, distributions can match for the wrong reason. 
Since our image simulations aim to emulate the COSMOS field, they must match the COSMOS data closely when the observing conditions in the simulations are similar to those in the VST observations, on an object-by-object basis. Such a comparison is therefore more stringent than simply comparing distributions.

\begin{figure}
\includegraphics[width=0.5\textwidth]{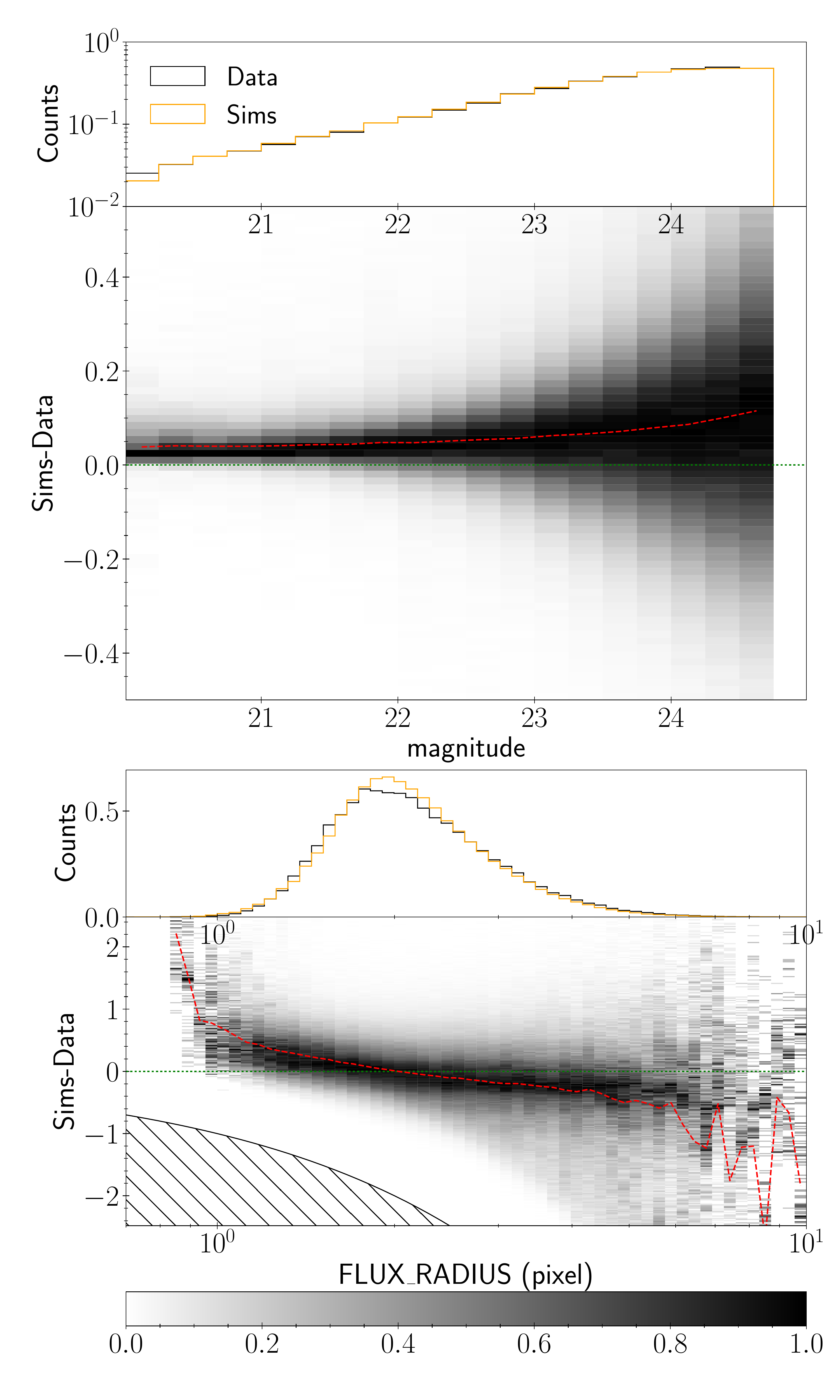}
\caption{Object-by-object comparison of \sextractor\ quantities. The upper panels show the unweighted histograms for measurements from the KiDS COSMOS field and our simulated data. The lower panels show 2-dimensional histograms with the difference between the simulations and the data along the vertical axis and the quantity of interest along the horizontal axis. Each vertical column is normalised such that each peak in the 1-dimensional slices of the 2-dimensional histogram is normalised to unity. This improves the histogram visually, allowing us to see the contrast across the full parameter range. The shaded region in the bottom figure indicates where the corresponding quantities in the simulations are negative. The red dashed lines show the median values, which indicate that the agreement between the simulations and the data is generally good.}
\label{fig:compare_individuals_heatmap_SE}
\end{figure}

\begin{figure}
\centerline{\includegraphics[width=\columnwidth]{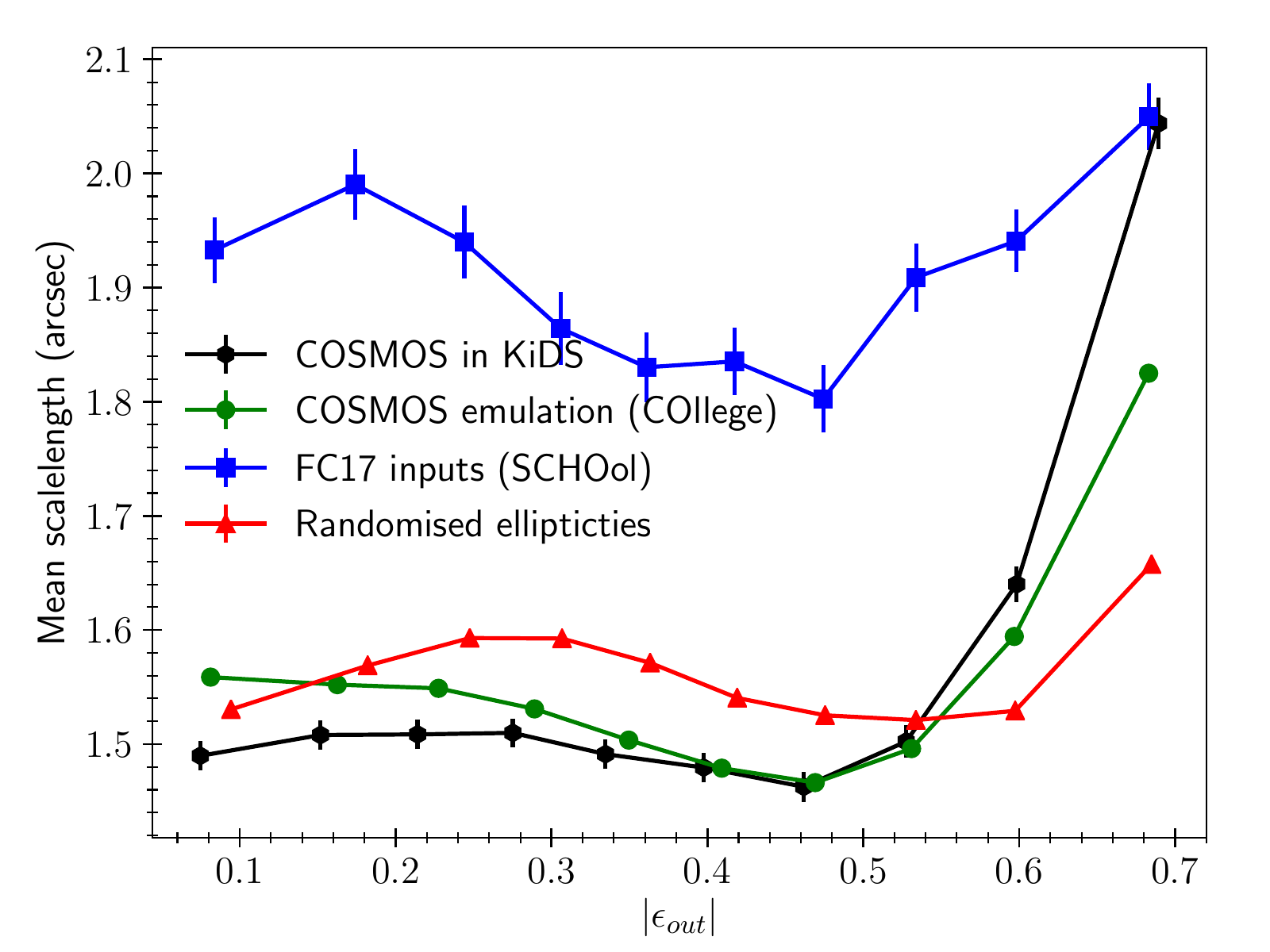}}
    \caption{ Mean \lensfit-measured scalelength, measured along the major axis as a function of \lensfit-measured ellipticity $\epsilon_{\rm out}$. The input in~\citetalias{FC17}, which is based on \lensfit\ priors, assumes that the scalelength is uncorrelated with the ellipticity. However, measurements on the KiDS data show a strong tendency, which our improved simulations capture very closely.}
\label{fig:size_e_comparison}
\end{figure}

\begin{figure*}
\centering
\includegraphics[width=1.05\textwidth]{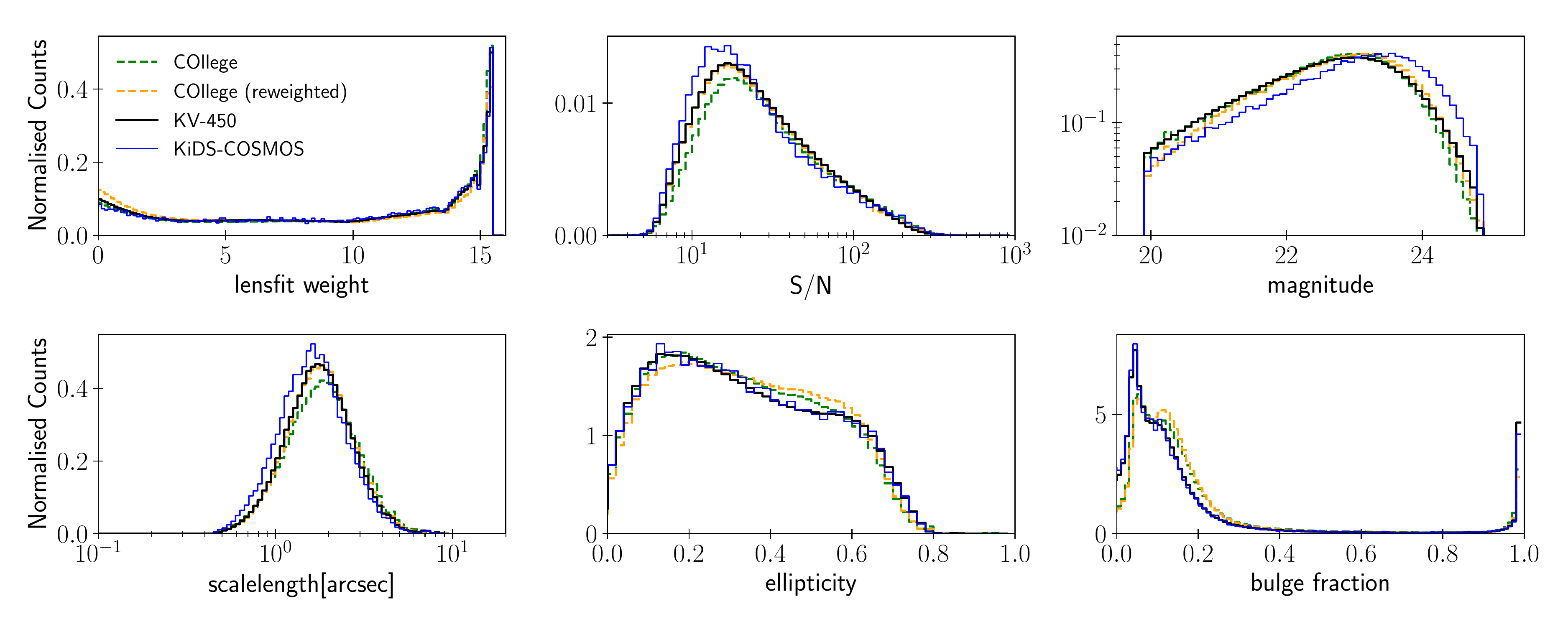}
\caption{Comparison of \lensfit -weighted normalised distributions of galaxy properties between KV-450 data (black), KiDS observations of COSMOS (blue), \college\ simulations (green) and \college\ simulations after re-weighting (orange) the simulated galaxies to match the joint distribution of signal-to-noise ratio and size in the KV-450 dataset (c.f.~\cref{sec:calibration_results}). No redshift cuts are applied here, but similar level of agreement can be seen for each tomographic bin individually.
}
\label{fig:compare_sims_data}
\end{figure*}

As a first step, we detect objects using \sextractor\ in both the simulated and observed stacked $r$-band image. Fig.~\ref{fig:compare_individuals_heatmap_SE} shows the difference between the magnitudes and half-light radii ({\tt FLUX\_RADIUS}) in the simulations and the data. The average difference in magnitude is generally small. The size estimates from \sextractor\ also agree well, but the overall distributions differ slightly. The apparent trend in the lower panel of Fig. ~\ref{fig:compare_individuals_heatmap_SE} is partly caused by noise in the images, as it traces the envelope of unphysical sizes indicated by the hatched region. 

The agreement between the \sextractor\ output from both the simulated and actual data is encouraging. In~\cref{app:lf_comparison} we present results of a similar comparison of several \lensfit\ output parameters, namely the weight, S/N, and the scale-length of the disc model. These parameters are used to calibrate the shear measurement bias, and the good agreement between the simulated and observed data on an object-by-object basis suggests that the resulting estimate of the bias should be robust. This is a significant improvement compared to \citetalias{FC17}, where the simulations contained too many small galaxies.

{The \lensfit\ measurements made on the COSMOS field of the KiDS data show that the average measured scale length depends on the observed ellipticity, as shown in Fig.~\ref{fig:size_e_comparison}. Our improved \college\ simulations match the observations fairly well. Comparison to simulations where we randomise the input ellipticities indicates that part of the sharp upturn for high values of $|\epsilon|$ is introduced by \lensfit\ itself. A similarly weak variation is seen for the \citetalias{FC17}-like simulations (the shift towards larger sizes was already noted earlier), where the input ellipticities did not depend on other parameters. However, the \lensfit\ measurement itself cannot fully explain the observed trend, which we therefore attribute to intrinsic correlations between galaxy size and ellipticity.} 
\newtext{~\cite{Hill2019} also observe the same trend between \galfit-measured size and ellipticity, specifically for star-forming galaxies}. These suggest that any sample selection which implicitly or explicitly places cuts on galaxy sizes also place a cut on galaxy ellipticity, and change the ellipticity distribution of the sample. This has an impact on shear calibration, and as we show in~\cref{sec:calibration_results}, this affects the estimate of multiplicative bias in a significant manner. Referring back to~\cref{sec:calibration_math},  because we include $r_e$ in $\vec{D}$ and $|\epsilon|$ in $\vec{h}$, $r_\text{sims}(\vec{h}|\vec{D})$ is rather flat for the simulations in~\citetalias{FC17} and when we randomise the ellipticity, but 
this is not so for the real data. The excellent agreement between our fiducial simulations and the data suggests that $\Delta r$ terms for our fiducial calibration simulations must be small. We do not expect the prior distributions to affect the \lensfit\ measurements significantly, and therefore we do not investigate the impact of the assuming an incorrect prior in \lensfit.

\subsection{Comparison to KV-450 data}
\label{sec:compare_sims_KV450}

Having established that we can emulate the VST observations of the COSMOS field, the next step is to use simulations to quantify the shear bias for each tomographic bin used in the KV-450 cosmic shear analysis. To do so, we need to simulate data with observing conditions sampled from the actual distribution in KiDS, because the bias could depend on the PSF size and sky background level, which in turn mildly affect the observed galaxy sample. Treating the input catalogue from COSMOS as a representative sample, we can then determine the bias in the shear estimates from \lensfit.~\citetalias{FC17} have shown that the multiplicative bias does not depend explicitly on the PSF properties. {Therefore, unlike the image simulations in DESy1~\citep{Samuroff2018,Zuntz2018}, we do not include any spatial variations of the PSF}.

We do not change the background noise level in the images, and consider the same 65 PSFs (i.e., 13 PSF sets, each set containing five PSFs corresponding to five exposures) used in \citetalias{FC17} for our fiducial simulations. A more comprehensive study, simulating the full survey in multiple filters to better capture the determination of photo-$z$s, is deferred to future work. As shown in Appendix~\ref{app:PSF_modelling} the resulting PSFs only coarsely sample the actual distribution of PSF properties observed in the KV-450 $r$-band images. Fortunately, the distribution of \lensfit\ output parameters does not change significantly when the PSF is varied.

Figure~\ref{fig:compare_sims_data} shows the marginal distributions of some of the observable properties of galaxies, with each galaxy weighted by its \lensfit\ weight. 
The simulated distributions (green dashed histograms) match the KV-450 data well (black histograms). We observe some differences at low S/N and small galaxy sizes, which may be result from the absence of variations in the noise levels. In reality the background levels vary between exposures, which will affect the fainter, smaller galaxies the most. Moreover, scattered light in the telescope optics causes additional spatial variations in the background in the KiDS data, whereas such complications are absent in our simulations. To account for these small mismatches we re-weight the simulations to improve the match with the data for the final shear calibration, presented in~\cref{sec:calibration_results}.

Initial comparisons of our simulation output and the KV-450 measurements showed good agreement for unweighted quantities, but failed to reproduce the weighted distributions. Further investigation revealed that the culprit was an error in the weight-recalibration of the KiDS-450 data set, not a problem with the simulations. This has been corrected in~\citetalias{Hildebrandt2019}, resulting in the excellent agreement presented in Fig.~\ref{fig:compare_sims_data}. Comparison to Fig.~3 in \citetalias{FC17} shows that the agreement with the data has improved dramatically. The realistic input catalogue plays a crucial role in this, but applying the correct weight-recalibration scheme (to both data and simulations) also affected the results for KiDS-450 analysis~\citepalias{Hildebrandt17} ever so slightly~\citepalias{Hildebrandt2019}. 

\section{Selection biases}
\label{sec:bias_in_sims}

The multiplicative bias in the shear estimated from a population of galaxies not only depends on our ability to determine the ellipticity from noisy data, but is also affected by subtle biases introduced during the detection and sample selection stage. We refer to these as selection biases, which cause biased shear estimates, even if the shape measurement itself is unbiased. As demonstrated in \citetalias{FC17}, selection biases cannot be ignored for current (and future) cosmic shear surveys. In this section, we quantify the impact of the various examples of selection bias introduced at different stages in the analysis \newtext{before proceeding to estimate in~\cref{sec:calibration_results} the multiplicative bias for each of the tomographic bins.}

\subsection{Bias due to source detection}
\label{sec:sextractor_bias}
The detection of objects by \sextractor\ can already introduce a selection bias if  the probability of detecting a galaxy depends on whether it is aligned (or anti-aligned) with the shear or the PSF. This was {first mentioned in~\cite{HS03} and discussed in the context of the KiDS data} in Section 4.2 of~\citetalias{FC17} and we revisit it here in this section.
To estimate the detection bias, we matched every detected source to the corresponding object in the input catalogue so we can use their true ellipticities (which include the shear), which are by definition unbiased. We discard a small fraction of false positive detections (noise peaks), which have no corresponding object in the input catalogue. Of the matched objects, we only remove the stars, thus assuming a perfect star-galaxy separation (although keeping the stars made no significant difference). This is motivated by the fact the contamination by stars is low after the determination of photo-$z$s and the classification of stars by  \lensfit. 

\begin{figure}
    \centering
    \includegraphics[width=\columnwidth]{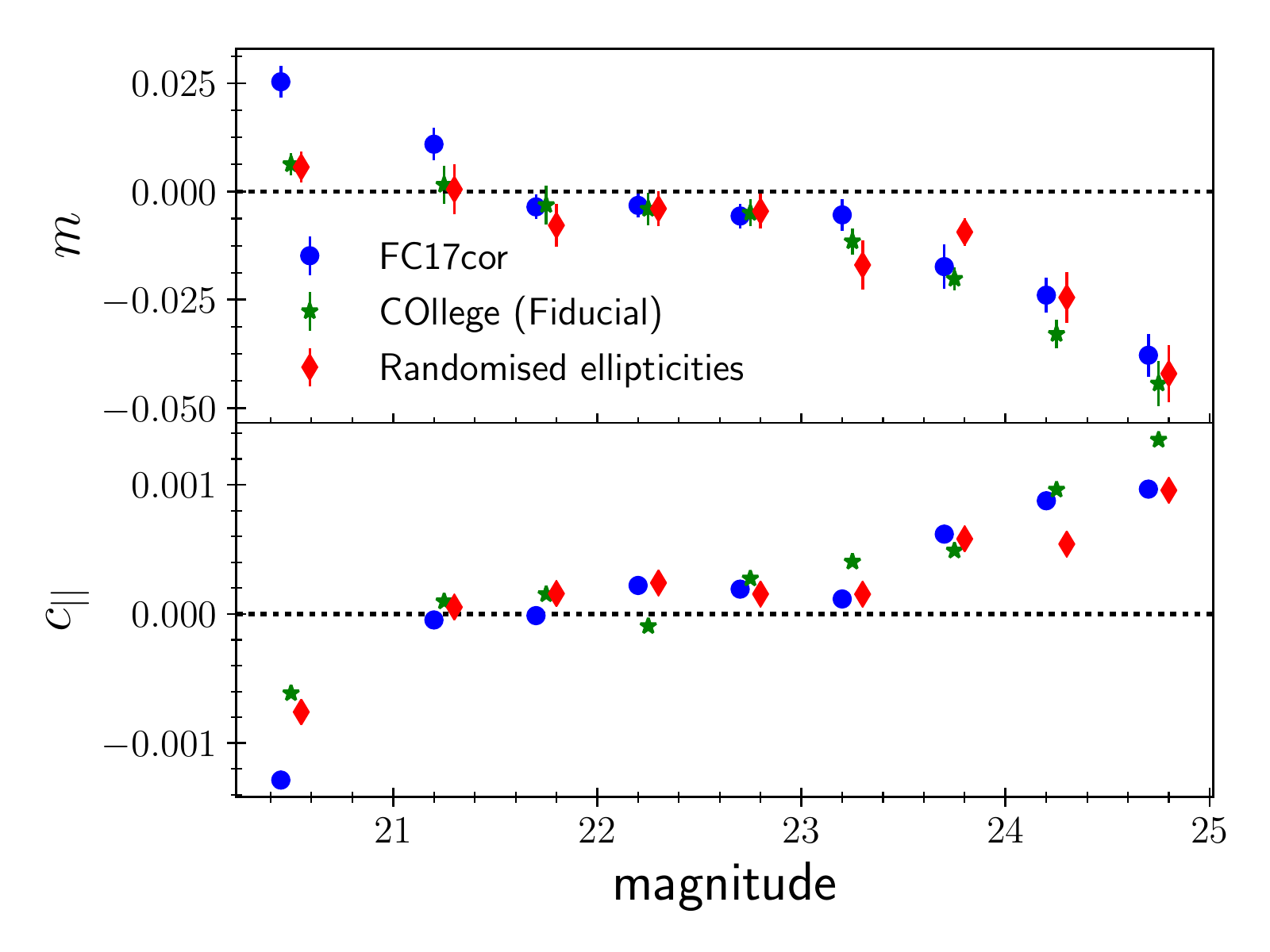}
    \caption{Shear selection bias introduced by \sextractor\ detection as a function of the measured magnitude.
    Upper panel: Multiplicative bias $m$ along the PSF direction and at $45^\circ$ to it. Lower panel: Additive bias in PSF frame, for PSFs with $\avg{|\epsilon_\text{PSF}|}\sim 0.025$.
    Faint galaxies can exhibit a shear bias as large as $-0.05$ after the detection step, prior to shape measurements. The points are slightly offset in the horizontal direction for visual clarity.}
    \label{fig:selection_bias_tomobin}
\end{figure}

For the full sample of detected galaxies we obtain an average multiplicative bias of about -0.025 (we find comparable values for both $m_1$ and $m_2$). As PSF anisotropy is a potential source of selection bias (see the lowel panel of Fig.~\ref{fig:selection_bias_tomobin}) we also determined the multiplicative selection bias in the frame parallel to the major axis of the PSF ($m_{||}$) and in the direction $45^\circ$ to it ($m_\times$). The results are similar to the $m$-values in the detector frame of reference, suggesting that PSF anisotropy is not the dominant cause. Instead, the negative value indicates that \sextractor, at least with the parameters chosen to analyse the KiDS data, preferentially detects objects that are not aligned along the direction of the shear. This may be attributed to \sextractor\ using a circular kernel to detect objects, and thereby having a slight preference to identify circular sources as opposed to elliptical sources at low signal to noise ratios. We emphasize that the detection bias precedes shape measurement, and is therefore independent of the shape measurement algorithm \newtext{in general. However, some methods such as~\cite{Bernstein2016} are beginning to incorporate their own corrections to account for detection biases (among other selection biases) that are somewhat algorithm-dependent}. The exact magnitude of the bias will depend on the observing conditions and the detection algorithm. 

The green stars in the upper panel of Fig.~\ref{fig:selection_bias_tomobin} show the multiplicative \sextractor\ detection bias as a function of magnitude for the KV-450 emulation. The bias is small for bright objects, but increases to almost $-0.05$ near the detection limit. In contrast, we find that the detection bias is negligible for the original \citetalias{FC17} analysis, because the detection was done only on the original image and not on the other three images where the galaxies were rotated (also see \cref{sec:comparison}). This causes an artificial cancellation of any net ellipticity, which does not occur in the real data. Thus, enforcing almost perfect shape noise cancellation by not using an independent detection catalogue or by assigning same weights to the rotated pairs of galaxies, as done for the HSC-DR1 shear calibration in~\citetalias{Mandelbaum2018a}, will miss the contribution to shear bias due to selection effects. We note that in \citetalias{FC17}, we detected a much smaller level of multiplicative and additive bias at the \sextractor\ stage, as a result of using a single detection image which cancelled any additive detection bias. If we instead follow our more realistic implementation of running \sextractor\ on each rotation of the \citetalias{FC17} images, we obtain the blue circles in Fig.~\ref{fig:selection_bias_tomobin}. These agree well with our results of the fiducial simulation, which suggests that the observed correlations between structural parameters and the ellipticity (see \cref{sec:input_cor}) are not important at this stage. We verified this explicitly by measuring the selection bias where the ellipticities were assigned randomly to the COSMOS galaxies. Those results (red diamonds) agree fairly well with the other points shown in Fig.~\ref{fig:selection_bias_tomobin}. {The relatively low sensitivity to the input galaxy population is fortunate, especially in view of the Stage IV cosmic shear surveys, since advancements in shape measurement methods can minimise the measurement biases and selection biases can be accounted for through image simulation.}

Although the multiplicative bias is unchanged when evaluated in the reference frame co-aligned with the PSF, the anisotropy of the PSF is expected to affect the object detection, resulting in additive bias. To evaluate this, we simply take the mean of true sheared ellipticities, after rotating them to the PSF frame, across all input shears to get additive bias terms $c_{||}$ and $c_\times$. This results in smaller errorbars on the $c$-terms compared to the $m$-terms. The lower panel of Fig.~\ref{fig:selection_bias_tomobin} shows that faint galaxies that are aligned along the PSF are preferentially detected, leading to an additive bias in the PSF frame. For galaxies fainter than $24^\text{th}$ magnitude show an additive bias $c_{||}$ at the level of $10^{-3}$  for our choice of PSF ellipticities (see Fig.~\ref{fig:psf_distribution}) which have a mean ellipticity $\avg{|\epsilon_\text{PSF}|} \sim 0.025$, while the cross-component term $c_{\times}$ is within $2\times 10^{-4}$ (not shown here). A naive calculation based on quadrupole moments suggests that PSF anisotropy introduces additive and multiplicative biases at a comparable level. These amplitudes are too small to have been detected when examining the multiplicative bias, and is anyway far lesser than the bias introduced by the detection kernel. Thus, we can safely ignore any coupling between shear and PSF ellipticity.

Before continuing, we briefly discuss the relevance of the selection bias from the detection step for the metacalibration approach. {The adopted procedure so far to deal with the noise correlations introduced in the metacalibration procedure is to add anti-correlated noise~\citep{Sheldon2017}. This worsens the S/N of all galaxies, and consequently reduces the limiting magnitude for shape measurements. As a result, the detection goes to fainter magnitudes than shape measurements are performed on, and the effect of shear bias due to detection should be minimal. This bias could still be captured by metacalibration, in principle, if the detection algorithm was re-run on the sheared images, to determine if the sheared galaxy would have been detected by the algorithm or not.}

\subsection{Bias after \lensfit\ selections}
\label{subsec:lf_selections}

To be consistent with the selection cuts imposed in the cosmic shear analysis, we apply the same \lensfit\ selection as was applied to the data and explained in Appendix D1 of~\citetalias{Hildebrandt17}. We reject objects for which  \lensfit\ measured sizes smaller than 0.5 pixels, as well as objects that would be excluded from the cosmic shear analysis based on their \texttt{fitclass} flags. Finally, we only keep those objects whose re-calibrated \lensfit\ weights are strictly positive. 

To estimate the resulting selection bias after the \lensfit\ selection criteria, we use the input (sheared) ellipticities and find that the net multiplicative bias for the remaining sample of galaxies is $+0.004$. This includes the \sextractor\ selection bias, which means that the bias due to \lensfit\ selections alone is $+0.029$, thereby largely cancelling the bias caused by \sextractor. The positive selection bias is likely due to the tendency of \lensfit\ to flag small objects anti-aligned with the shear as stars. 

\subsection{Bias from \lensfit\ weights}
\citetalias{FC17} show that selection bias from \lensfit\ weights have a strong magnitude dependence. We observed a similar dependence with our new inputs and also noticed that this selection bias is sensitive to the joint distributions of galaxy properties, since for~\citetalias{FC17}-like simulations, the bias due to \lensfit\ selections alone is found to be as high as about +0.055 with the correctly recalibrated weights in the \citetalias{FC17} simulations. Thus, the approximate cancellation of the \sextractor\ and \lensfit\ selection biases is  coincidental and the two selections cannot be guaranteed to be complementary under all circumstances.

Although the usage of the true (sheared) ellipticities, rather than the measured ellipticities,  should mimic the performance of a perfect algorithm, measurement bias is still present in a subtle manner through the recalibrated weights from \lensfit. To quantify the resulting multiplicative bias, we compute the weighted average of true (sheared) ellipticities. The net multiplicative bias with lensing weights is about 0.015 for the full sample of galaxies. Comparison to the mean bias from the previous section implies that the bias due to the recalibrated \lensfit\ weights alone is little over 0.01. This shows relatively low-sensitivity to the joint distributions in the input catalogue, with~\citetalias{FC17}-like simulations exhibiting multiplicative bias of about 0.01 caused by the \lensfit\ weight themselves.

\section{Calibration results}
\label{sec:calibration_results}
Having established that our image simulations provide a good representation of the KiDS data, we now proceed to determine the multiplicative biases for KV-450 for the different tomographic bins. These are used to correct the cosmic shear correlation functions used in~\citetalias{Hildebrandt2019} to infer cosmological parameters. Before we do so, however, we first compute the mean bias for the full sample of simulated galaxies. We find that for the full sample of simulated galaxies, $m_1[q_\text{sims}]=-0.008 \pm 0.001$ and $m_2[q_\text{sims}]=-0.005 \pm 0.001$. These uncertainties represent the deviation of the data points from the linear fit (Eq.~(\ref{eq:linear_bias_model})), and underestimate the true uncertainty. For the calibration of the cosmic shear signal, the uncertainties are derived in a more rigorous way through bootstrapping.

\citetalias{Hildebrandt17} examined the impact of residual multiplicative bias on the cosmological parameter estimates and concluded that a error in $m$ of 0.05 would be acceptable. Based on the results presented in  \citetalias{FC17}, an overall systematic uncertainty of 0.01 was adopted by~\citetalias{Hildebrandt17}. The error was assumed to be the same for all four tomographic bins in~\citetalias{Hildebrandt17}, because redshift information for the simulated source was lacking. In contrast, emulation of the COSMOS field allows us to calibrate each tomographic bin separately. The resulting bias estimates, and their associated uncertainties for each bin are therefore (largely) independent, which effectively increases the overall uncertainty in cosmological parameters when all bins are considered. An overall tolerance of 0.05 implies that we can accommodate errors of about 0.02 per tomographic bin, which we adopt as our systematic error budget. We show in~\cref{sec:sensitivity_analysis} that we can control our multiplicative bias errors to within 0.02. In this paper, we do not constrain the multiplicative bias more than the requirement. Hence our results are considerably more conservative than the \citetalias{FC17} estimates.

\subsection{Bias for tomographic bins}
\label{sec:bias_tomo}

Although the mean multiplicative bias for the full galaxy sample is small, it is not guaranteed to be so for a particular subset of galaxies, and residual multiplicative biases that are greater than 0.05
are still possible, particularly for small galaxies with low signal-to-noise ratios. Furthermore, variations in observing conditions across the survey will modify the distributions of sizes and S/N values from pointing to pointing, thus changing the bias. Hence, as discussed in Eq.~\ref{eq:zeroth_estimator}, we cannot simply assume that $m[p_\text{real}]=m[p_\text{sims}]$, but we need to account for the differences between the observations and the simulations. Finally, the simulations 
match the observations very well, but not perfectly and small corrections need to be made to estimate $m[p_\text{real}]$ from Eq.~\ref{eq:master_correction} or equivalently from Eq.~\ref{eq:zeroth_estimator}.

To do so, we re-weight the simulated catalogue following `Method C' of ~\citetalias{FC17}. In principle the shear bias is a function of many galaxy properties, denoted by $\vec{ D}$ in~\cref{sec:calibration_math}, but following  \citetalias{FC17},~\citetalias{Zuntz2018} and~\citetalias{Mandelbaum2018a} we focus on the two dominant observables, namely the S/N and a resolution parameter $\mathcal{R}$ that quantifies how the observed shape is affected by PSF convolution. As in~\citetalias{FC17} (Eq.~(7)), we define the resolution parameter $\mathcal{R}$ as
\begin{equation}
    \mathcal{R} = \frac{r^{2}_\text{psf}}{\left(r^2_{ab}+r^2_\text{psf}\right)}
\end{equation}
where $r_{ab}$ is the circularised galaxy size defined in~\cref{sec:input_cor} and $r^2_\text{psf}:=\sqrt{P_{11}P_{22}-P_{12}^2}$, where $P_{ij}$ are the weighted quadrupole moments, measured with a circular Gaussian function of size 2.5 pixels. {For any given galaxy, $\mathcal{R}$ assumes a value between 0 and 1. We remind the reader that $\mathcal{R}$ is large for poorly resolved galaxies and small for well-resolved galaxies.}

The deviation from $m[p_\text{sims};s_\text{sims}]$, as given in Eq.~\ref{eq:master_correction} is approximated by the summation over all S/N-$\mathcal{R}$ bins of $b_i[r_\text{sims};t_\text{sims}] \Delta w_i$, where $i$ denotes a generic S/N-$\mathcal{R}$ bin, $b_i[r_\text{sims};t_\text{sims}]$ is the multiplicative bias in that bin obtained from simulations and $\Delta w_i$ is the difference between the fraction of weight between data and simulations. We divide the simulated galaxies into \newtext{an irregular} $20\times 20$ grid of S/N and $\mathcal{R}$, with each bin containing the same sum of \lensfit\ weights \newtext{(similar to Fig. 9 of~\citetalias{FC17})}. We compare the ratio of the sum of \lensfit\ weights in the KiDS data to those in the simulations and find that the ratios are generally close to unity, implying our results are robust. We also explicitly verified the robustness of our results to the binning, and found no significant differences.

\begin{table*}
    \centering
    \caption{Residual multiplicative bias for the five tomographic bins}
    \begin{tabular}{ccccccc}
    \hline \hline
    Tomographic bin & Bin & {Median} & RMS & {$m$ (with-$z$)} & $m$ (no-$z$) & $m$ (FC17cor) \\ {definition} & {label} & {magnitude} & {ellipticity} \\
    \hline
        $0.1 < z_B \le 0.3$  & B1 & 22.34 & 0.264 & $-0.013 \pm 0.008$ & $-0.036 \pm 0.003$ & $0.021 \pm 0.004$\\
        $0.3 < z_B \le 0.5$  & B2 & 23.00 & 0.258 & $-0.010 \pm 0.006$ & $-0.028 \pm 0.003$ & $0.017 \pm 0.004$\\
        $0.5 < z_B \le 0.7$  & B3 & 23.71 & 0.274 & $-0.011 \pm 0.006$ & $-0.008 \pm 0.003$ & $0.023 \pm 0.004$\\
        $0.7 < z_B \le 0.9$  & B4 & 23.85 & 0.237 & $+0.007 \pm 0.006$ & $+0.006 \pm 0.003$ & $0.029 \pm 0.004$\\
        $0.9 < z_B \le 1.2$  & B5 & 24.06 & 0.237 & $+0.006 \pm 0.007$ & $+0.009 \pm 0.003$ & $0.045 \pm 0.004$\\
    \hline
    \end{tabular}
    \label{tab:residual_bias_tomo}
\tablefoot{The bias values in the second column correspond to our fiducial calibration where we split the galaxy sample first into tomographic bins (with-$z$); for the $m$ (no-$z$) results the bias as a function of size and S/N is determined before the split into tomographic bins; $m$ (FC17cor) lists the results if the \citetalias{FC17} simulations had been analysed with the correct weight recalibration.}
\end{table*}

\begin{figure}
\begin{center}
\includegraphics[width=\columnwidth]{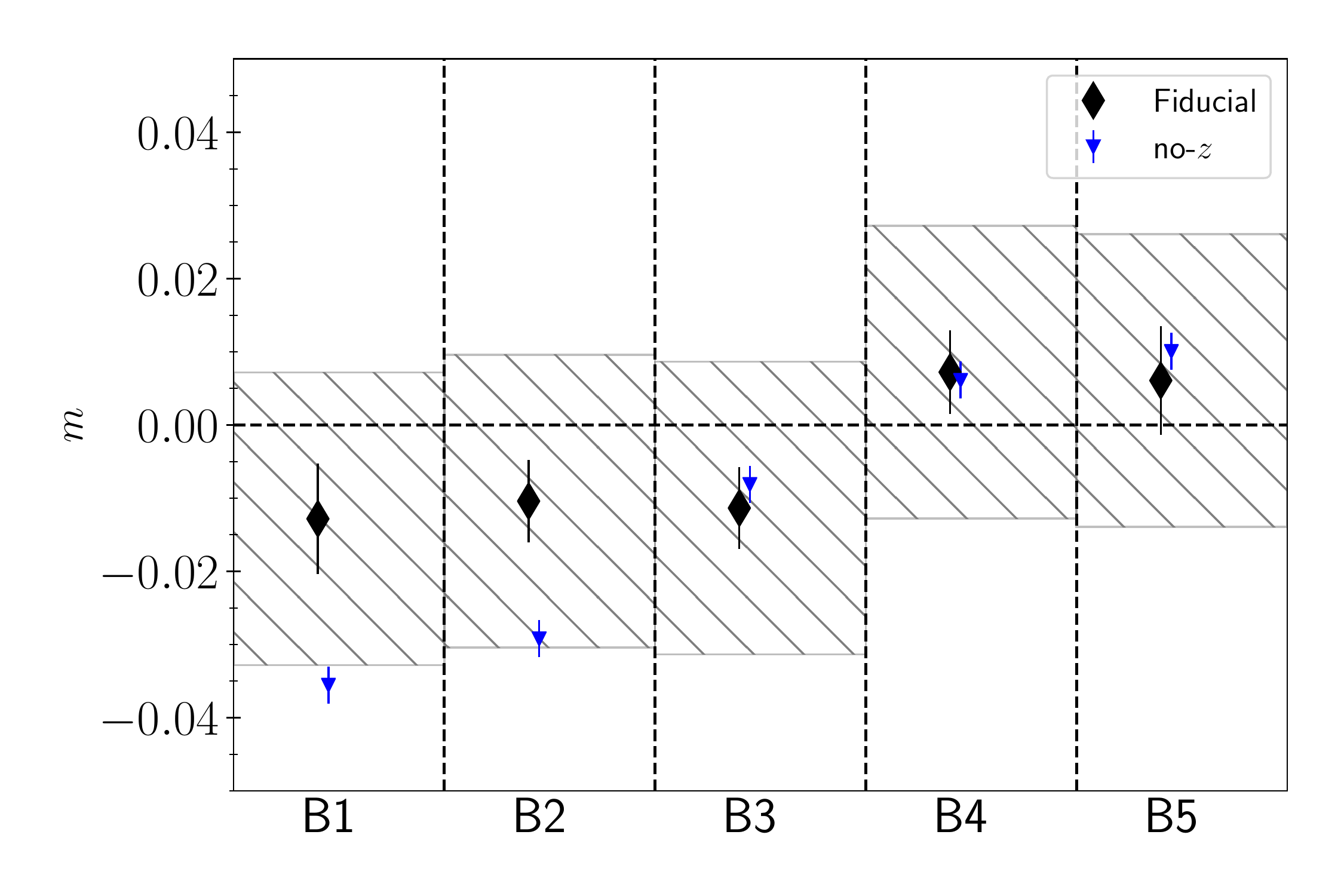}
\end{center}
\caption{Multiplicative bias calculated using the re-weighting technique as a function of tomographic bins used in the KV-450 cosmic shear analysis. The black diamonds correspond to the fiducial bias values obtained after tomographic splitting of the simulated galaxy population, while the blue downward pointing triangles correspond to the bias values obtained without such a splitting. 
The hatched regions indicate the $\pm 0.02$ region around the fiducial values.}
\label{fig:money_plot}
\end{figure}

To determine the residual shear calibration bias we divide the galaxies first into their respective tomographic bins. Hence we take advantage of the fact that we have assigned photo-$z$s to the simulated galaxies in the COSMOS field. We then re-weight the samples based on the observed distributions of $\mathcal{R}$ and S/N for each bin separately. As we use the redshift estimates for individual galaxies, we refer to this implementation as `with-$z$'. This implementation captures the different distributions of galaxy properties in each tomographic bin most faithfully. For this reason we adopt it as our fiducial calibration.  Unless otherwise mentioned, we will use this redshift implementation to obtain the shear calibration biases.

To highlight the importance of redshift information in the simulations, we also consider a calibration scheme where we use the full sample of simulated galaxies in the re-weighting step. As before, the resulting distributions of $\mathcal{R}$ and S/N match the observations, but the bias estimates include galaxies outside the tomographic bin. This method, which is most similar to the approach taken by~\citetalias{FC17}, does not use the redshift information, and we refer to this as `no-$z$'. 

The results for both these implementations are listed in Table~\ref{tab:residual_bias_tomo} and shown in Fig.~\ref{fig:money_plot}. In both implementations we determine the statistical uncertainties in the bias through a bootstrap procedure. To do so, we select galaxy samples at random, and for each random sample we obtain the bias values. The standard deviation of $m_1$ and $m_2$ over the bootstrap realisations are taken to provide $1\sigma$ uncertainties on the multiplicative biases.
The shear bias estimates with the `no-$z$' implementation are within our two per cent tolerance per tomographic bin, but only barely for the two lowest redshift bins. Thus, the naive way of re-weighting the simulated galaxies to match the galaxies in tomographic bins, from redshift-agnostic simulations, as in \citetalias{FC17}, \citetalias{Zuntz2018} for the \imtshape\ catalogue and~\citetalias{Mandelbaum2018a}, does not give an accurate estimate of the residual biases. 

Our result exposes an important limitation of redshift-agnostic image simulations. So far, redshift estimation and shear calibration have been treated as two independent problems, at least in practice, within cosmic shear analyses. In~\cref{sec:calibration_math}, we argued that the bias surface must be sensitive to the overall joint distribution of the galaxy properties, which may differ from one tomographic bin to another. Thus, if the sample of galaxies in a tomographic bin is not representative of the overall sample used for re-weighting, the terms involving $\Delta r$ and $\Delta t$ in Eq.~\ref{eq:master_correction} become large enough so that they lead to residual biases that are not captured by the re-weighting procedure used for calibrating shear. If the galaxy properties for a given size and S/N value vary as a function of redshift, we expect the `no-$z$' approach to be (slightly) biased. Due to the complex selection function, and possibly in combination with a redshift-dependent intrinsic ellipticity distribution, the ellipticity distributions in the different tomographic bins may differ. 
~\cite{Viola2014} show that the bias in ellipticity, in general, depends on the ellipticity itself and this is known to be true for \lensfit\ as well. But because we do not (and should not) characterise the bias in terms of the measured ellipticities, the mixture of galaxy populations between the data and the simulations may differ, even though the S/N-$\mathcal{R}$ distributions in the two are made identical.

We characterise the input ellipticity distribution using per-component root-mean-square (RMS) values of the ellipticity, defined as
\begin{equation}
    e_\text{rms} := \sqrt{\frac{\sum_k w_k |\epsilon_\text{in}|_k^2}{2\sum w_k}},
\end{equation}
where the $w_k$ are \lensfit\ weights and the summation is over all the galaxies in the sample, and over all input shears and rotation, so the the weights are averaged out. In Table~\ref{tab:residual_bias_tomo}, we notice that the RMS ellipticity show some differences. {For comparison, the RMS ellipticity values when the input ellipticities were scrambled were found to be $0.274 \pm 0.002$ for the different tomographic bins, indicating that the differences we see in Table~\ref{tab:residual_bias_tomo} are due to intrinsic differences in the ellipticity distributions. Different values of RMS ellipticity in the different tomographic bins, when the inputs are based on COSMOS, are not surprising given the dependence of mean colour on the ellipticity, as shown in Fig.~\ref{fig:e_vs_col}. We also noted (not shown here) the the distributions themselves were visually different for the tomographic bins, particularly at low redshifts.~\cite{Kannawadi2015} show that some of these differences may be due to overdensities or underdensities along the line-of-sight, which alter the ellipticity distributions. Therefore, the input ellipticity distributions may not be typical of galaxies at that redshift, leading to a systematic error of $\sim 0.01$ in shear calibration bias~\citep{Kannawadi2015}. However, as we show later in Fig.~\ref{fig:ellip_dist}, the input catalogue with randomised ellipticities do not describe the KV-450 data very well, suggesting that our input ellipticities are indicative of the true ellipticity distributions, and that the differences are not dominated by cosmic variance alone. The uncertainty in the ellipticity distribution is one of the major source of systematic error in the multiplicative bias for low redshift bins}. In this work, we are relatively insensitive to intrinsic variations in $p(|\epsilon|)$ across tomographic bins since we impose the same selection as in the data on assigned $z_B$ values. This is a strong motivation for future cosmic shear analysis to build a fully self-consistent shear calibration with realistic, multi-band image simulations, so that the sample  selection in the simulations and in the data are done from their colours and magnitudes.

{The use of redshift information, however, has additional advantages that are worth considering. For instance, in redshift-agnostic, single-band simulations, we rely entirely on the shape measurement pipeline, in this case \lensfit, to perform the star-galaxy classification. In multi-band observations, this may be mitigated thanks to the various photometric cuts on the data. In the absence of photometric information in the simulations, stars misclassified as galaxies by \lensfit\ that may have been rejected based on their photometry contaminate the simulated galaxy sample. Removing such stars from the simulated catalogues based on the ground truth can affect the bias estimate itself. Moreover, in our case, the redshift distribution for COSMOS exhibited a small bump at around $z_B \sim 5$ (not shown). This spurious galaxy population is eliminated in our fiducial scheme, but not in the other. In order to ensure that the difference between the `with-$z$' case and the no-$z$ case does not arise because of this photometric selection, we re-did the `no-$z$' calibration, but now restricting the sample to $0.1 < z_B \le 1.2$ but no further tomographic splitting. We find that the bias values shift by an amount $\le 0.006$, too small to explain the difference in the first two tomographic bins}.

{Finally, we also present results in Table~\ref{tab:residual_bias_tomo} and in the upper right panel of Fig.~\ref{fig:sensitivity_analysis} for the~\citetalias{FC17} simulations (labelled as FC17cor) after we corrected our earlier error in calculating the \lensfit\ weights, and when we detect galaxies in each of the rotated images}. For a fair comparison, the `no-$z$' case must be used as the reference. We find that the bias shifts upwards by 0.02-0.04 compared to our fiducial results (`with-$z$'). To put our results in the context of~\cref{sec:calibration_math}, we attribute the differences in the multiplicative biases between the `with-$z$' (fiducial) and `no-$z$' cases to the $\Delta t$ terms of Eq.~\ref{eq:master_correction}. The differences in the bias values between the fiducial simulations and `FC17cor' are due to a combination of $\Delta r$ and $\Delta t$ terms. This highlights the importance of realistic image simulations that reproduce the actual observations and contain redshift information for the sources. We discuss more on this in~\cref{sec:sensitivity_analysis}. We note that, fortuitously, the bias values provided in \citetalias{FC17} match the current estimates in the first three bins, and are 0.02 lower for the fourth bin. The tomographic samples used in KiDS-450 and KV-450 analyses are somewhat different due to new photo-$z$s, despite having the same global set of galaxies. However, we note that this difference affects the residual multiplicative bias only mildly. Although our results are fundamentally different from that of \citetalias{FC17}, the final impact on the cosmic shear analysis is expected to be small because the errors in the analysis partially cancelled the actual bias.

\subsection{Additive bias}
Although the image simulations are realistic and resemble the data well, they do not capture all potential sources of bias that are present in the actual observations. For instance, we did not include
\begin{enumerate}
    \item artefacts, such as cosmic rays, asteroids, satellite trails, binary stars etc.,
    \item camera distortions, pointing errors and astrometric corrections,
    \item spatial variation of PSFs and PSF modelling errors,
    \item detector non-idealities such as variation in pixel response, charge trailing etc.,
\end{enumerate}
to list a few. A thorough study of instrumental effects of the OmegaCAM detectors is under way (Hoekstra et al., in prep). Ignoring these effects is not expected to significantly change the multiplicative correction terms, but could have an impact on the additive bias. \newtext{In particular, Hoekstra et al., (in prep) find that detector effects primarily affect $c_1$ terms.} For this reason, additive shear bias corrections are derived empirically using the data themselves in~\citetalias{Hildebrandt2019}(See Appendix D4 in~\citetalias{Hildebrandt17} for another example).

It is nevertheless interesting to compare the level of additive biases in the image simulations to that observed in the data. \newtext{To separate the contribution of PSF leakage to additive bias, we restricted our analysis to the first five PSF sets but performed more simulations with all the PSFs rotated by $90^\circ$ so as to produce ten PSF sets with zero average PSF ellipticity.} We find $c_1$ to be \newtext{small ($\lesssim 5\times 10^{-4}$)} for the different tomographic bins, whereas $c_2$ is in the range $[5,10] \times 10^{-4}$. In the PSF frame, we found $c_{\times}$ to be consistent with zero and $c_{||}$ in the range $[-12, -3]\times 10^{-4}$. Similar results were also obtained in \citetalias{FC17}. 

\begin{figure}
\includegraphics[width=\columnwidth]{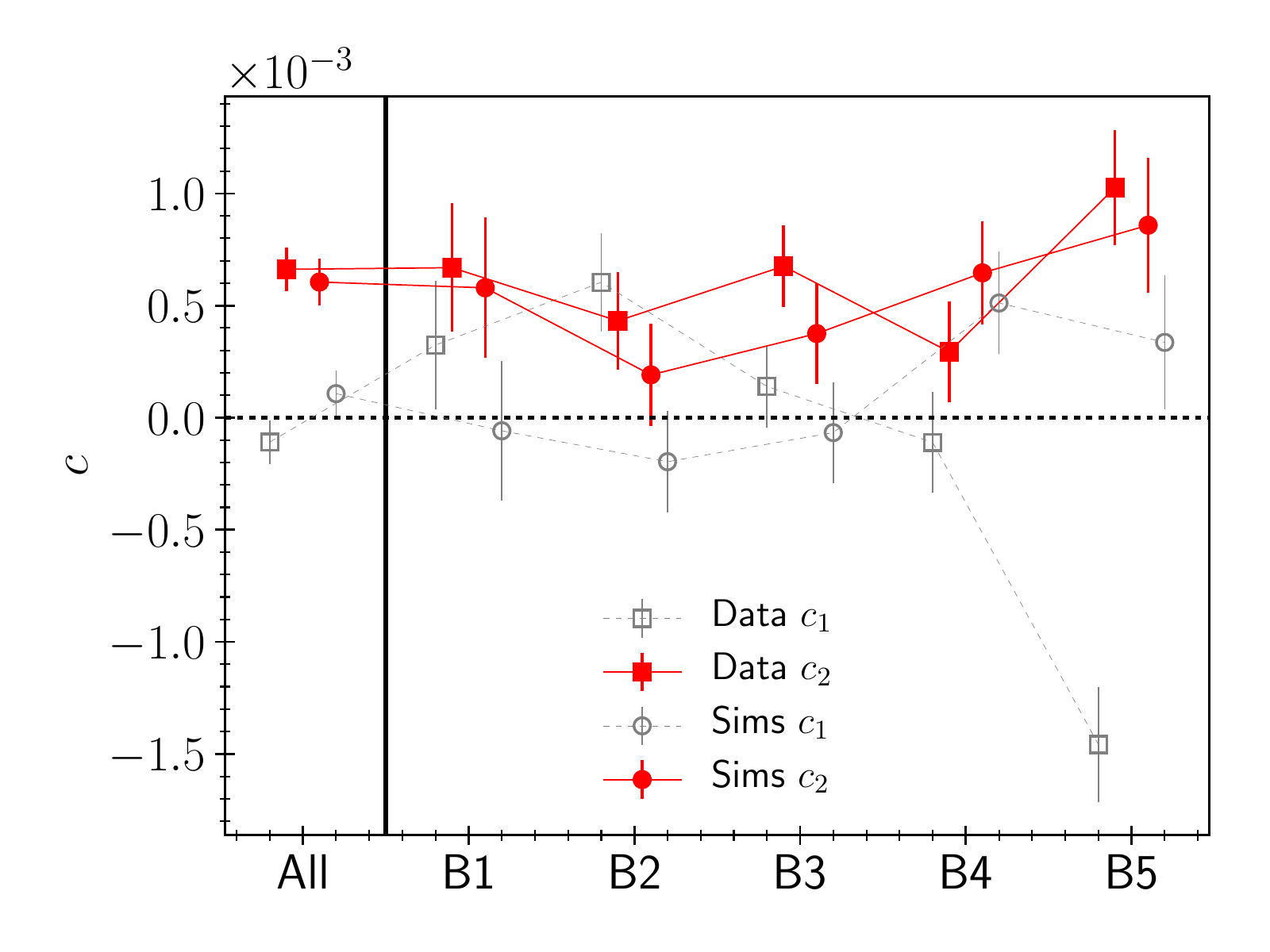}
\caption{\newtext{Additive bias, estimated as the mean ellipticity, for the entire galaxy sample (All) and for the five different tomographic bins (B1 through B5). Identical selection criteria has been applied to both the KV-450 data and to the simulations. The $c_1$ terms are expected to differ due to the detector effects in the OmegaCAM data which are excluded in the simulations.}}
\label{fig:cterms}
\end{figure}

Interestingly, \citetalias{Hildebrandt17} also observed a statistically significant additive bias in the KiDS data, in particular for $c_2$ (see Fig. D6 of \citetalias{Hildebrandt17}), which is corroborated by the simulations. \newtext{Fig.~\ref{fig:cterms} shows that the amplitude of the $c_2$ terms are in agreement between the data and the simulations. }
We find similar level of $c_2$ when we simulate images with a circular PSF. We did not find any significant $c_2$ immediately after the \sextractor\ detection step, or by replacing the \lensfit-measured ellipticities with the input ellipticities, while continuing to impose the same cuts that are applied on the data. This suggests that the \lensfit-measured ellipticities give rise to the $c_2$ terms at the level of $ \lesssim 10^{-3}$. Similar levels of additive biases were found in~\citetalias{FC17} as well, where we verified that the incorrectly recalibrated weight in ~\citetalias{FC17} has little effect on the additive bias terms. We note that such small but statistically significant additive terms were also found in the DESy1 shape catalogue~\citepalias{Zuntz2018} and on small angular scales in the HSC-DR1 shape catalogue~\citep{Mandelbaum2018b}. The similarity in the level of $c-$terms between the KV-450 data and the simulations indicates that other sources of shear bias are small.

We also explored how the additive bias depends on the galaxy properties. In~\cref{sec:bias_in_sims} we already found that a magnitude dependent (or S/N-dependent) additive bias along the direction of the PSF can arise at the detection step, which is particularly significant for faint galaxies. After the shape measurement step, we find a strong dependence of $c_2$ as a function of the resolution $\mathcal{R}$, at a per cent level for poorly-resolved galaxies, but tending to zero quickly for well-resolved galaxies. In contrast, the amplitude of $c_1$ is much lower $<0.003$ and exhibited only mild dependence on S/N and $\mathcal{R}$. While $c_1$ took both positive and negative values depending on the galaxy properties, we observed $c_2$ to be always positive. \newtext{The excellent agreement between the $c_2$ terms, given the strong dependence on galaxy properties, indicate that the galaxy population in the simulations is indeed representative of the data.}

{In the PSF frame, $c_{||}$ showed a strong dependence on both S/N and $\mathcal{R}$, with $c_{||} \lesssim -0.01 $ for galaxies with $\mathcal{R} > 0.7$ or S/N $< 10$. For each sub-population of galaxies split finely in bins of S/N or $\mathcal{R}$, the amplitude of the cross-component additive bias $|c_{\times}| <  10^{-3}$, with no particular trend as a function of the observables.  }
\section{Sensitivity analysis}
\label{sec:sensitivity_analysis}

\begin{figure*}
    \includegraphics[width=0.49\textwidth]{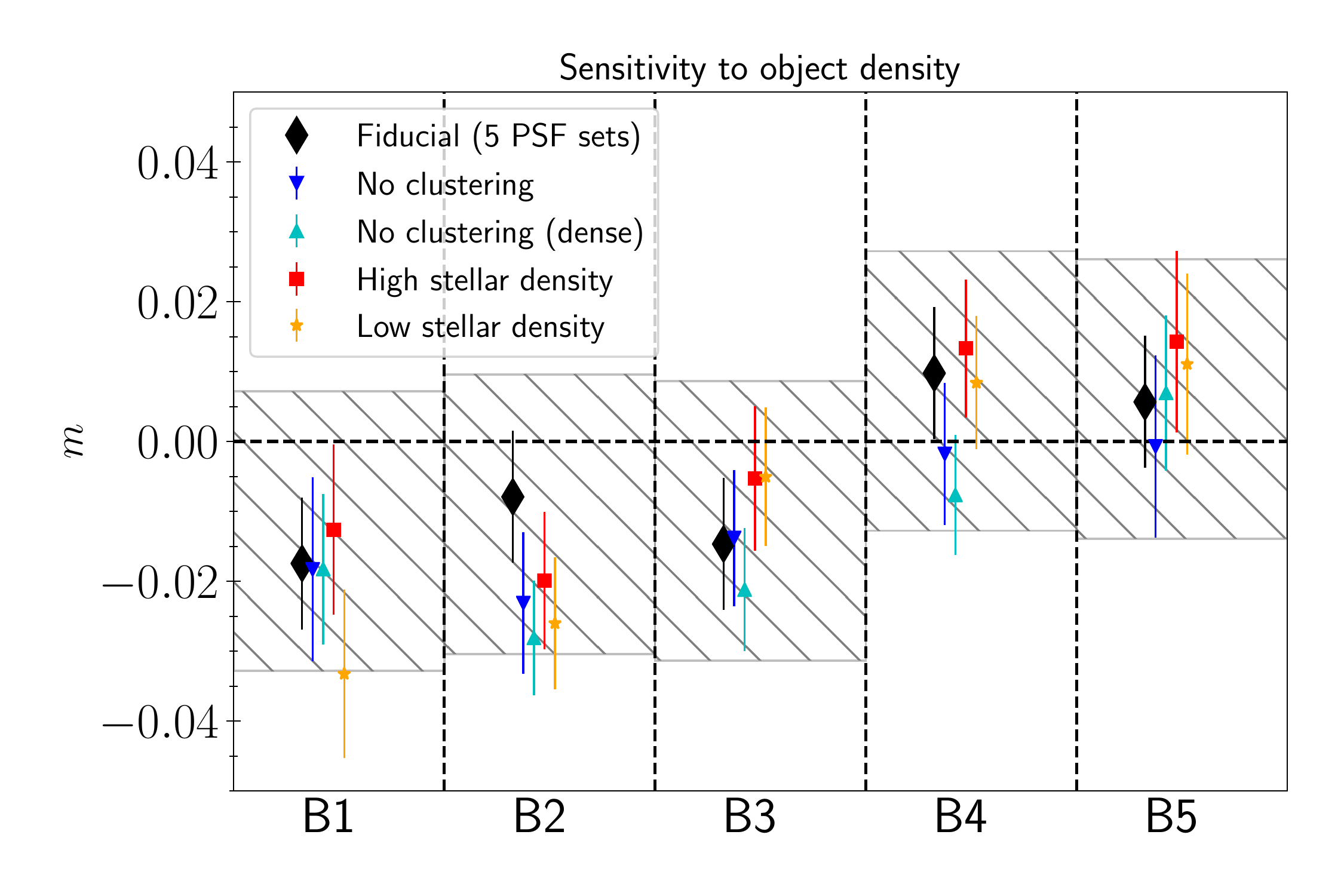}
    \includegraphics[width=0.49\textwidth]{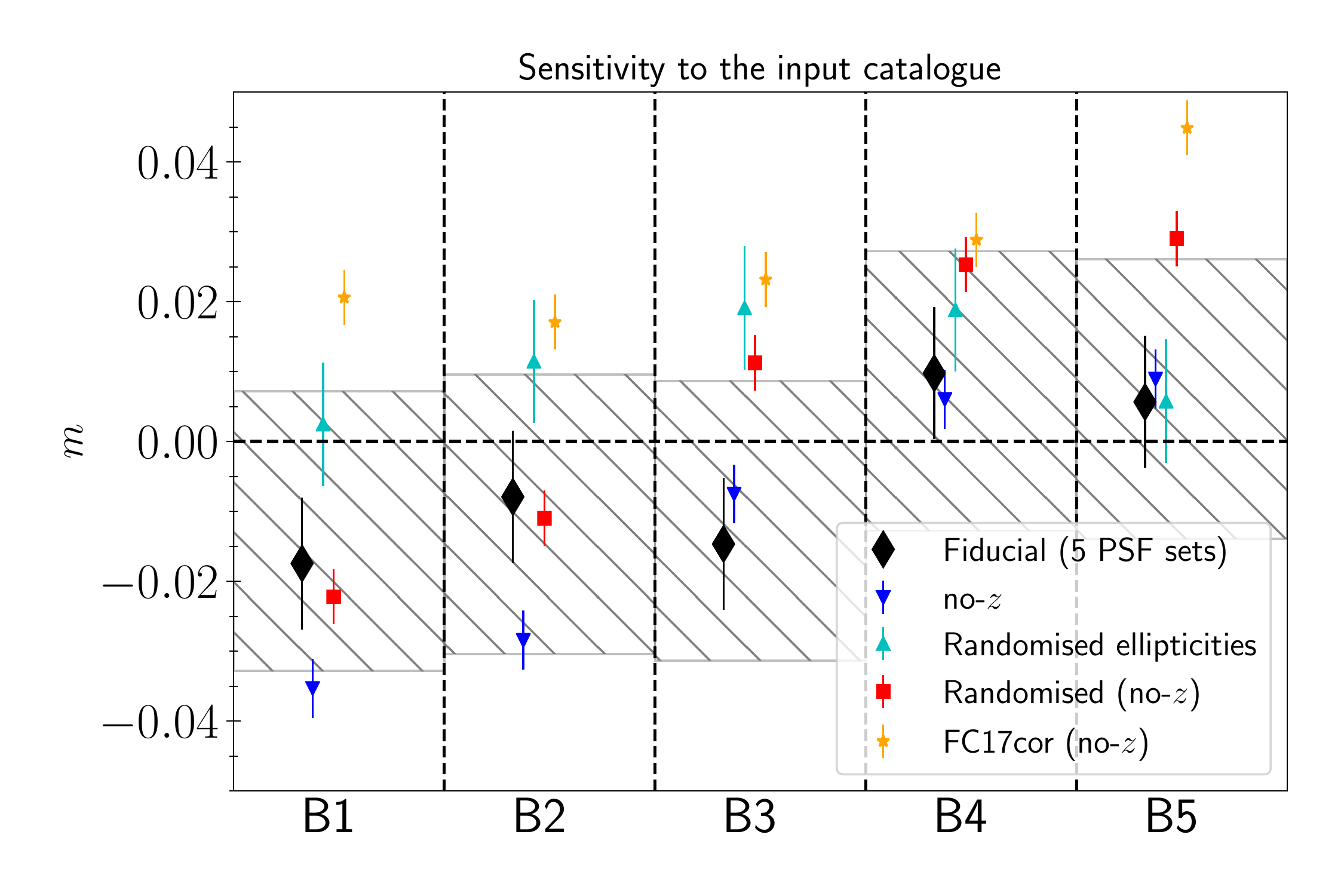}
    \; \includegraphics[width=0.49\textwidth]{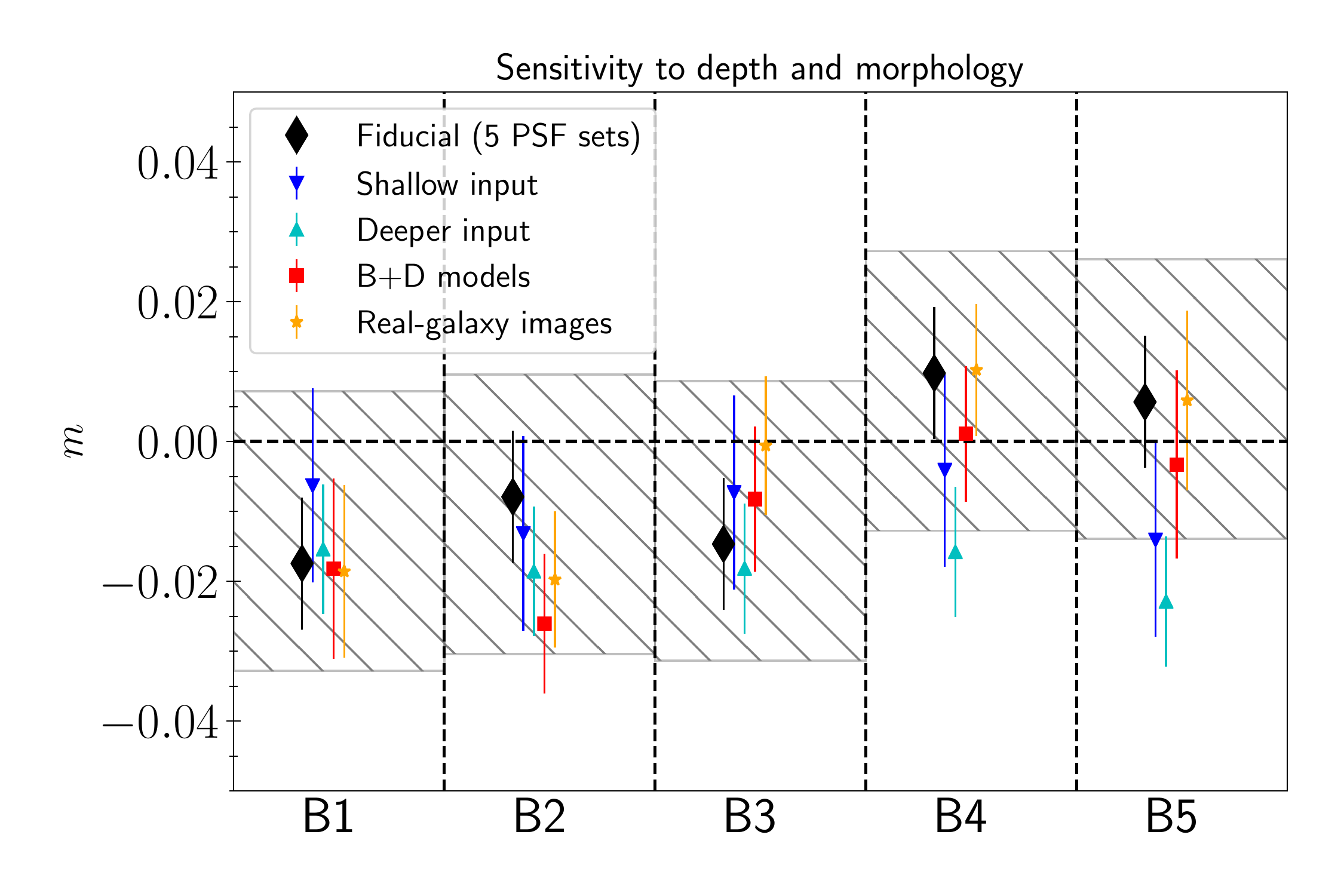}
   \, \includegraphics[width=0.49\textwidth]{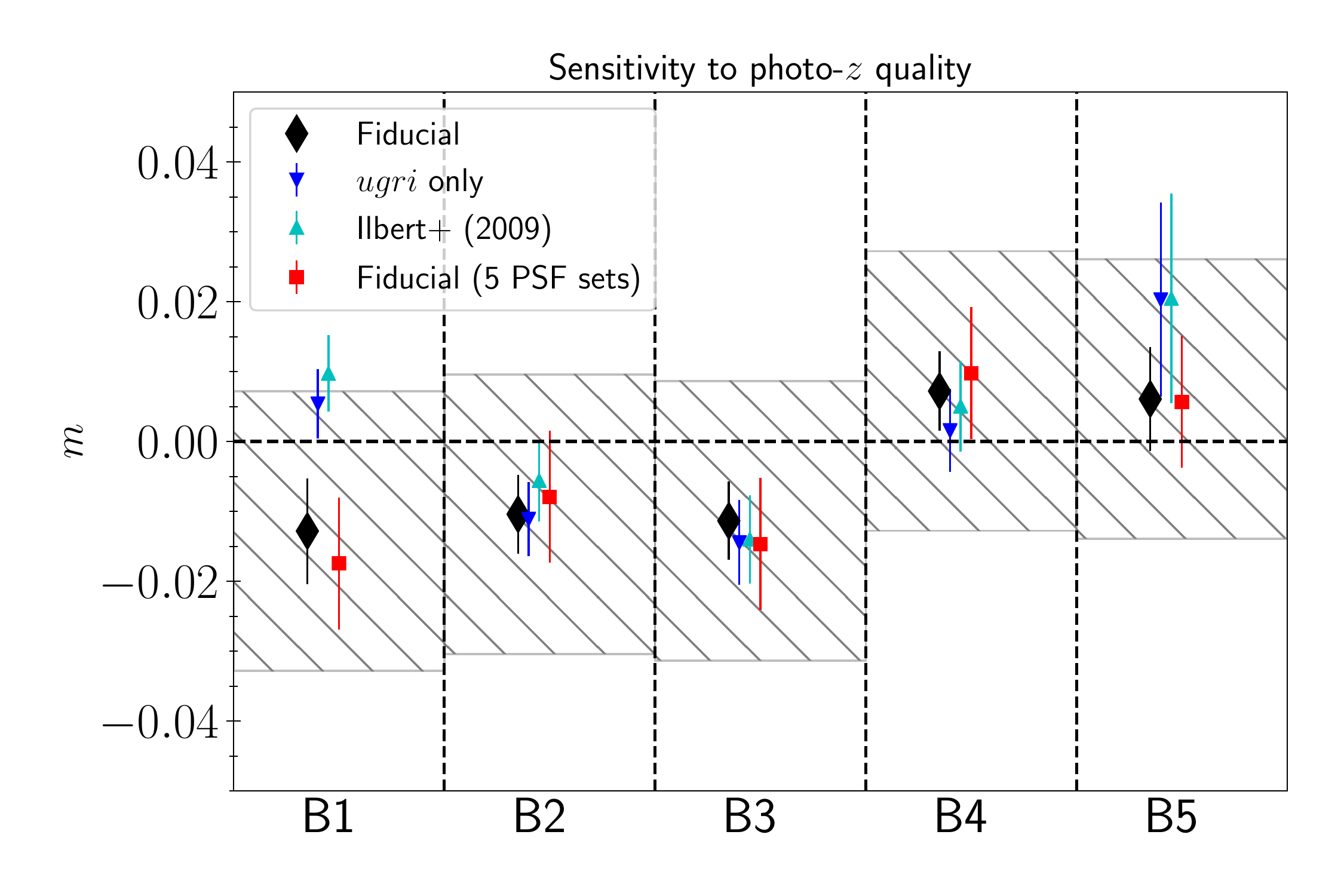}
\caption{Residual biases calculated with different simulations indicating the robustness of our calibration to different assumptions about the input to the simulations. {\it Top-left}: Sensitivity to galaxy clustering and number density (\cref{sec:clustering} and~\cref{sec:star_density}). {\it Top-right}: Sensitivity to input ellipticity correlations with galaxy parameters (\cref{sec:input_dependence}). {\it Bottom-left}: Sensitivity to the magnitude completeness of the input catalogue (\cref{sec:magnitude_dependence}) and to galaxy model in the simulations (\cref{subsec:model_bias}). {\it Bottom-right}: Sensitivity to the quality of photo-$z$s (~\cref{sec:redshift_dependence}). The hatched regions in the all panels are indicate the $\pm 0.02$ region around the fiducial values of each panel (black diamonds).}
\label{fig:sensitivity_analysis}
\end{figure*}

In~\cref{sec:compare_sims_data} and~\cref{sec:compare_sims_KV450} we showed that our image simulations match the actual data very well and that the re-weighting factors are close to unity for subsets of galaxies split by size and S/N, {with fluctuations of the order of $10^{-4}$ or less}. Consequently, the estimates of the shear biases presented in Table~\ref{tab:residual_bias_tomo} should be accurate. Nonetheless it is worthwhile to explore the robustness of our results and attempt to quantify the potential systematic uncertainties that may still be present. For instance, we simplified the galaxy morphologies by representing them with \sersic\ profiles. Our input catalogue is incomplete at the faintest magnitudes, but the missing galaxies may still affect the estimate of the multiplicative bias. Varying star densities can affect the results~\citep[e.g.][]{Hoekstra15,Hoekstra2017}. In this section we therefore explore the sensitivity of the shear measurement bias to various assumptions and simplifications made in the simulations. These results help to assess the robustness of the calibration presented in the previous section. In the language described in~\cref{sec:calibration_math}, these tests correspond to various evaluations of the $\Delta$ terms from different simulations. As these tests can become computationally expensive very quickly, we use only 5 of the 13 PSF sets. This results in only minor changes in the mean residual bias values ($<0.005$; see lower right panel of Fig.~\ref{fig:sensitivity_analysis}), and hence they remain a good representation of the data. The smaller volume of simulations naturally results in a larger statistical uncertainty, but we note that we are interested in determining the change in the mean values of the multiplicative biases of the different tomographic samples when we vary the inputs. The errors are tightly correlated among the different simulations within a tomographic bin, {as there is a significant overlap among the input samples of galaxies in the different simulations and because the noise realisations in the images are identical in many of these simulations, unless explicitly mentioned otherwise}. Hence, the shifts are not driven by noise. Our main objective is to ensure that uncertainties in the input quantities do not change the bias by more than 0.02. The results in this section indicate that they are indeed controlled under 0.02, but it appears that we cannot impose a tighter limit on the overall uncertainty at this time.

\subsection{Impact of clustering and galaxy density}
\label{sec:clustering}

The images of galaxies that are nearby on the sky blend together, making the measurement of their shapes challenging. \citetalias{Mandelbaum2018a} show in their Fig. 16 that blending can introduce multiplicative shear biases as large as 0.1. The impact of blending for KV-450 is expected to be much less, as the data are relatively shallow in comparison to the HSC observations. More representative for our data, ~\cite{Samuroff2018} show that the bias in the tomographic bins  due to neighbour contamination is about $\sim 0.05$ for DESy1 results with the \imtshape\ catalogue. In particular, their Fig. 10 shows how the bias depends on the distance to the nearest neighbour. The exact level of bias depends not only on the shape measurement algorithm, but also on how galaxies in the neighbourhood in the data are accounted for. 

\citetalias{FC17} show that the impact of blending of galaxies for KiDS is negligible, by computing the multiplicative bias with simulations where the fiducial number density was lowered by 50\%. The change in the multiplicative bias is found to be small (less than 0.002), as the self-calibrating \lensfit\ accounts for the adjacent galaxies internally and is therefore fairly robust. Although this suggests that blending can be ignored, it is worth noting that the galaxies were placed randomly, whereas in reality galaxies cluster. Our fiducial simulations, by virtue of emulating the COSMOS field, do capture the realistic clustering of galaxies present in the field. To quantify any contribution to the multiplicative bias caused by clustering, we perform another set of simulations with the same input catalogue but with random positions for the galaxies. As shown in the upper left panel of Fig.~\ref{fig:sensitivity_analysis}, the change in multiplicative bias is greater than $1\sigma$ for the B2 and B4, and negligible compared to the statistical error in the other three bins.

It is possible that the galaxy density in the COSMOS field is atypical. To explore the impact of this, we create new simulations where the number density is increased by a factor $f$ relative to the simulations with random galaxy positions\footnote{Since the masks in the COSMOS field are not incorporated in the simulations without clustering, the number density is reduced by a small fraction.}. For lower densities ($f<1$), we select a different random subset from the COSMOS sample for every input shear and input PSF. For higher densities ($f>1$), we include the full sample once and then, we select a random subset to account for the rest of the galaxies. We find that the change in the biases were negligible (not shown) for $f=0.8$, and within the tolerance limit for $f=1.5$. We emphasise that a global increase in galaxy density by $50\%$ over the full COSMOS area exceeds realistic variations, and thus resembles an extreme scenario. The large differences found in B2 and B4 become even more significant, indicating the possibility that the COSMOS sample exhibit lower clustering in those redshift bins.

\subsection{Impact of stellar density}
\label{sec:star_density}

The density of stars varies across the sky as a function of Galactic coordinates. Given the large area covered by KiDS, we expect some variation in star-galaxy blending. In the simulations the density of the stars is held fixed. As already mentioned in~\cref{sec:sim_setup}, the stellar magnitude distribution is obtained from the Besan{\c c}on model\footnote{{http://model.obs-besancon.fr/}}~\citep{Robin2003,Czekaj2014} corresponding to right ascension $\alpha$ of $175^\circ$ and declination $\delta$ of $0^\circ$. As already pointed out by \citetalias{FC17}, this region has a star density than is higher than the average for the entire KV-450 footprint. To test our sensitivity to the star density, we increase the density of stars further by a factor of two. 

The results are presented as red squares in the upper left panel of Fig.~\ref{fig:sensitivity_analysis}. These show that the change in multiplicative bias is within $1\sigma$, with the exception of B2, which is where the misclassified stars in the data usually end up. The change is nevertheless within 0.01.
We note that the stars do not contaminate the galaxy catalogue due to the photometric selection based on their input redshift. Our hypothesis for the change is that the increased stellar density leads to increased blending with galaxies, particularly the faint ones, leading to an increased rejection of faint galaxies. This explains the smaller sample size in B5, leading to larger error bars and a change in the values towards positive values in four of the tomographic bins. We did not explore this further, as the star density is very high whilst the changes are small enough to be of no concern for KV-450.

To investigate the possibility that we are mis-estimating our bias due to higher-than average stellar density, we repeated the simulations with only half the numbers of stars, removing them randomly. We find from the upper left panel of Fig.~\ref{fig:sensitivity_analysis} that both B1 and B2 are sensitive to a reduction in stellar density, especially B1, which marginally exceeds our error budget of 0.02. While other tomographic bins do not show any monotonic behaviour for the central values as a function of stellar density, B1 appears to exhibit a monotonic behaviour. Therefore, considering the two cases to bracket realistic variation of stellar densities with sufficient margin of error, we do not expect the bias to shift by more than 0.02.

\subsection{Dependence on the input catalogue}
\label{sec:input_dependence}
The residual multiplicative bias estimates from the fiducial simulations are in good agreement with the published values of the same in~\citetalias{FC17}. However, we emphasize that this is merely a lucky coincidence. Had we only corrected the weight recalibration, and used the input catalogue used in~\citetalias{FC17} to calibrate shear, we would have obtained a correction quite different from our fiducial results as indicated by the orange stars in the upper right panel of Fig.~\ref{fig:sensitivity_analysis}. We attribute this shift to the fact that ellipticities are correlated with other galaxy properties that affect the shear bias. As argued in~\cref{sec:input_cor}, we have strong reasons to believe that such correlations are real, and thus crucial to include in the simulations.

As mentioned in~\cref{sec:comparison}, there were a few improvements and corrections made to the simulation pipeline. We obtained multiplicative bias estimates using the input catalogues used in~\citetalias{FC17}. For the ~\citetalias{FC17} input, $m[q_\text{sims}]$ for the population of the galaxies that enter the cosmic shear analysis is measured to be $(3.22\pm0.42)\times 10^{-2}$ and $(3.14\pm0.27)\times 10^{-2}$ for $m_1$ and $m_2$ respectively. This is evidently very different from what we obtained from the fiducial simulations, which had bias values consistent with zero. Due to the redshifts being absent, we are able to obtain bias for the tomographic bins using only the `no-$z$' method. 
The upper right panel in Fig.~\ref{fig:sensitivity_analysis} suggests that the multiplicative bias can differ by $0.02$ depending on which input catalogue is chosen. {However, based on a careful comparison with the data, as in Fig.~\ref{fig:size_e_comparison} for example, we can rule out the input catalogue of~\citetalias{FC17} as not a good representation of real galaxy population, and therefore the relative large change in the bias is not surprising.}

This effect is not expected to impact the shear calibration of DESy1~\citepalias{Zuntz2018} and~\cite{Samuroff2018}, and HSC-DR1~\citepalias{Mandelbaum2018a} shape catalogues significantly, which use postage stamps from the HST COSMOS sample. {Although this is primarily done with the intention to capture any model bias, it also ensures that any dependencies on size-ellipticity correlations (see~\cref{sec:input_cor},~\cref{sec:compare_sims_data}) are also captured implicitly. We also see this again in~\cref{subsec:model_bias}, where we repeat the simulation where we substitute a fraction of the input with real galaxy images and see no significant differences.}

\begin{figure}
    \centering
    \includegraphics[width=\columnwidth]{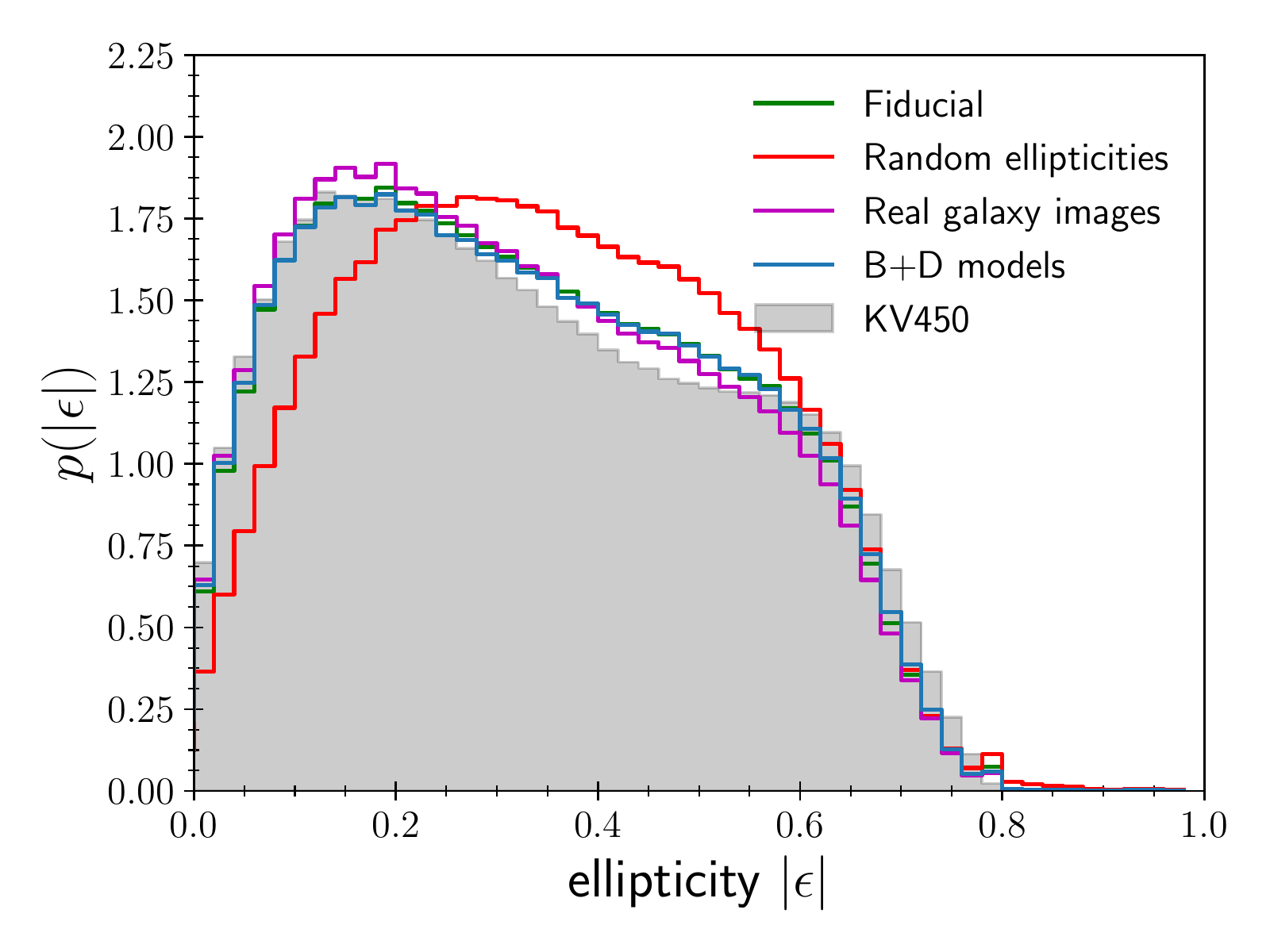}
    \caption{Weighted distribution of ellipticities as measured by \lensfit. The grey histogram shows the distribution for the KV-450 data, which is reproduced fairly well using our fiducial simulations (green) that use \sersic\ profiles. Similar distributions are obtained when we use real galaxy images from COSMOS (blue) or instead represent galaxies by bulge+disc models (purple). The simulation where the input ellipticities were scrambled gives a very different distribution (red) and is inconsistent with the data.}
    \label{fig:ellip_dist}
\end{figure}

To quantify the importance of capturing ellipticity correlations, we created another set of simulations where we scrambled the input ellipticites randomly to make them uncorrelated. We find a significant change in the multiplicative bias values approaching those from the~\citetalias{FC17}-like simulations. From Fig.~\ref{fig:ellip_dist}, we also note that in the simulations with ellipticities scrambled, the resulting weighted ellipticity distribution does not resemble that from the KV-450 data whatsoever. Thus, this also acts as a sensitivity test to the ellipticity distributions, albeit indirectly.

\subsection{Impact of the input magnitude distribution}
\label{sec:magnitude_dependence}
\begin{figure}
    \centering
    \includegraphics[width=\columnwidth]{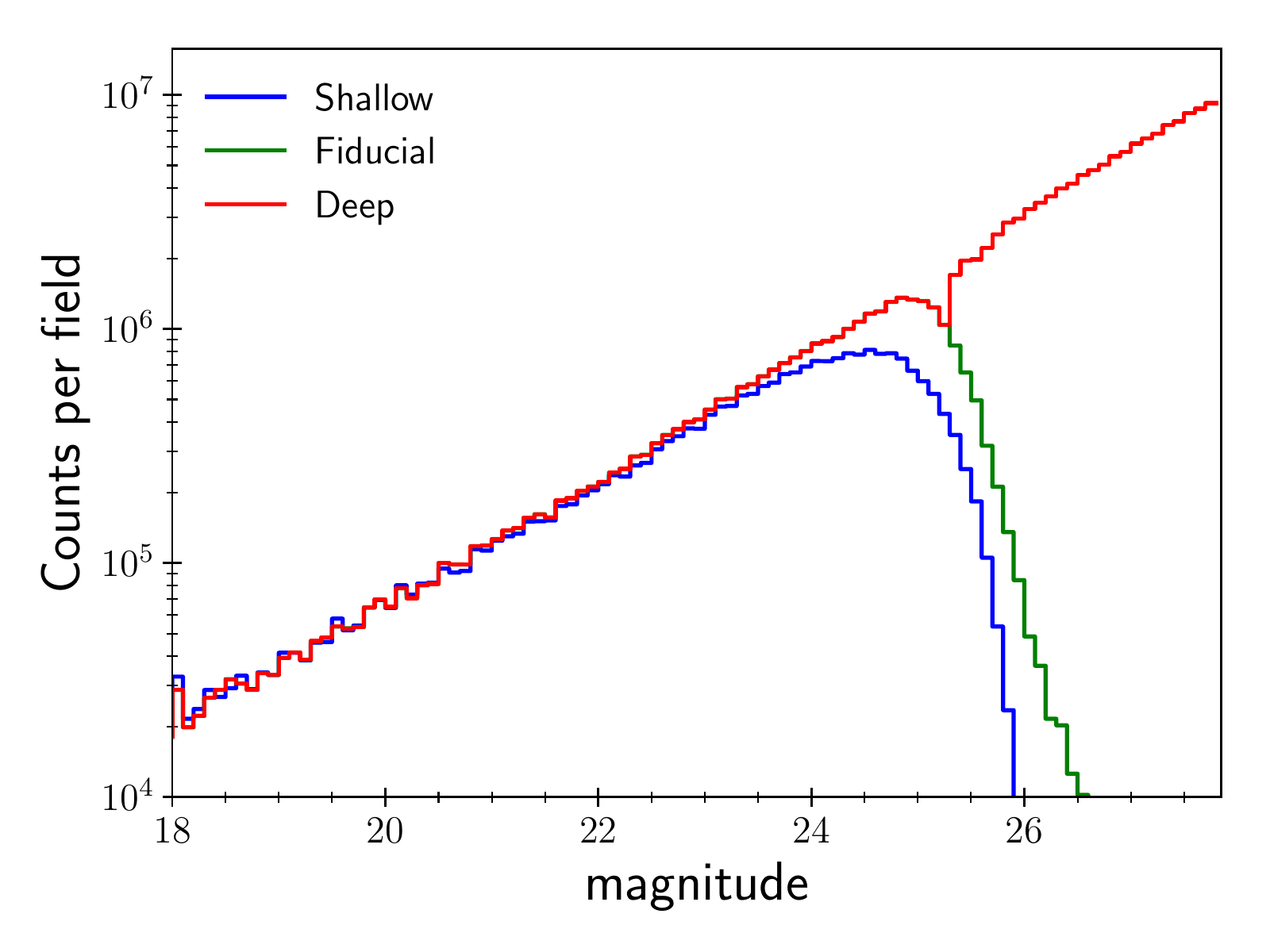}
    \caption{Distribution of the (input) galaxy magnitudes for the fiducial, shallow and deep input catalogues.}
    \label{fig:input_dist}
\end{figure}

In order to assess the impact of the incompleteness of the ACS-GC catalogue at the faint end, we take the ratio between the input magnitude histograms of the simulation used in~\citetalias{FC17} and the ACS-GC catalogue and obtain a weight for the magnitude bins for magnitudes greater than 25. We create copies of the faint galaxies several times, a number roughly equal to their magnitude weight and randomly place them in the simulated images. We do not expect the clustering properties of such galaxies to be of major concern, because their main impact is to introduce noise in the background around the bright galaxies that are used in the lensing analysis. We refer to this input catalogue as the `deep' catalogue and show its input magnitude distribution in Fig.~\ref{fig:input_dist}. The faint galaxies that are detected in the ACS-GC catalogue might not be a fair subset of the population of faint galaxies. From Fig.~\ref{fig:goodness_matching}, we already saw that the faint galaxies are preferentially small due to their higher surface brightness for a given total magnitude (small galaxies are not missing in the KiDS catalogue, because they are detected). Including a large number of such compact galaxies is expected to affect the detection catalogues themselves and drive the multiplicative bias to become even more negative as opposed to including a fair population. This therefore is a worst case scenario and is very conservative.

We also created another set of simulations where only the galaxies that were detected in the KiDS-COSMOS catalogue were included. We call this the `shallow' input catalogue.  We know from~\cite{Hoekstra2017} that {missing faint undetected galaxies in the image simulations can shift the multiplicative bias towards positive values}. From our work, we know that the exclusion of faint galaxies around the limiting magnitude based on their non-detection can introduce much larger selection biases (negative multiplicative bias), similar to the levels found in Fig.~\ref{fig:selection_bias_tomobin}. Therefore, we expect the residual multiplicative bias to become more negative overall.

We find in Fig.~\ref{fig:sensitivity_analysis} that for the first three tomographic bins, all three input catalogues give values for the multiplicative bias that are consistent within $1\sigma$ uncertainty. For the last two tomographic bins, the multiplicative bias derived from the shallow and deep catalogues are significantly more negative than the fiducial ones, as expected. Both these cases exaggerate the effects of faint galaxies, and we can expect the true variations to be much smaller than our error budget of $0.02$.

\subsection{Impact of morphology and model bias}
\label{subsec:model_bias}

We use \sersic\ models based on HST observations to describe the surface brightness profiles of galaxies, which differs from the model that \lensfit\ uses to fit the data. Such a mismatch may introduce
model bias \citep{Voigt2010, Zuntz13, Kacprzak2014}, but \cite{Miller2013} have argued that this should be subdominant for the analysis of ground-based data. This is supported by the performance of \lensfit\ on realistic simulated galaxies in the GREAT3 challenge~\citep{Mandelbaum2015} and additional tests presented in~\citetalias{FC17}.

We explicitly verified that the model bias is subdominant in this work as well. This was done by matching the ~\cite{Griffith2012} catalogue with the \texttt{COSMOSCatalog} available in \galsim~\citep[see][for details]{COSMOS_Alexie,great3}. The \texttt{COSMOSCatalog} contains the best-fit \sersic\ and bulge+disc parameters for 87717 
galaxies in the HST COSMOS field and a flag indicating which of the two parametric models is a better fit. As a consistency check, we found that the \sersic\ parameters in both the catalogues are largely in agreement with each other. Whenever the two component model was deemed to be a better fit, we replaced those galaxies in our fiducial simulation by the equivalent bulge+disc model. This resulted in about 10\% 
of the input galaxies being replaced. The residual multiplicative biases do not change significantly (see lower left panel of Fig.~\ref{fig:sensitivity_analysis}) for the tomographic bins. We ran another simulation where we replace the \sersic\ models with realistic galaxy images available via \texttt{COSMOSCatalog}. This resulted in about 30\% 
of the input galaxies being replaced. {Although the fraction of galaxies for which the \sersic\ models were replaced is small, this predominantly happens for bright galaxies, which carry relatively larger weights}. In both these cases, the ellipticity distributions are also consistent with our fiducial simulations as seen in Fig.~\ref{fig:ellip_dist}. As seen in the lower left panel of Fig.~\ref{fig:sensitivity_analysis}, no significant change in the multiplicative biases is found.  Thus, we once again conclude that there is no evidence of any significant model bias.

\subsection{Impact of redshift quality}
\label{sec:redshift_dependence}
As we have shown in~\cref{sec:calibration_results}, sample selection based on photometric redshifts modifies the ellipticity distribution, and therefore, the bias in the shear estimate. We divided the simulated galaxies into tomographic bins using photo-$z$ estimates that were derived from VST and VISTA (and CFHT) observations of the COSMOS field. The seeing of the COSMOS field, at least through the $r-$band filter, is better than most of the pointings (see Fig.~\ref{fig:psf_distribution}). As a result, the quality of the nine-band photo-$z$s for galaxies in the COSMOS field is likely to be better than most of the other fields, and hence may not be representative of the KV-450 data set. In our simulations, we have varied the seeing conditions for shape measurements, but in all cases, each galaxy is assigned the same redshift. The results from the `no-$z$' case may be seen as a limiting case of extremely poor photo-$z$s, which exceeds the tolerance limit for B1. We therefore have to investigate how sensitive the inferred biases are to the quality of the photo-$z$s.

To assess the impact of tile-to-tile variation of $n(z)$, we generate tomographic samples in the simulations based on the $ugri$ redshifts as in KiDS-450. These are of poorer quality compared to the nine-band redshifts used in this work. We see from the lower right panel of Fig.~\ref{fig:sensitivity_analysis} that the change in the bias is the largest for B1 and B5, and are well within the statistical errorbars for the other three tomographic bins. This is in line with what we would expect given the results of ~\cite{Wright2019}. They find that in the case of four-band photo-$z$s, B1 contains a significant population of intermediate to high redshift (spectroscopic) interlopers $(z_\text{spec} \gtrsim 0.5)$ that are not present in the nine-band photo-$z$ case. Similarly, they find that the four-band photo-$z$s are unable to define B5 robustly. Thus the change from nine-band to four-band photo-$z$s has the maximal impact on B1 and B5 naturally.

We also incorporate redshifts of significantly higher quality by assigning the COSMOS-30 redshifts from~\cite{Ilbert2008}. We find that the tomographic bins defined with COSMOS-30 redshifts give consistent bias with the exception of B1, although it is within the allowed limit of two per cent. It is intriguing that the the tomographic samples defined based on four-band photo-$z$ and COSMOS-30 redshifts have shear biases that are in very good agreement with each other, especially in B1 and B5. We note that this agreement is merely coincidental, despite the intersection of the tomographic samples based on four-band photo-$z$ and COSMOS-30 being similar to, or smaller than the intersection of the samples based on the other two pairs of redshifts. It is important to recognise that in all cases, the shape catalogues are the same and we only change our selection function used to define the tomographic bins.

More recent redshift estimates for galaxies in the COSMOS field are provided in the COSMOS2015 catalogue~\citep{Laigle2016}. However, unlike the ones available in~\cite{Griffith2012} which detects objects in the optical band ($F814W$),~\cite{Laigle2016} use near-infrared images from $YJHK_s$ bands to detect objects. This introduces selection effects for faint galaxies, particularly at high redshifts and does not allow for a fair comparison for our purposes, and therefore we do not show it here. We nevertheless  cross-matched these two HST-based catalogues and assigned the galaxies in our simulations more accurate redshifts. We found that the multiplicative biases were largely consistent in the first four bins, but a relatively large bias  (about +0.068) was seen in B5. We investigated this further and found that when 30-band photo-$z$s from COSMOS2015 are used, large galaxies were preferentially left out from B5. We attribute the increase in the multiplicative bias due to relative abundance of small galaxies (i.e., high $\mathcal{R}$) which exhibit high positive bias on average. 
We emphasise once again that this does not affect the KV-450 shear calibration, but is only intended to show the importance of including accurate and unbiased redshift information in the simulations that are used to calibrate the shear.

{A similar approach was tested for DESy1~\citepalias{Zuntz2018}, where they defined sub-samples in their simulations using 30-band redshifts and show that the bias obtained is different in the presence of redshift errors. Moreover, the difference in the bias increased with increasing redshifts, and exhibited a maximum difference of $\sim 0.025$ (see Fig. 15 of~\cite{Zuntz2018}). The DESy1-like redshift assignments were stochastic to mimic $n(z)$ in the different tomographic bins and did not capture any correlation with the galaxy properties. In contrast, we have selected our tomographic sample in a more realistic way in our fiducial simulations. We find in the lower right panel of Fig.~\ref{fig:sensitivity_analysis} larger shifts at low redshift bins compared to the high redshift bins. We also see from Fig.~\ref{fig:money_plot} that the difference in the bias values is largest at low redshifts when we include no redshift information (no-$z$) } 

If the change in ellipticity distributions is purely due to an evolving population of galaxies, then accounting for the difference in distributions is less important if the redshift errors are large enough to dilute any small differences. But as redshift errors shrink, or equivalently, as the tails of the $n(z)$ in the different tomographic bins vanish, which will be the case for future surveys, accounting for the difference in ellipticity distributions becomes important.

\section{Summary and Conclusions}
\label{sec:discussion}
Measuring the gravitational lensing shear field in a sufficiently unbiased way is absolutely necessary for an accurate estimation of the cosmological parameters. This is particularly of importance when different probes agree with the standard cosmological paradigm of $\Lambda$CDM, but disagree on the exact values of the model parameters. Various systematic errors and biases have to be checked to ensure that the results are accurate when precise measurements become possible with the ongoing and future surveys. 

In this paper, we have calibrated the shear measurements from the Kilo-Degree Survey using simulations with increased realism in comparison to previous studies. We have also provided a mathematical framework to understand the limitations of the calibration procedure itself and to provide context to our results, and point out where the additional systematic uncertainties come from. It is useful to classify the main results of this paper into those that are only applicable to our self-calibrating \lensfit\ analysis of KiDS, and those that are generic and relevant to all ongoing and future weak lensing surveys. We first summarise our results for KiDS and then list the more generic results.

By emulating the COSMOS field first, and extending it to a larger area by varying the observing conditions, we have simulated a mock KiDS+VIKING-450 dataset. A key improvement with respect to the previous study is the use of the catalogue from~\cite{Griffith2012} as our input to the simulation, which contains structural parameters for galaxies in the COSMOS field based on HST observations. The simulations are found to represent the observed data extremely well. By cross-matching our input catalogue to the catalogue obtained from the KiDS observation of the COSMOS field, we have assigned a nine-band photo-$z$ to each simulated galaxy. We have split the simulated galaxies, based on their assigned photo-$z$s, into tomographic samples of width $\Delta z_B = 0.2$, as used in the KV-450 cosmic shear analysis~\citepalias{Hildebrandt2019} and calibrated the shear for the individual tomographic galaxy samples. The self-calibrating version of \lensfit\ used to measure shear removes almost all of the raw biases internally and shows only small residual biases that may be considered to be consistent with zero within the allowed tolerance limit ($\sim 2$ per cent). We emphasise that an almost unbiased shear estimate is obtained internally for the tomographic bins, without any baseline correction applied from the simulations.

Compared to the previous calibration study undertaken by~\citetalias{FC17}, the uncertainty in the shear calibration has increased from 0.01 to 0.02. However, the values of the multiplicative bias for the first four tomographic bins are consistent with that reported in~\citetalias{FC17} within the allowed tolerance limits, and therefore do not call into question the accuracy of cosmological results from the KiDS-450 cosmic shear study~\citepalias{Hildebrandt17}. While the uncertainties on shear estimates have doubled, the use of nine-band photometry facilitates the use of a fifth high-redshift tomographic bin, thereby making the constraints on the $S_8$ parameter in the KV-450 cosmic shear analysis of~\citetalias{Hildebrandt2019} more robust. 

Through an extensive suite of image simulations, we have performed a variety of sensitivity tests and show that the shear calibration biases are controlled within the error budget of 0.02 per tomographic bin. In a few exaggerated test cases where the biases vary by more than 0.02 from the fiducial values, we show that such scenarios can be deemed unrealistic by comparing the properties of the simulated galaxies with the observed galaxies. It is worth noting that  the single largest systematic uncertainty in the $S_8$ estimate comes from the uncertainty in the shear calibration.

Several sources that affect the shear calibration have come to light during this work that require us to (not) make strong assumptions about the population of galaxies and model the astrophysical variations in the survey footprint. Further developments are being carried out on shape measurement algorithms for subsequent data releases of KiDS with the aim to make the shear measurements less sensitive to the properties of the simulated galaxy population and foreground contributions. Many of our results are applicable to other weak lensing surveys and other shape measurement methods. We conclude by listing the generic take-away messages below.

We show that the KV-450 dataset itself exhibits mild selection biases arising from the \sextractor\ detection step that are significant for faint galaxies. Because the detection bias occurs prior to measuring galaxy shapes, it is independent of the shear measurement algorithm, and is expected to be generic for all ongoing and upcoming lensing surveys. In the KV-450 tomographic bins, we estimate the contribution to multiplicative bias to be at worst $-0.005$, which is non-negligible, particularly as the survey volume increases. The magnitude-dependence, or equivalently, the S/N dependence of detection bias implies that, irrespective of the performance of the shear measurement algorithm, the image simulations used to infer the bias in shear must have a realistic distribution of magnitude or S/N to capture these biases accurately. However, because object detection bias appears to be rather insensitive to the types of galaxy  morphology simulated and the presence of any correlations between galaxy ellipticity and size,  image simulations do not have to be ultra-realistic in order to accurately calibrate object detection selection bias. They could be used in combination with metacalibration that can remove not all, but most of the raw shear biases.

The multiplicative bias for a sample of galaxies inferred from the simulations depends on the fidelity of the input catalogue and various other assumptions in the simulations.  To constrain the uncertainty in multiplicative bias from simulations further, it may be required to emulate the entire dataset, not just by including instrumental variations, but also by including the astrophysical variations such as stellar and galaxy densities. Increasing the realism of the input galaxy catalogue to emulate future surveys seems challenging, as it requires us to input the unknown. Large cosmological simulations, such as the Euclid flagship simulation\footnote{{http://sci.esa.int/euclid/59348-euclid-flagship-mock-galaxy-catalogue/}} could help in this regard.
By requiring the cosmological results obtained from two different shape catalogues calibrated using the same simulations be in agreement, it is possible to self-consistently verify if the simulations match the data to the desired accuracy. We base this claim based on our framework. Alternatively, shape measurement methods must be improved to reduce their sensitivities to the galaxy properties.

We argue that it is important to include the observable parameters that are used to define any galaxy sample. For a tomographic cosmic shear analysis, this corresponds to the photo-$z$ of the same quality as in the data~\citepalias{Zuntz2018}, or equivalently galaxy colours from the same set of filters. By assigning photo-$z$s to our input catalogue based on the matching to the observed data, we explicitly show that biases estimated for the low redshift bins can differ by more than 0.02, and hence calibrating shear simply based on signal-to-noise ratio and size distributions alone is insufficient. For a fully consistent high accuracy tomographic cosmic shear analysis, it will be essential to create multi-band image simulations in order to be able to apply photo-$z$ selections consistently to the simulations and the data.

\section*{Acknowledgements}
\newtext{The authors are grateful to the referee James Bosch for improving the structure of the paper}.
We thank Reiko Nakajima for providing the PSF model parameters extracted from {\sc PSFEx}, \newtext{Simon Samuroff for clarifying some aspects of the image simulations used in the DESy1 analysis} and the developers of \galsim\ for making their example scripts available. This work has made extensive use of the following Python packages: {\sc NumPy} (\url{www.numpy.org}), {\sc SciPy} (\url{www.scipy.org}) and {\sc AstroPy} (\url{http://www.astropy.org}). All plots in this paper are produced using {\sc matplotlib}~\citep[\url{https://matplotlib.org};][]{Hunter2007}.

The authors acknowledge support from: the Netherlands Organisation for Scientific Research (NWO) under grant numbers 639.043.512 (AK, HHo) and 614.001.103 (MVi); the European Research Council under grant numbers 279396 (MVi, RH), 647112 (CH) and 770935 (HHi); STFC under grant ST/N000919/1 (LM); the US Department of Energy under Award Number DE-SC0018053 (RH); the Deutsche Forschungsgemeinschaf under grants Hi 1495/2-1 (HHi) and Hi 1495/5-1 (HHi); the Alexander von Humboldt Foundation (KK).


{\it Author contributions:} All authors contributed to the development and writing of this paper. The authorship list is given in three groups: the lead authors (AK, HHo, LM, MVi), followed by two alphabetically ordered groups. The first group includes the key contributors to both the scientific analysis and the data products. The second group consists of those who have either made a significant contribution to the data products, or to the scientific analysis.

\bibliography{mendeley_master_ref}

\appendix
\section{Practical Estimator}
\label{app:estimator}
\subsection{Estimate from samples}
We can express the integral of $b[r_\text{sims};t_\text{sims}]$ over $w_\text{real}$ as a sum of integrals over mutually exclusive partitions of $\mathcal{D} = \bigcup\limits_{k} \mathcal{D}_k$, such that $p'_\text{sims}(\vec{D}) > 0$ and $s'(\vec{D}) > 0$ for every $\vec{D} \in \mathcal{D}$.

 \begin{equation} \begin{split}
 I &= \int\rmd\vec{D}  w_\text{real} (\vec{D}) b[r_\text{sims};t_\text{sims}](\vec{D}) \\ & = \sum_i \int\limits_{\mathcal{D}_i} \rmd \vec{D} w_\text{sims}(\vec{D}) b[r_\text{sims};t_\text{sims}](\vec{D}) \frac{w_\text{real}(\vec{D})}{w_\text{sims}(\vec{D})} 
 \end{split}. \end{equation}
We rewrite each of the three probability distributions with the Dirac delta function as
  \begin{equation} \begin{split}
  I = \sum_i \iint\limits_{\mathcal{D}_i \times \mathcal{D}_i} \rmd \vec{D}_0 \rmd \vec{D}_3  &  \delta(\vec{D}_3-\vec{D}_0) w_\text{sims}(\vec{D}_3) b[r_\text{sims};t_\text{sims}](\vec{D}_3) \\
  & \times \frac{ \int\limits_{\mathcal{D}_i} \rmd \vec{D}_1 \delta(\vec{D}_1-\vec{D}_0 ) w_\text{real}(\vec{D_1})  }{.\int\limits_{\mathcal{D}_i} \rmd \vec{D}_2 \delta(\vec{D}_2-\vec{D}_0 ) w_\text{sims}(\vec{D_2})}
  \end{split}. \end{equation}
To obtain a practical estimator for the above integral, the Dirac delta is replaced with a top-hat function which may be considered as smoothed delta function. For every $\vec{D} \in \mathcal{D}_i$ and $\vec{D}_0 \in \mathcal{D}_j$, we define the smoothed delta function as
 \begin{equation}
\delta(\vec{D}-\vec{D}_0) \longrightarrow \Pi(\vec{D}-\vec{D}_0) =  \frac{\delta_{ij}}{A_i} = \begin{cases} 1/A_k \;\;\; \text{if } \vec{D}, \vec{D}_0 \in \mathcal{D}_k \\ 0 \;\;\;\;\;\;\;\;\, \text{otherwise} \end{cases},
\end{equation}
where $\delta_{ij}$ is the Kronecker delta and $A_i := \int\limits_{\mathcal{D}_i} \rmd \vec{D}$, is the `area' of the partition $\mathcal{D}_i$. 

\begin{align}
\hat{I} &= \sum_i \int\limits_{\mathcal{D}_i} \rmd \vec{D}_0 \frac{w_{i,\text{real}}}{w_{i,\text{sims}}} \int\limits_{\mathcal{D}_i} \rmd \vec{D}_3 \frac{1}{A_i} w_\text{sims}(\vec{D}_3) b[r_\text{sims};t_\text{sims}](\vec{D}_3) \\
&= \sum_i \frac{1}{A_i} \int\limits_{\mathcal{D}_i}\rmd \vec{D}_0 \frac{w_{i,\text{real}}}{w_{i,\text{sims}}}\avg{b_i[r_\text{sims};t_\text{sims}]} \\
&= \sum_i \frac{w_{i,\text{real}}}{w_{i,\text{sims}}}\avg{b_i[r_\text{sims};t_\text{sims}]},
\label{eq:practical_estimator}
\end{align}
where $w_{i,\text{real}} = \int\limits_{\mathcal{D}_i}^{}\rmd\vec{D} w_\text{real}(\vec{D}) = \int\limits_{\mathcal{D}_i}^{}\rmd\vec{D} p'_\text{real}(\vec{D})s'(\vec{D})$ and similarly for the simulations.

\subsection{Bias in the estimator}
The estimator derived in Eq.~\ref{eq:practical_estimator} uses galaxy samples taken from the observed data and from the simulations. Thus,  the shot noise due to discrete galaxy counts can make the estimate biased even if the underlying populations are identical. In this sub-section, we show that the systematic bias can be neglected in comparison to the statistical errors if the binning includes a sufficient number of galaxies.

Let two random variables $N_{i,\text{sims}}$ and $N_{i,\text{real}}$ denote the number of galaxies (or sum of the lensing weights) in the simulations and observed data respectively that fall in the $i^\text{th}$ bin. The samples are independent, hence we can write the average of ratios as
\begin{equation}
\avg{ \frac{N_{i,\text{real}}}{N_{i,\text{sims}}} } = \avg{N_{i,\text{real}}} \avg{\frac{1}{N_{i,\text{sims}}}} = \frac{{n_{i,\text{real}}}}{{n_{i,\text{sims}}}} + r_i,
\end{equation}
where the $n_i$'s refer to the averages and $r_i$ is the deviation of the average from the ratio of averages. If $p_\text{sims} = p_\text{real}$, $\avg{N_{i,\text{sims}}} = \avg{N_{i,\text{real}}}$. The ratio of the averages, however, is generally not equal average of the ratios. 

For simplicity, consider $N_{i,\text{sims}}$ to be a random variable that follows a Poisson distribution with a mean $n_{i,\text{sims}} \gg 1$. Then
\begin{equation} \begin{split}
\avg{\frac{1}{N_{i,\text{sims}}}} &\approx \avg{\frac{1}{N_{i,\text{sims}}+1}} = \frac{1-e^{-n_{i,\text{sims}}}}{n_{i,\text{sims}}} \\ &= \frac{1}{\avg{N_{i,\text{sims}}}} - \frac{e^{-\avg{N_{i,\text{sims}}}}}{\avg{N_{i,\text{sims}}}}.\end{split}
\end{equation}
This result is fairly independent of the distribution of $N_{i,\text{sims}}$ if its mean is much higher than unity. In this limit, the difference between the average of the inverse and inverse of the average scales down exponentially as the mean increases. The resulting bias in the estimator $ \sum_{i} r_i \avg{ b_i[r_\text{sims};t_\text{sims}]}$, is
\begin{equation}
r_i \sim \frac{\avg{N_{i,\text{real}}}}{\avg{N_{i,\text{sims}}}}e^{-\avg{N_{i,\text{sims}}}} \ll \sqrt{N_{i,\text{sims}}}.
\end{equation}
Thus, the bias in the estimator is negligible when there are sufficient numbers of galaxies in each bin.

\section{Detailed comparison between the KV-450 data and simulations}
\label{app:lf_comparison}
\begin{figure}
\includegraphics[width=0.5\textwidth]{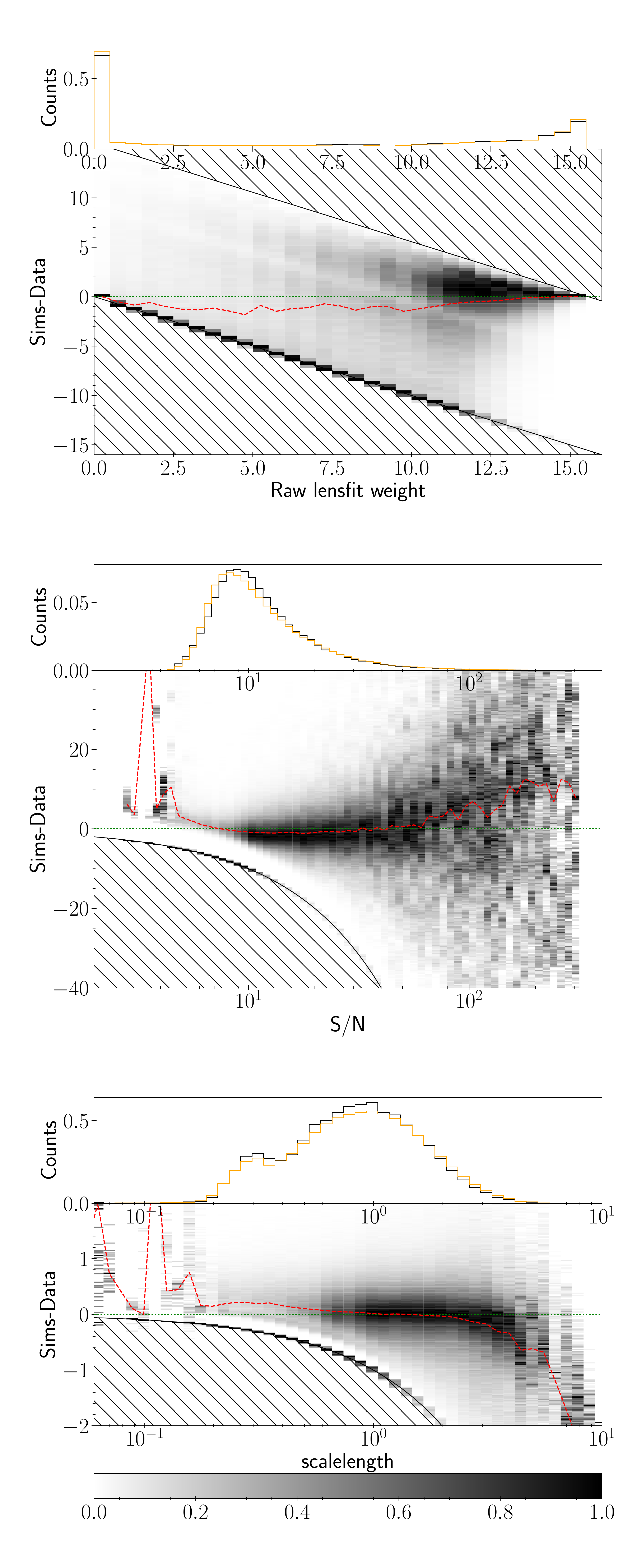}
\caption{Object-by-object comparison of \lensfit\ quantities, similar to Fig.~\ref{fig:compare_individuals_heatmap_SE}. See text in~\cref{app:lf_comparison} for details.}
\label{fig:compare_individuals_heatmap_LF}
\end{figure}
We compared the output quantities of the \lensfit\ catalogue for the COSMOS tile with our simulations when we used similar PSFs. The upper panes in Fig.~\ref{fig:compare_individuals_heatmap_LF} show the unweighted histogram of \lensfit\ weights, while the lower panes show the weighted histogram of the signal-to-noise ratio and the \lensfit-measured scale length along the major axis of the galaxy. The lower panels in each of the three figures show 2-dimensional histograms with the difference in $X$ between the simulations and data along the vertical axis and the $X$ in the data along the horizontal axis. The red lines indicate the median value of the differences. As in Fig.~\ref{fig:compare_individuals_heatmap_SE}, each vertical column is normalised such that each peak in the 1-dimensional slices of the 2-dimensional histogram is normalised to unity. This improves the histogram visually, allowing us to see the contrast across the full parameter range.  The shaded regions correspond to unphysical regions, due to the corresponding quantities in the simulations taking negative values.

As the recalibration procedure for the COSMOS tile is carried out at a catalogue level with the rest of the galaxies in the G12 patch of the KV-450 dataset, the weight depends more than just on the galaxy properties. A one-to-one comparison might not be fair in this case. Although the weights of the individual object are not exactly in agreement with each other, the overall weight distribution, and the weighted distribution of other observables appear to be in good agreement.

The scatter in the S/N increases with S/N. This is so because the estimate of there noise term $N$ in the ratio $S/N$ itself is noisy which is amplified by the high value of the signal $S$. For a random scatter of $N$ around its true mean, the median (and mean) are biased towards positive values. However, as there are very few galaxies at such high values of S/N, it does not affect our sample significantly. Similarly, there is good agreement between the measured sizes, particularly for the bulk of the galaxies. 


\section{PSF models}
\label{app:PSF_modelling}

\begin{figure*}
\includegraphics[width=\textwidth]{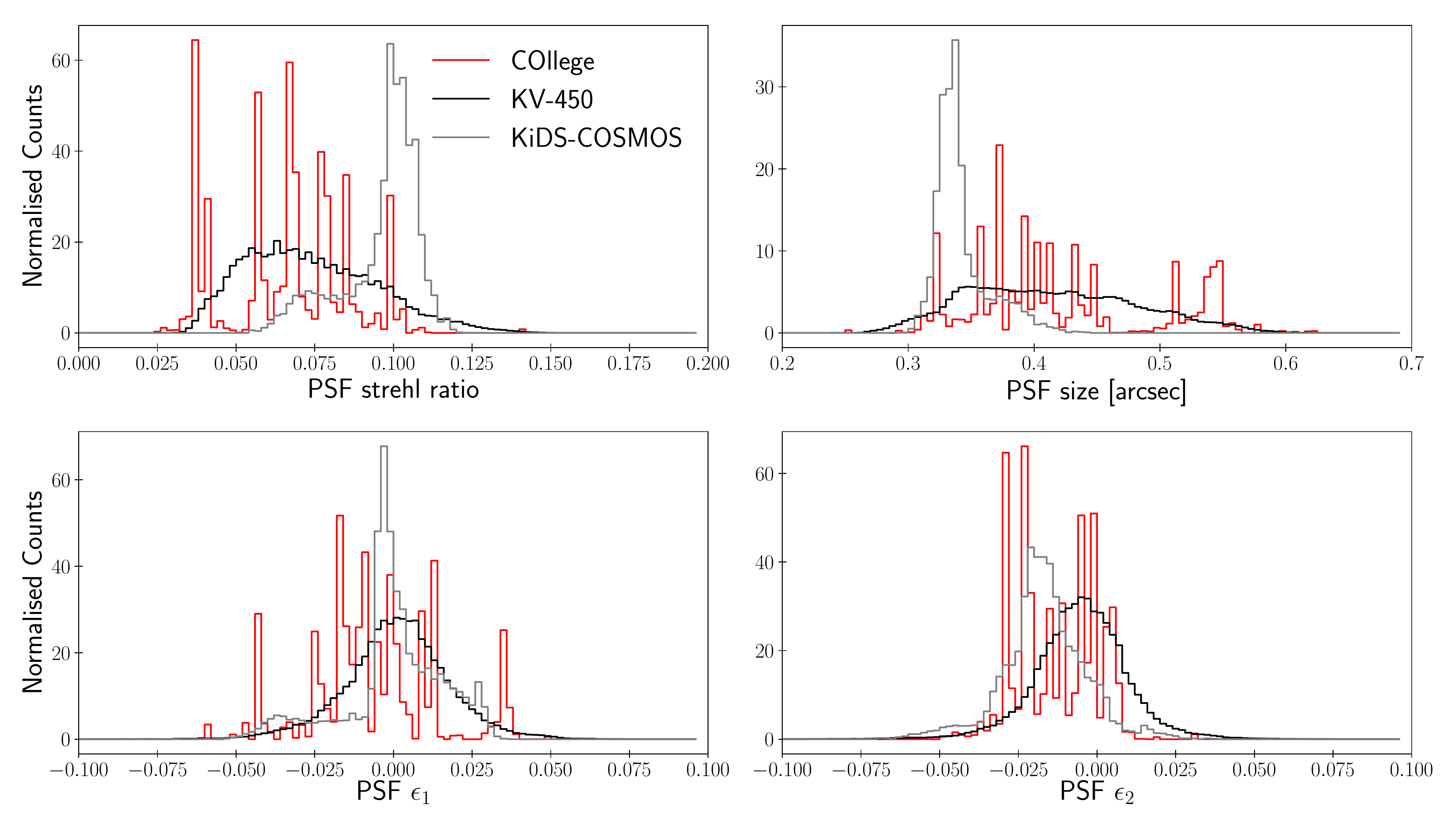}
\caption{Histograms of the PSF parameters for the simulations using the \college\ pipeline (red), entire KV-450 data (black) and for the COSMOS field alone (grey).}
\label{fig:psf_distribution}
\end{figure*}

The PSF in the observations is modelled by fitting an elliptical Moffat profile. Although the atmosphere at the VST site is fairly stable, to account for the time-dependent seeing conditions and spatial variation of the PSF, the model parameters are obtained for each exposure and for each CCD separately by running {\sc PSFEx}~\citep{psfex} on KiDS-DR1/2 data. To capture the realistic temporal variation in the PSF, we select a series of PSF parameters that correspond to five consecutively observed exposures. These are used for the different exposures that make up a single pointing. However, any spatial variation is ignored altogether and all galaxies in a given simulated exposure are convolved with the same PSF for simplicity. To account for long-term variations in seeing conditions that occur throughout the survey, we chose the PSF parameters from 13 KiDS pointings, resulting in a total of 65 different PSFs. These correspond to the same 65 PSFs used in \citetalias{FC17}, which themselves are a subset of the observed PSFs\footnote{A benign error was found after the~\citetalias{FC17} analysis where the PSF ellipticity from {\sc PSFEx} was not propagated correctly to the ellipticity of the PSF in the simulations, but increased by almost a factor of 2. Since the ellipticities themselves are small, the effect is minor and the simulations are fully self-consistent. We therefore did not correct this error in this analysis}. 
We compared the residual multiplicative bias in the tomographic bins by running the \citetalias{FC17} pipeline with only 25 out of the 65 PSFs, and the results were essentially the same.

For reference Fig.~\ref{fig:psf_distribution} compares the distribution of the PSF parameters in the simulations (red) and in the actual data (black). We also show the distribution of parameters for the VST $r$-band observations of the COSMOS field (grey).

To quantify the PSF size and shape we do not use the Moffat model parameters, but instead we use quantities derived from weighted quadrupole moments (see Eq.~(6) in \citetalias{FC17}) with a Gaussian weight function of size 2.5 pixels. The pseudo-Strehl ratio is defined as the fraction of light in the central pixel of the PSF and is available in the \lensfit\ output catalogue. The distributions of the PSF parameters in the simulations are spiky because we only use a small set of PSFs. The broadening around the peaks is caused by the gaps between the chips; at those locations only four exposures contribute.

\end{document}